\RequirePackage{fix-cm}
\documentclass[smallextended]{svjour3}       

\usepackage{booktabs} 
\usepackage{graphicx}
\usepackage{amsmath}
\usepackage{algpseudocode}
\usepackage[ruled]{algorithm2e} 

\SetCommentSty{mycommfont}
\usepackage{amssymb}
\usepackage{subcaption}
\usepackage{color}
\usepackage{array}
\usepackage{multirow}
\usepackage[space]{grffile}

\makeatletter
\newcommand\subparagraph{%
	\@startsection{subparagraph}{5}
	{\parindent}
	{3.25ex \@plus 1ex \@minus .2ex}
	{-1em}
	{\normalfont\normalsize\bfseries}}
\makeatother
\usepackage{titlesec}
\let\subparagraph\relax

\usepackage{titlesec}

\titleformat{\paragraph}
{\normalfont\normalsize\bfseries}{\theparagraph}{1em}{}
\titlespacing*{\paragraph}
{0pt}{3.25ex plus 1ex minus .2ex}{1.5ex plus .2ex}

\graphicspath{{./}}

\begin{document}

\title{Using network motifs to characterize temporal network evolution leading to diffusion inhibition}


\author{Soumajyoti Sarkar         \and
        Ruocheng Guo \and
        Paulo Shakarian 
}

\institute{Soumajyoti Sarkar \at
		Arizona State University \\
	\email{ssarka18@asu.edu}           
	\and
Ruocheng Guo \at
Arizona State University \\
\email{rguosni@asu.edu} 
\and
Paulo Shakarian\at
Arizona State University \\
\email{shak@asu.edu} 
}


\date{Received: date / Accepted: date}

\maketitle

\begin{abstract}
Network motifs are patterns of over-represented node interactions in a network which have been previously used as building blocks to understand various aspects of the social networks. In this paper, we use motif patterns to characterize the information diffusion process in social networks. We study the lifecycle of information cascades to understand what leads to saturation of growth in terms of cascade reshares, thereby resulting in expiration, an event we call ``diffusion inhibition''. In an attempt to understand what causes inhibition, we use motifs to dissect the network obtained from information cascades coupled with traces of historical diffusion or social network links. Our main results follow from experiments on a dataset of cascades from the Weibo platform and the Flixster movie ratings. We observe the temporal counts of 5-node undirected motifs from the cascade temporal networks leading to the inhibition stage. Empirical evidences from the analysis lead us to conclude the following about stages preceding inhibition: (1) individuals tend to adopt information more from users they have known in the past through social networks or previous interactions thereby creating patterns containing triads more frequently than acyclic patterns with linear chains and (2)  users need multiple exposures or rounds of social reinforcement for them to adopt an information and  as a result information starts spreading slowly thereby leading to the death of the cascade.
Following these observations, we use motif based features to predict the edge cardinality of the network exhibited at the time of inhibition. We test features of motif patterns by using regression models for both individual patterns and their combination and we find that motifs as features are better predictors of the future network organization than individual node centralities.    
 
\end{abstract}

\section{Introduction}

Information sharing among users is a common phenomenon in social media and such information can be in the form of text, photos or links to other sources. When such a sharing process goes on for a long time, we observe a cascade of reshares. Modeling information diffusion by studying such cascades in an attempt o predict whether the cascade would go viral or reach a certain size has been a very popular research problem in the social network and machine learning community \cite{domingos2005,Kempe2003,Shakarian2012,Guo16}.
The problem has been studied mainly in two dimensions: (i) modeling the network diffusion process and then using learning algorithms to predict the diffusion links in future and its related problems like inferring the final size of the cascade and the associated temporal attributes \cite{Cui2013,Gomez2011} (ii) using social network analysis to study the network structure of the cascades and thus using network topology to understand the diffusion process \cite{Shak2014,Kitsak10,Pei14}. However one of the relatively less discussed topics of information diffusion in the cascading mechanism has been the area of diffusion inhibition, specifically to identify or model the characteristics of the network that could indicate when the cascade would stop progressing in terms of reshares. In this paper, we define ``inhibition'' time as the final time phase during which the cascade starts to decay in terms of reshares and is unable to regain momentum thereafter leading to a complete stop in reshares. 

As opposed to a lot of previous research which focuses on individual nodes as a factor of influence in the information diffusion process \cite{Gomez20103,Gomez2011}, in this paper, we focus on groups of nodes as drivers of the cascade progress to understand how chained influence adds up to impact the diffusion process. We observe time periods in which the cascades grow and then characterize the dynamics of node interactions using interaction patterns to reason what ultimately hampers the growth of the cascade leading to a complete stop in reshares. We use \textit{network motifs} \cite{Milo824}, which are defined as induced subgraphs in a network occurring more frequently than would in a random network with same network properties, to study the pattern of interactions among the nodes in a temporal cascade setting. We use motifs to analyze the stages of the cascade lifecycle where we define an ideal cascade lifecycle as being generated by a sequence of phase transitions starting with a slow growth phase to attaining the maximum growth rate and finally attaining its saturation point to finally die. 

Studying what causes inhibition in cascades can be important in
determining when an incipient cascade would start decaying and in
situations that demand controlling cascade failure, can be instrumental in removing bottlenecks in the diffusion process. However, identifying the inhibition phase of the cascade early-on has a few challenges due to the way we define ``inhibition'' as being the last stage before expiration, in this paper: there is no ground truth data available that would provide us with the exact time of cascade decay without adopting retrospective analysis to observe the entire cascade lifecycle. We describe the procedure to determine the inhibition periods for the population of cascades used in this study and the related assumptions in Section~\ref{sec:steep_inhib}. Once we determine these approximate inhibition periods of the cascades, we use network motifs to analyze the network structures in the periods preceding inhibition. To this end, the main contributions and conclusions of this paper are as follows:

\begin{itemize}
	\item We use undirected network motifs to study the structural properties of the network obtained from the cascade links and the historical  diffusion links. Subsequently, we observe that compared to the steep region of growth in the cascade lifecycle where there is more abundance of linear chain motifs like \includegraphics[scale=0.04]{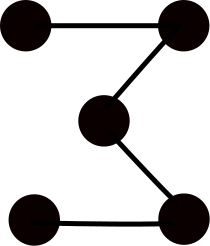}, in the inhibition phase there is lesser concentration of linear chains and more significant motifs patterns like \ \includegraphics[scale=0.04]{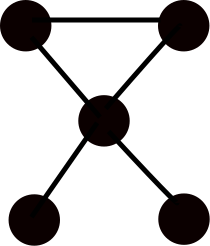} \ and \ \includegraphics[scale=0.04]{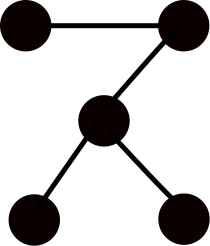} \  and \includegraphics[scale=0.04]{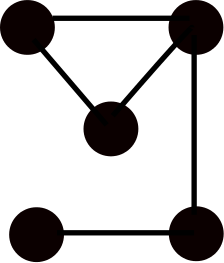} emerge that drive the cascade progress. It suggests that in the phase of high adoption, individuals do not wait to share messages just from their social network links or within some circles, rather they adopt them fast from people who shared them recently even if they may not know them. On the contrary, in the inhibition phase, people tend to share messages either from one central person or someone they have known or interacted with in the past.
	
	\item We analyze the formation of patterns with triads towards the inhibition stage. More frequent appearances of such patterns in the vicinity of the inhibition stage compared to the phase of maximum growth suggest that as the cascade approaches inhibition, users need multiple exposures or rounds of social reinforcement for them to adopt an information. As a result,  information starts spreading slowly thereby leading to the death of the cascade.
	
	\item To evaluate the appearance of motifs as predictors of the network structure towards the inhibition stage, we predict ahead of time the edge cardinalities of the temporal network in the inhibition stage.  We compare the performances of motifs on the prediction problem with network node centralities in an attempt to observe how structural patterns compare with individual nodal functions. We observe that the best performance obtained from a combination of motif patterns outperforms those from the combination of centralities. It suggests that motifs as network micro-structures can explain the future network diffusion process better than individual nodal features by being better predictors of the network topology.
	
\end{itemize}

\noindent \textbf{Roadmap}: The rest of the paper is organized as follows: we introduce notations, the concept of information cascades and inhibition intervals formally in Section~\ref{sec:tech_prelim}. We describe the dataset used for our analysis and the temporal networks used for analysis of the temporal evolution of the cascades in  Section~\ref{sec:dataset}. Following this, we describe the notion of Network Motifs in Section~\ref{sec:motifs}. We then present our main analyses on diffusion towards the inhibition stage by studying motif based dynamics in Section~\ref{sec:motif_feat}. Finally, we present a prediction study to understand the predictive power of motifs in forecasting the network structure organization in Section~\ref{sec:prediction}. We end by  performing an explanatory analysis of the motif dynamics  on another dataset in Section~\ref{sec:flixster_exp} to see the extent to which  the dynamics of motif patterns generalizes across different media platforms.

\section{Related Work}
Cascade growth prediction \cite{Cheng2014} has been an area of work that has attracted widespread attention in the past few years owing to its applications. Modeling the trajectory of the cascade growth has been studied in \cite{Cui2013}, where the authors build a model to understand the underlying diffusion process and in the event predict the future course of the cascade lifecycle. Using network diffusion models to understand the cascade properties has also been a crucial component in many studies \cite{Guo16}. Numerous methods have been highlighted in \cite{Shak2015} surrounding diffusion models for social networks. One aspect of cascades that  
has gained popularity has been to predict the viral spread of the cascade in terms of reshares. To this end, \cite{Shuang13} used Hawkes model to measure time-varying social influence and to model viral network diffusion. Applications of Hawkes process dates back to studies describing self-exciting processes of earthquakes \cite{Ogata1998}. Following their methodology and that used in \cite{Zhao15,Shuang13}, used Hawkes intensity and a data likelihood based approach to come up with an approximate inhibition time for each cascade, a brief summary of which has been later described in this paper. Previous work on what limits the size of the cascade has been studied in \cite{SteegGL11} where the main motivation was to see how repeated exposures limited the final size of the cascade.  Since then, there has been relatively few studies that focus on what causes cascades to completely stop and are there any specific attributes in the early stages that could tell whether the cascade would experience decayed growth

Graphlets based characterization of directed networks has been used to understand the connectivity patterns in networks as done in \cite{Sarajlic16}. Similarly, cliques \cite{Katona11} and dense communities \cite{Zhao2011} have been used to find cohesive subgroups within networks to group similar users. However in most of these subgraph based analyses, an open research problem is to extract subgraphs or patterns that are considered significant where ``significance'' is defined using some statistical measure. The problem is difficult due to the combinatorial explosion issue \cite{Hocevar14,Ciriello2008} of the possible patterns of subgraphs within a group of nodes and the computational complexities of subgraph matching algorithms\cite{Ullmann1976}. In this paper, we focus on using network motifs to assess the structural properties of the network that help us understand how nodes collaborate with each other during different phases preceding the end of the cascade lifecycle. The idea of using network motifs to characterize networks goes back to the study \cite{Alon2002}, where authors use the feedforward loop (FFL) motif pattern and the Single Input Module (SIM) pattern to understand how transription factors regulate the genes in \textit{E.Coli}. Such motifs have also been used in studying collaboration networks of authors \cite{Chakraborty2015}, biological genes \cite{Alon2007}, human communication networks \cite{Liu2013} and thereby such studies give a hint as to why motifs form core components in studying time varying networks. Along this line of work, temporal motifs, which forms the basis of one of our measures used in this paper, have been studied in the context of communication networks \cite{Kovanen2013} in which the authors use the concept of motif patterns distinguished by times of events to study homophily in networks. Studying \textit{intermediate-scale} substructures in networks, which represent the denser part of the original network structure has been an area of interest in many applications in network theory . Finding the $k$-core of the network \cite{Dor2006}  or dense communities in a network has been considered useful in understanding the flexibility and adaptability of the network . Previous studies have shown that it is these set of densely connected nodes that act as influential nodes in the network diffusion process and that the extent to which network diffusion occurs depends largely on the core structure of the network \cite{Kitsak10}.

In this paper, we use network features in the stages preceding inhibition to predict the attributes of the inhibition network structure of cascades. Such problems of structure prediction has been a subject of research mainly from a diffusion network inference perspective \cite{Gomez2011}. Using topological features of the network structure has been used in \cite{Fire2011} to predict the future links of the network. In this paper, we adopt motif based network features to observe whether the motifs' appearances are indicators of how the nodes would organize themselves to form the network during inhibition.

\section{Preliminaries and Notations}
\label{sec:tech_prelim}

\subsection{ Information cascades}
We will use the symbol $C$ to denote an arbitrary information cascade (i.e. a microblog that  spreads in the social network). Formally, a cascade can be represented by a sequence of $3$-tuples denoted by $(u, v, t) $ which indicates that the microblog was reshared by $v$ from the user $u$ at time $t$. We denote the sequence of reshare times for $C$ as $\tau_C$ = $\langle 0, \dots, t, \ldots T_C \rangle$ ordered by time, where $T_C$ denotes the time difference between the first and last posting. Here $t \in \tau_C$ denotes the reshare time offset by the starting time for that cascade. We will drop $C$ from all notations when they are applicable for all cascades. We will often use the notation $\tau'$ to denote a subsequence of $\tau$. In our work, we create the sequence of subsequences $\tau$ = $\langle \tau'_1, \ldots, \tau'_{\mathcal{Q}} \rangle$, ordered by the starting time of each subsequence, where we denote $ \mathcal{Q}$ to be the number of subsequences for $C$, which would vary for different cascades. The method of splitting the cascade $C$ into a sequence of such subsequences $\tau'$ has been described in detail in Section~\ref{sec:dataset}. We slightly abuse the terms \textit{interval} and \textit{subsequences} in this paper - an \textit{interval} is considered here as a generic term for a range of time points not subject to any constraints whereas we define \textit{subsequences} formally later in Section~\ref{sec:dataset} as a contiguous subset of $\tau$ and are subject to a set of constraints.

\begin{table}[!t]
	\centering
	\renewcommand{\arraystretch}{1}
	\caption{Table of Symbols}
	\begin{tabular}{|p{2cm}|p{8cm}|}
		\hline 
		{\bf Symbol} & {\bf Description}\\ 
		\hline\hline
		C           & Information cascade \\
		\hline
		$t $ & A reshare time instance
		\\
		\hline
		$\mathcal{R}_C $ & Total number of reshares in $C$
		\\
		\hline
		$T_C $ & Total span of cascade C in minutes (time difference between the first and the last reposting)
		\\
		\hline
		$S_C $ & Cumulative size of cascade $C$ equal to the total number of individuals who participated in $C$.
		\\
		\hline
		$S_{t, C} $ & Size of cascade $C$ or number of individuals after the reshare at time $t$.
		\\
		\hline
		$\tau_C $ & Sequence of reshare times within a cascade ordered by time. 
		\\
		\hline
		$\tau'_C $ & Subsequences within $\tau$ for the cascade $C$.
		\\
		\hline
		$\mathcal{Q}$ & Number of subsequences in a cascade.
		\\
		\hline
		$V^{\tau'}_C $ & Nodes which participated in the cascade $C $  in the subsequence $\tau'$. \\
		\hline
		$E^{\tau'}_C$ & The social interactions between pairs of individuals denoted by $e$=$(i,j)$ in the subsequence $\tau'$. \\    
		\hline
		$k$ & Node size of motif pattern  \\ 
		\hline
		$\mathcal{M}_k$ & Global family of motif patterns of size $k$ \\ 
		\hline
		$M_k$ & A motif pattern $\in \mathcal{M}_k$  \\  
		\hline
		$\{M_k\}$ & Set of motif instances belonging to family of the pattern $M_k$  \\
		\hline
		$m$ & A motif instance generalized for any pattern $M$.  \\   
		\hline 
	\end{tabular}
	\label{tab:table0}
\end{table}

\begin{figure}[t!]
	\centering
	\hfill
	\minipage{0.5\textwidth}%
	\includegraphics[width=6.5cm, height=3.5cm]{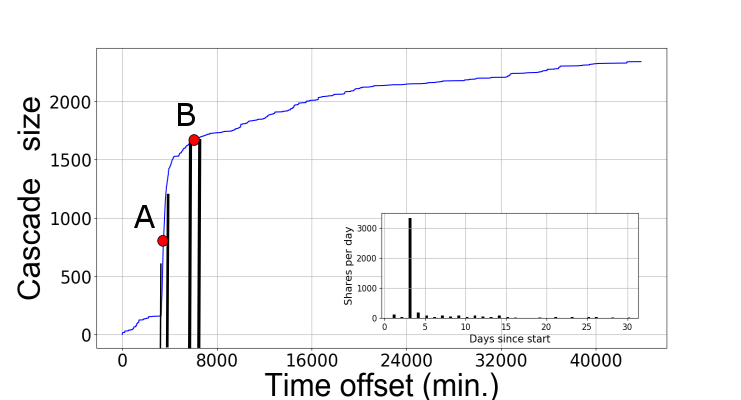}
	\hspace*{3.25cm}\subcaption{}
	\label{cascA}
	\endminipage 
	\hfill
	\minipage{0.5\textwidth}
	\includegraphics[width=6.5cm, height=3.5cm]{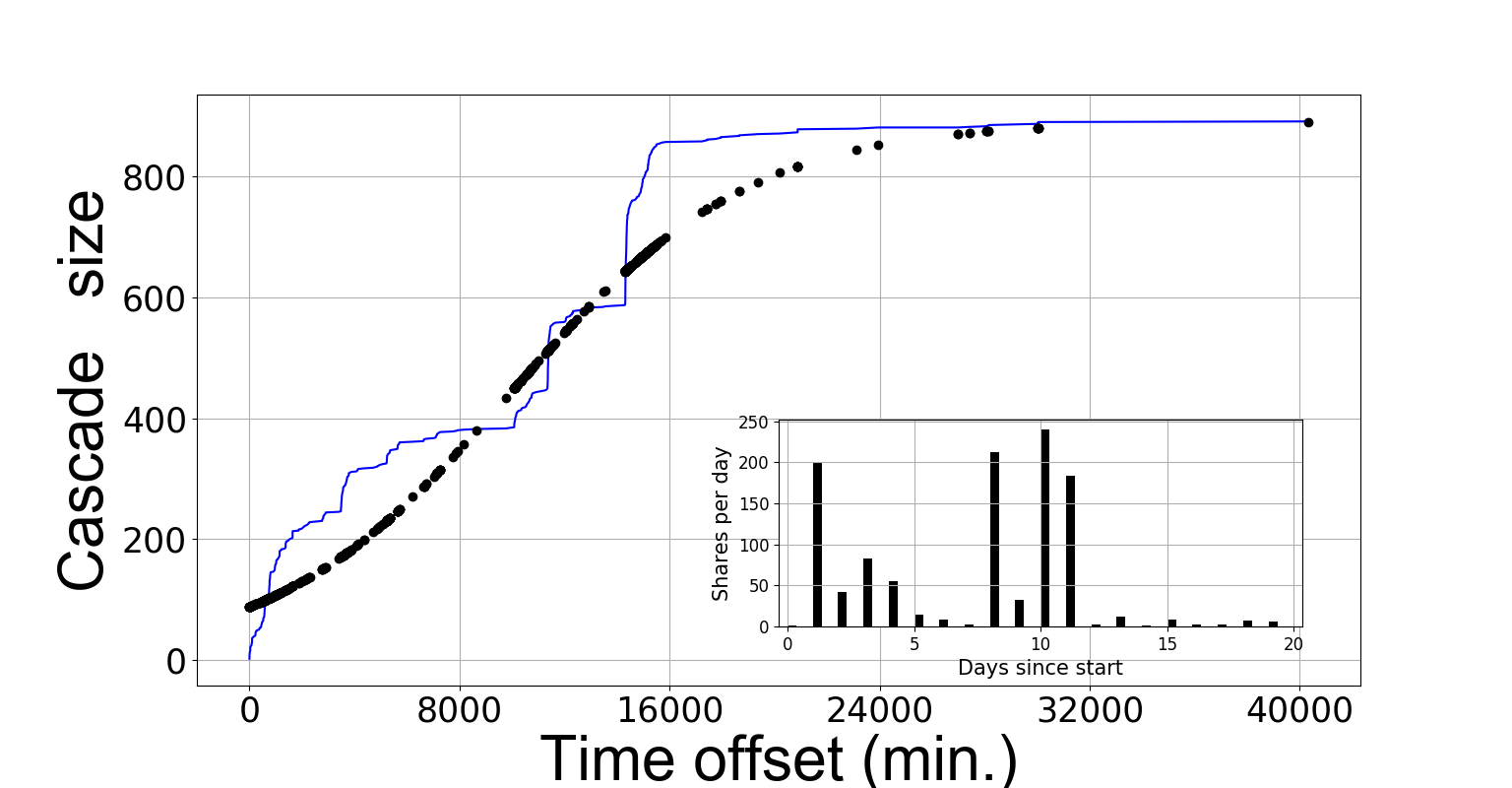}
	\hspace*{3.25cm}\subcaption{}
	\label{cascB}
	\endminipage 
	\hfill
	
	\minipage{0.5\textwidth}%
	\includegraphics[width=6.5cm, height=3.5cm]{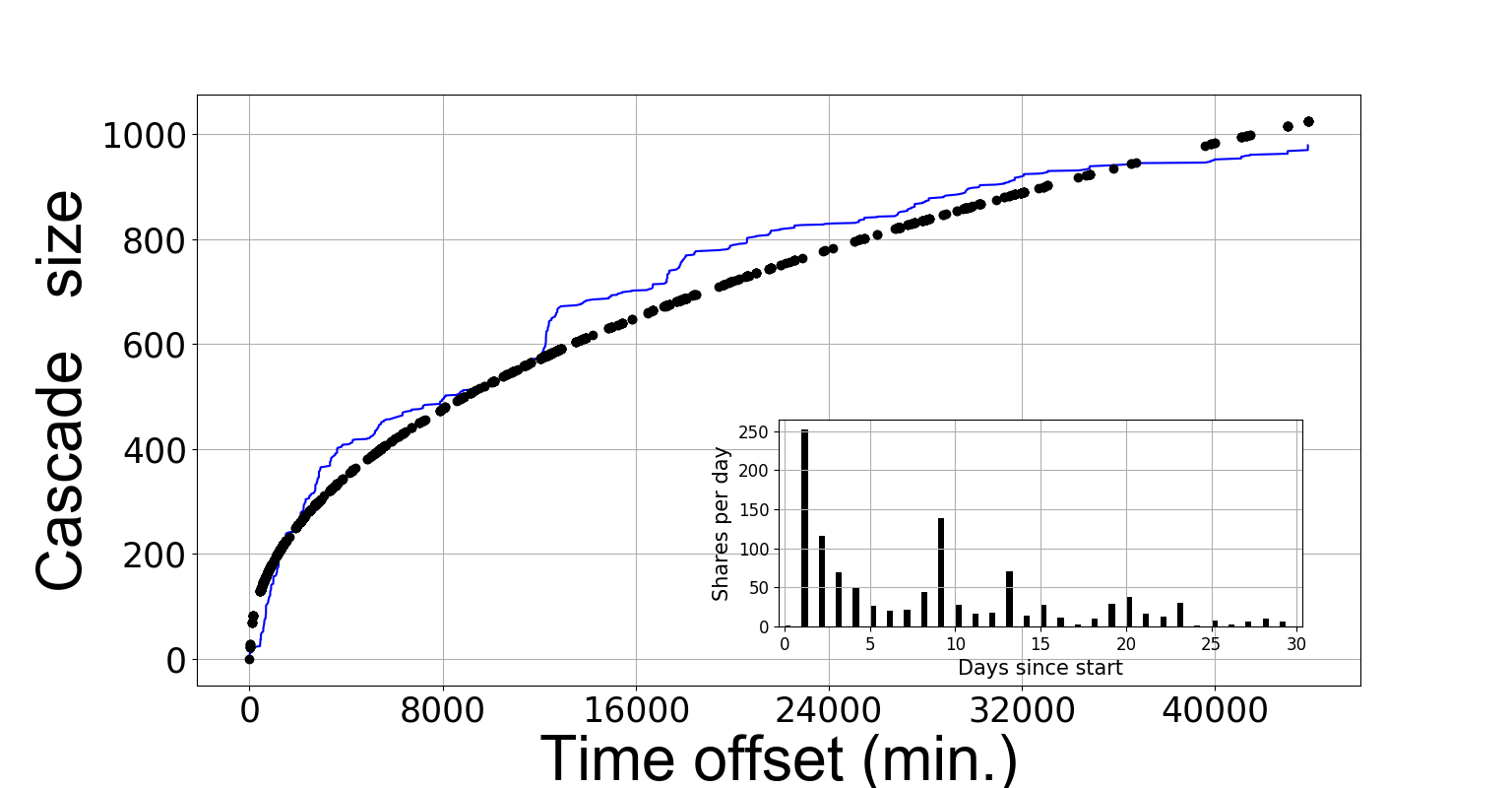}
	\hspace*{3.25cm}\subcaption{}
	\label{cascC}
	\endminipage 
	\hfill
	\minipage{0.5\textwidth}
	\includegraphics[width=6.5cm, height=3.5cm]{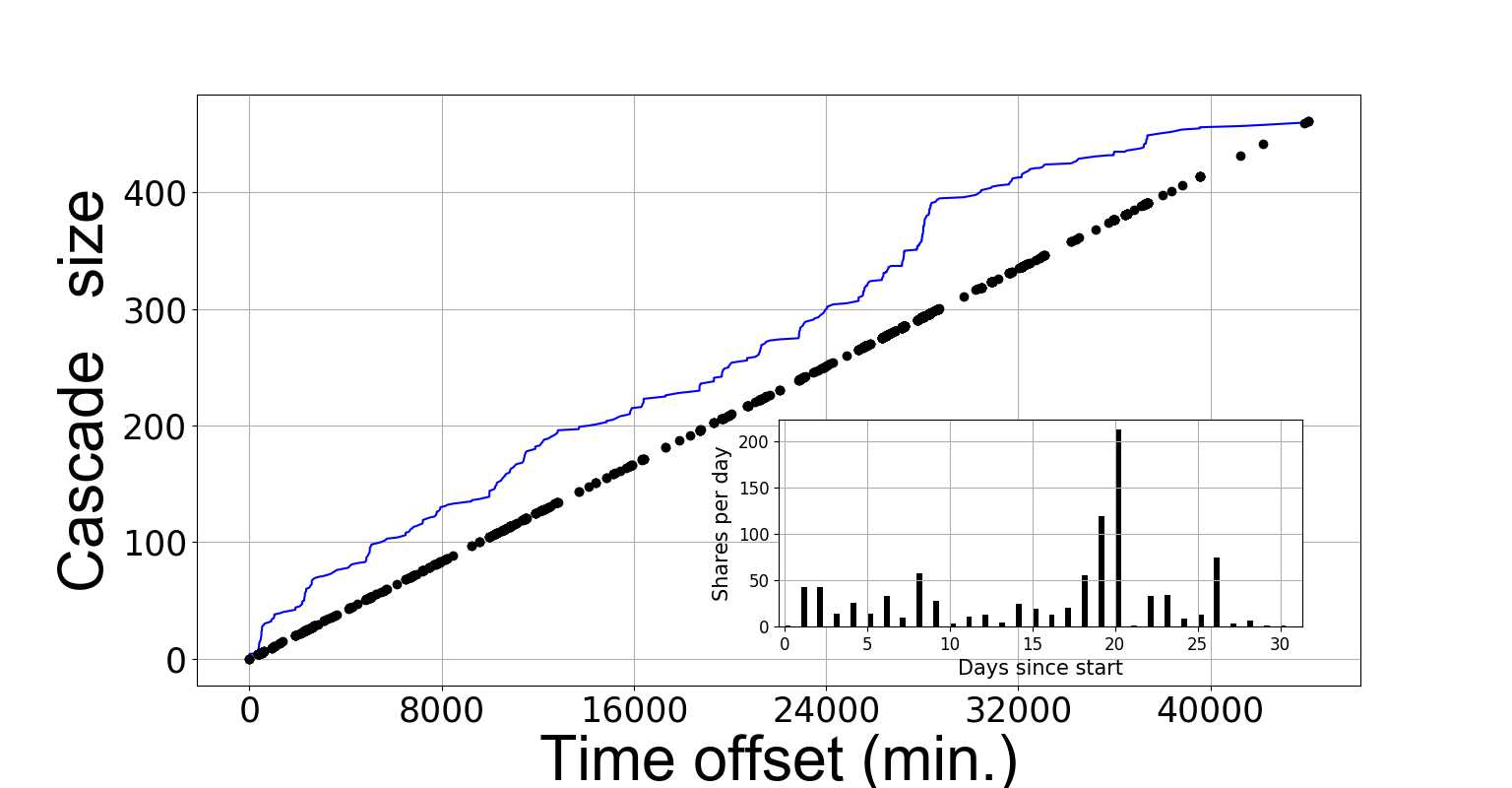}
	\hspace*{3.25cm}\subcaption{}
	\label{cascD}
	\endminipage
	\hfill
	\caption{(a) \textit{Growth curve} depicting the progress of a Type I cascade. Points A and B marked in red denote $t_{steep}$ and $t_{inib}$ respectively and the subsequences bounding them denote $\tau'_{steep}$ and $\tau'_{inhib}$ respectively. (b) Type II curve with multiple phases of step growth fitted to and ideal logistic function. Note that this is not a perfect fit as would be obtained from Type I cascades (c) Type III curve fitted to a concave increasing function (d) Type III curve fitted to a straight line curve. The blue curve denotes the cascade curve while the black dotted curve denotes the estimated curve fit. The subplots inside depicts the number of shares per day over its lifecycle for that example cascade. These cascades are part of the Wiebo dataset described in Section 4 and which has been used as the main dataset for the experiments in addition to the Flixster dataset used for explanatory analysis. }
	\label{fig:types_cascades}
\end{figure}

Rules for identifying cascade stages mapping the intervals of maximum growth which we call henceforth the ``steep" interval and the ``inhibition" interval are generally not well defined in the context of information cascades. In order to be able to compare the diffusion processes happening at different stages of the cascade lifecycle, the first problem we address is to identify 2 subsequences: $\tau'_{steep}$ during which the cascade experiences the maximum rate in reshares and $\tau'_{inhib}$ during which behavior change occurs in the context of cascade adoption after which the cascade fails to regain any momentum in growth and finally expires. We adopt a 3-step procedure to detect these 2 intervals (Appendix A1) and we use them to compare the diffusion process from the perspective of network structures exhibited during and preceding these intervals. However, in information cascades we do not generally see  smooth transitions in all cascades leading to the \textit{steep} and \textit{inhibition} intervals. This is explained by empirical observations where we find three types of Growth curves shown in the plots in Figure~\ref{fig:types_cascades}, each of which depict cumulative cascade size over time $t$. Apart from Type I cascades as illustrated in Figure~\ref{fig:types_cascades}\subref{cascA}, which depicts an ideal logistic S-shaped growth pattern, most transitions in the cascade lifecycle are not smooth. One such example is given by Type II cascades illustrated by Figure~\ref{fig:types_cascades}\subref{cascB} which are characterized by multiple temporal patterns of growth with recurring attention \cite{Cheng2016}, a problem which has been previously studied in the context of time series where convex and concave patterns are used to fit the phases within the lifecycle \cite{Yu15,Xie2007}. Since our focus in this paper is in understanding the \textit{steep} and \textit{inhibition} subsequences from a network analysis perspective, we avoid such rigorous pattern fitting mechanisms and instead use point processes to model cascade generation and identify the subsequences mapping the two aforementioned events (described in the Appendix A1). In Figure~\ref{fig:types_cascades}\subref{cascA} we show an example plot of \textit{Growth curve} and show three major types we identified empirically:

\begin{enumerate}
	\item \textit{Type I cascades}: Cascades which follow an ideal logistic function as shown in Figure~\ref{fig:types_cascades}\subref{cascA}.
	\item \textit{Type II cascades}: These cascades exhibit a step-like pattern of growth shown in Figure~\ref{fig:types_cascades}\subref{cascB}.
	\item \textit{Type III cascades}: These cascades do not follow the logistic function as shown in Figures~\ref{fig:types_cascades}\subref{cascC} and \subref{cascD} and one of the many reasons for these cascade curves is that they probably do not complete their lifecycle within the period of 1 month that we have considered for each cascade.
\end{enumerate}

The motivation behind defining these three types of \textit{Growth Curves} lies in the way we define \textit{inhibition} intervals.
Intuitively, the inhibition interval is a period whereafter the cascade fails to regain any surge in adoption rate. We aim to tag only a single period as $\tau'_{steep}$ and a single period as $\tau'_{inhib}$, therefore we define these three types and only consider Type I cascades.
As shown in Figures~\ref{fig:types_cascades} \subref{b}, \subref{c} and \subref{d}, other types of cascade have multiple regions with slackness in growth, making it difficult to tag only one $\tau'_{inhib}$.
In addition, Type II and Type III only make up minority of the whole set of cascades and capture most anomalies due to time scaling issues.
Therefore, we only consider Type I cascades and assume a logistic fit to the growth
curve. In the next section, we formally define the inhibition interval used for  social network analysis in this paper.
 \\

\subsection{ Inhibition Interval}
\label{sec:steep_inhib}
The problem of identifying the inhibition period in a cascade is difficult since there has been no literature that formally defines it. The problem is aggravated by the lack of ground truth for validation since unlike cascade growth outbreak prediction \cite{Cui2013,Guo16} where authors use a size threshold to classify cascade epidemics, there are no standard threshold mechanisms to label the inhibition period in a cascade. We first formally define what we mean by an \textit{inhibition interval} below. In order to define an inhibition interval for a cascade, we introduce the intuition behind the constraints used: (1) given the time point at which the rate of reshare is maximum which we term $t_{steep}$, we posit that there would be a time difference $\Delta TG$ before inhibition arrives, and along with this time gap (2) we find that cascades start to saturate when its grows beyond a certain size since $t_{steep}$ occurs, so we use a growth ratio $g$. We note that $t_{steep}$ and $t_{inhib}$ are two reshare time indices in $\tau_c$ but we use it in order to highlight their importance in the cascade lifecycle. We understand that this would not be the only way to define and infer when inhibition occurs, but we use these constraints to pick a set of time points that suggest the cascade would expire soon.  

Let $t_{steep}$ be a point within the subsequence having the highest resharing rate among all $\tau'$ for a cascade $C$. Relative to this point, the inhibition time is defined as the first time point $t$ $\in [t_{steep}, T_C$] such that the following constraints hold: $t$ - $t_{steep}$ $\geq \Delta TG$ and $\frac{S_t}{S_{t_{steep}}}$ $\geq$ $g$ (refer to Table~\ref{tab:table0} for symbols), where $g$ and $\Delta TG$ are threshold parameters which are calculated in an inference process based on a data likelihood approach and where $t_{steep}$ is calculated using Hawkes intensities which approximates the point at which the highest slope for the \textit{growth curve} occurs, described in Appendix Section A1. We refer to this $t$ as $t_{inhib}$ and the subsequences containing $t_{steep}$ and $t_{inhib}$ as $\tau'_{steep}$ and $\tau'_{inhib}$ respectively. Similar to our previous notations, $\tau'_{steep}$ and $\tau'_{inhib}$ are simply two subsequences in $\tau$ but we use these indices to highlight their importance.

As described before, since there are no mechanisms to know the exact $\tau'_{inhib}$ or $\tau'_{steep}$, we need inference algorithms to identify the approximate time intervals. We use the procedure described in \cite{Sarkar2017} which is based on the works of \cite{Zhao15,Riz2017} that uses Hawkes intensities to model the diffusion process. We identify the inhibition interval in a three-step process using few manually selected cascades containing all the three types, the technical details of which are provided in Appendix section of \cite{Sarkar2017}. This process of calculating the threshold parameters $\Delta TG $ and $g$, is described intuitively below:
\begin{enumerate}
	\item We calculate the Hawkes intensity at each reshare time point $t$ as a function of the number of past interactions of the participating users for the current reshare, and the distribution of times taken by the users to adopt the cascade $C$. We convert this curve into \textit{Hawkes interval curve}  by  summing the intensities of time points in each interval.
	\item\label{step2} We then identify intervals with local maxima (which are candidates for the steep interval) and local minima (which are candidates for the inhibition interval). The maximum among all the local maxima is tagged as $t_{steep}$. 
	\item Based on ideas from \cite{Yang2013}, to compute $t_{inhib}$ we then use a maximum-likelihood approach to filter the points in step~\ref{step2} and obtain the parameters that we use to infer $t_{inhib}$ of the new cascades.
\end{enumerate}

\begin{figure}[!t]
	\centering
	\hfill
	\begin{minipage}{0.48\textwidth}%
		\includegraphics[width=5.5cm, height=4cm]{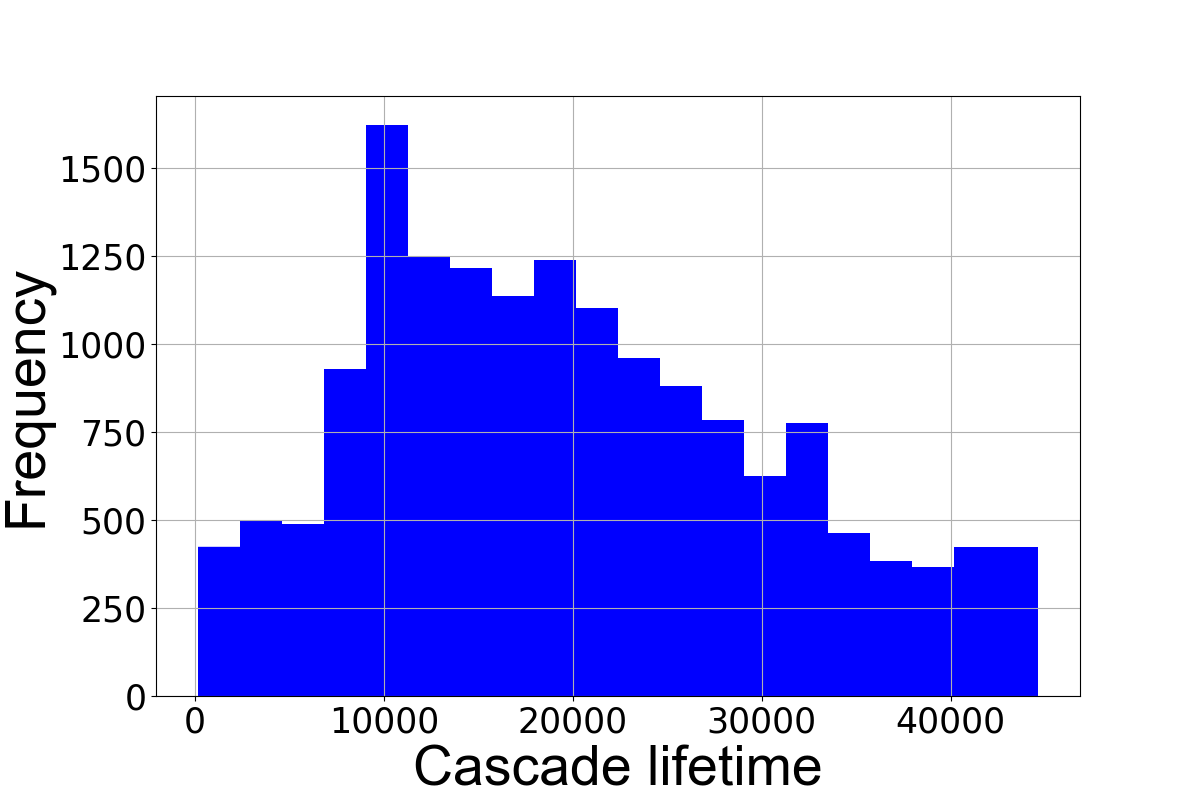}
		\subcaption{}
		\label{hist_net:a}
	\end{minipage}
	\hfill
	\begin{minipage}{0.4\textwidth}
		\includegraphics[width=5.5cm, height=4cm]{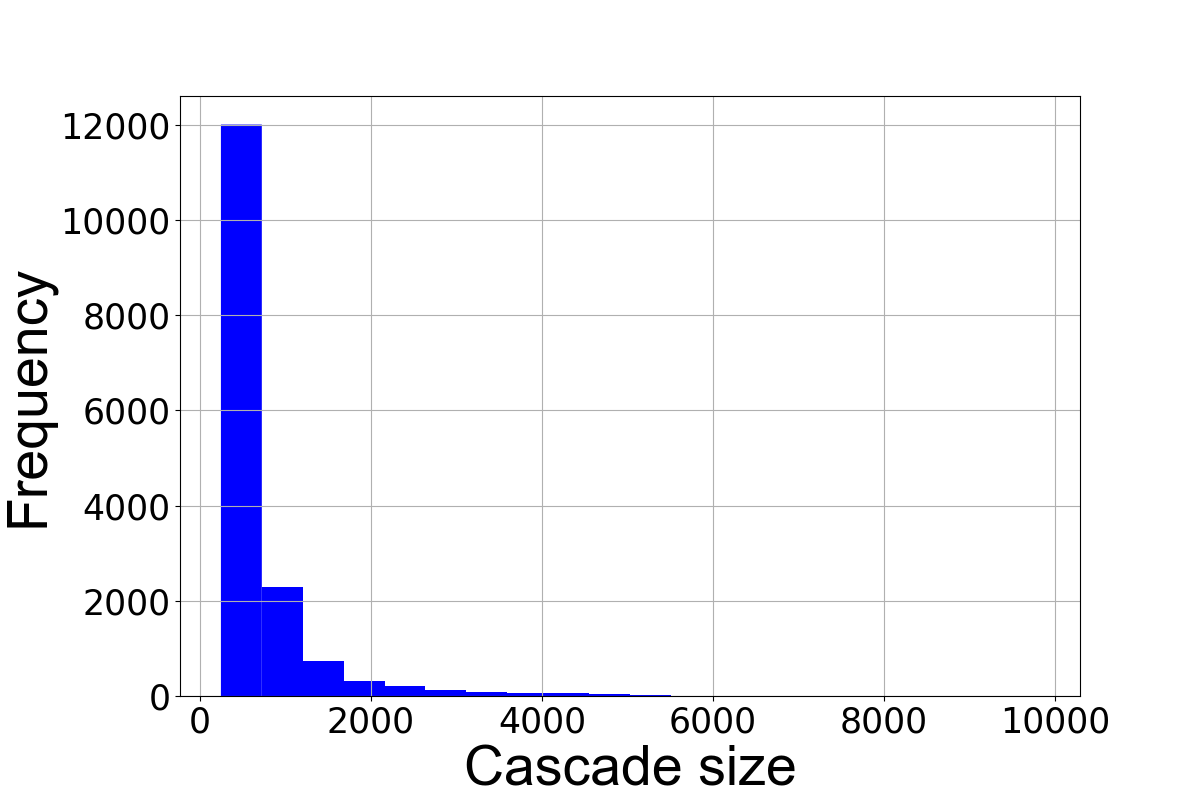}
		\subcaption{}
		\label{hist_net:b}
	\end{minipage}
	\hfill
	\caption{Histogram of (a) Cascade lifetimes in minutes (b) Cascade size.}
	\label{fig:hist_net}
\end{figure}

\section{Dataset and Temporal Network Construction}
\label{sec:dataset}
In this work, our primary analysis and our conclusions have been drawn from the the microblogging dataset provided by WISE 2012 Challenge\footnote{http://www.wise2012.cs.ucy.ac.cy/challenge.html} as has been previously used  in \cite{Guo16}. However, towards the end of this paper, we draw some comparisons of the observations from the dynamics of interactions in this microblogging platform with a movie rating platform to see the extent to which these observations can be generalized across social media platforms.  The Weibo dataset provides us with user data and the reposting information of each microblog along with the reposting times which enables us to form the cascades for each microblog separately.  From the corpus of cascades which spanned between June 1, 2011 and August 31, 2011, we only work with cascades with more than 300 nodes. Since we are considering subsequences preceding  $\tau'_{inhib}$ for our analysis, we discard cascades of smaller sizes in our experiments. Figures~\ref{fig:hist_net}\subref{hist_net:a} and \subref{hist_net:b} show the histograms for the cascade lifetimes measured by $T_C$ and the cascade sizes measured by $S_C$. As seen in Figure~\ref{fig:hist_net}, although the lifetimes follow a skewed Gaussian distribution, most cascades survive for less than 600 reshares having accounting for the skewed distribution.

Amongst the corpus of cascades, the number of Type I cascades is 5924 while the total number of Type II and Type III cascades is 1483. The total number of cascades of Type I is roughly around 80 \% of the total number of cascades that are more than size 300.
We separate Type I cascades from Type II and Type III cascades by using a time threshold mechanism described in Section {\color{red}4} of \cite{Sarkar2017}. The reason for using this threshold to label Type I cascades instead of using more complex curve fitting methods is two-fold: firstly, since the shape of curves even within Type I cascades vary based on when $t_{steep} $ occurs, it is difficult to manually select a set of Type I cascades to estimate parameters by MLE for a logistic model that are representative for all Type I cascades and secondly we observed that the Type I cascades are mainly characterized by situations where $t_{steep} $ occurs within a very short time after the cascade starts and cascades where $t_{steep}$ occurs after a certain amount of time do not exhibit the Type I pattern. As shown in Figure~\ref{fig:hist_net} \subref{hist_net:a}, the mode for the lifetimes occur at around 10000 minutes. \\

\begin{table}[!t]
	\centering
	\renewcommand{\arraystretch}{1}
	\caption{Properties of Reposting Network and Cascades}
	\begin{tabular}{|p{5cm}|p{4cm}|}
		\hline 
		{\bf Properties} & {\bf Reposting Network}\\ 
		\hline\hline
		Vertices           & 6,470,135 \\
		\hline
		Edges & 58,308,645 \\
		\hline 
		Average Degree & 18.02    \\       
		\hline \hline 
		Number of cascades & 7,479,088 \\
		\hline
		Number of cascades over 300 & 7407\\
		\hline
	\end{tabular}
	\label{tab:table2}
\end{table}

\noindent \textbf{Social Network:} A \textit{social network} is represented as a directed graph $G=(V,E)$ where $V$ is a population of individuals and the edge $(i,j)\in E$ refers to individual $i$ having the ability to influence individual $j$. We determine these relationships from the historical repository of diffusion links between nodes that occur prior to the cascades we use for our analysis in this paper. In our work, we denote the social network information by an undirected network $G_D$=$(V_D$, $E_D)$ where $V_D$ denotes the individuals involved in the historical diffusion process and an edge $e \in E_D$ denotes that information has been shared between a pair of individuals ignoring the direction of propagation. The diffusion network  $G_D$=$(V_D$ , $E_D)$ is created by linking any two users who are involved in a microblog reposting action within the period May 1, 2011 and July 31, 2011.  Similar to most social networks, this network also exhibits a power law distribution of degree \cite{Guo16}. Table~\ref{tab:table2} shows the statistics of the diffusion network and the corpus of cascades used in our experimental study. \\

\noindent \textbf{Cascade Network:} A \textit{cascade network} produced by a set of individuals participating in a set of reshares is denoted by  $G^{\tau}_C$ = $(V^{\tau}_C, E^{\tau}_C)$ where $C$ is the identifier for the cascade, $V^{\tau}_C$ denotes all the individuals who participated in the diffusion spread of $C$ in its entire lifecycle spanned by $\tau$. An edge $e=(u, v) \in$ $E^{\tau}_C$ denotes that either $v$ reshared $C$ from $u$ indicated by the presence of $(u, v, t) \in \tau$ at some time $t$ or the interaction happened in the past denoted by the presence of $e$ $\in$ $E_D$, that is to say we add the influence of the propagation links from our historical diffusion network obtained from cascades in the past. We note that these historical links can be substituted by social network links like friend links, follower-followee links. The addition of links is to understand how the effect of homophily, social influence, mutual preferences exhibited by external social network information couples with the current cascade propagation process. The ignorance of an independent cascade model \cite{Kempe2003} this way would help us observe how nodes form groups having known each other previously in the present cascade scenario. This construction of the cascade network structure would eliminate the default tree structure of cascades and introduce cycles within the network. However, all our analysis has been done on individual cascades using their respective network $G^{\tau}_C$. As mentioned before, we drop the subscript $C$ from the notations when the analyses are applicable for all cascades. \\

\begin{figure}[]
	\centering
	\includegraphics[width=12.5cm, height=3cm]{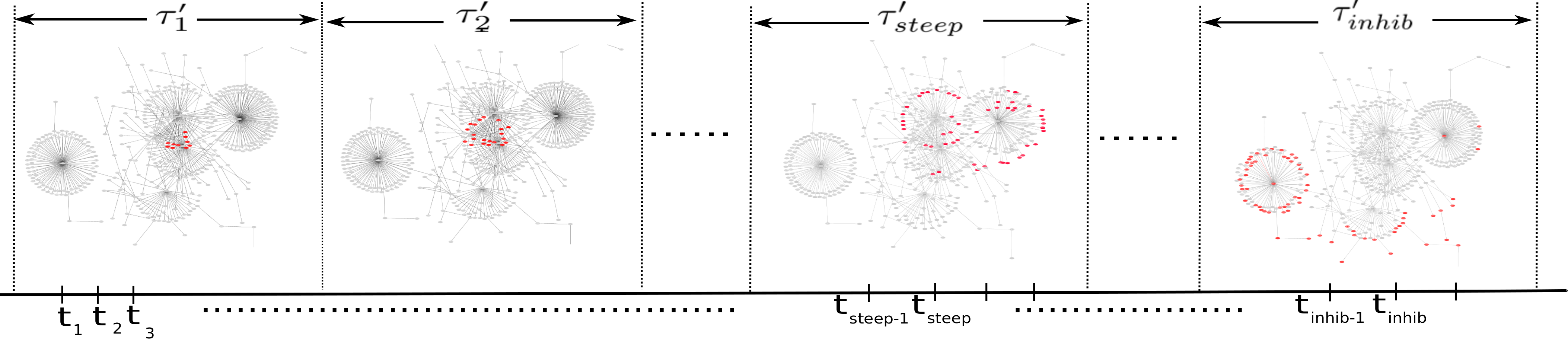}
	\caption{Example showing how the network is partitioned over subsequences for network analysis. Observe that $\tau'_{steep}$ and $\tau'_{inhib}$contain $t_{steep}$ and $t_{inhib}$ respectively. Each $\tau'$ contains equal number of nodes for analysis and as the cascade progresses different nodes get activated marked in red. Although each $\tau'$ looks uniform in the time span, we note that each $\tau'$ may differ in the time range depending on how much time it took for $|V^{\tau'}|$ nodes to form the network.}
	\label{fig:net_evolve}
\end{figure}

\noindent \textbf{Temporal cascade network:} In context of $G^{\tau}$,  we denote $G^{\tau'}$ = $(V^{\tau'}, E^{\tau'})$ as the subgraph of $G^{\tau}$, where $V^{\tau'}$ denotes the individuals who reshared a cascade $C$ in the time subsequence $\tau'$ and  $E^{\tau'}$ denotes the set of edges $e=(u, v)$ where the resharing from $u$ to $v$ happened in the time subsequence $\tau'$ or there was an interaction in the historical diffusion period indicated by the presence of $e$ in $E_D$.  For the subsequences, the following conditions hold: (1) $|V^{\tau'_i}| = |V^{\tau'_j}|$ and (2) $E^{\tau'_i} \cap E^{\tau'_{j}}$ = $\emptyset$, $\forall i \neq j, \in [0, \mathcal{Q}]$. We note that the condition $|\tau'_i|$ $\neq |\tau'_j|$ may or may not hold for any $i \neq j$, that is to say the time range spanned by the subsequences in itself may differ depending on the time taken by $G^{\tau'}$ to form the network. In our work, we select and keep $|V^{\tau'}|$ fixed for every cascade in our corpus. Since we analyze the subsequences in a sequence that depicts the evolution of the cascade over time, the advantage of selecting this subsequence node set size a-priori is that we can avoid retrospective observation of the entire cascade lifecycle and incrementally progress with the network analysis using motifs over intervals until we reach $N_{inhib}$. This gives us the freedom to be agnostic about the final cascade size $|V^{\tau}|$ or $T_C$ for selecting the subsequence size $|V^{\tau'}|$ or span $|\tau'|$. Since we do not fix $\mathcal{Q}_C$, the number of subsequences for $C$ and let it vary based on $|V^{\tau'}_C|$, we do not fix a specific index $inhib$ $\in [0, \mathcal{Q}]$ for $\tau'_{inhib}$  for all cascades.

\begin{table}[!h]
	\centering
	\renewcommand{\arraystretch}{1}
	\caption{Properties of $N_{inhib}$ for all cascades - Weibo dataset}
	\begin{tabular}{|p{4cm}|p{2cm}|p{2cm}|p{2cm}|}
		\hline 
		{\bf Properties} & {\bf Mean} & {\bf Max.} & {\bf Min.} \\ 
		\hline\hline
		Number of nodes & 78.59 & 80 & 64 \\
		\hline
		Number of edges & 82.22 & 153  & 56 \\
		\hline 
		Average Clustering coefficient & 0.09 & 0.51 & 0.0  \\       
		\hline 
		Pagerank & 0.012 & 0.015 & 0.012 \\
		\hline		Average Vertex betweenness & 0.014 & 0.073  & 0.0005 \\
		\hline
		Average Nodal degree & 2.39 & 4.02 & 1.40 \\
		\hline
	\end{tabular}
	\label{tab:table_inhib}
\end{table}

A \textit{temporal representation} of a cascade is denoted by a sequence of overlapping subsequences $\mathcal{N}$ = $\langle N_1, \ldots, N_{\mathcal{S}} \rangle$ such that the following conditions hold: $V^{N_i}$ = $V^{\tau'_{i-1}}$ $\cup$ $V^{\tau'_i}$, and $E^{N_i}$ = $E^{\tau'_{i-1}}$ $\cup$ $E^{\tau'_i}$ $\forall$  $i \in [1, \mathcal{Q}]$. We perform network analysis on each $N $ where we drop the index subscript when we generalize the analysis for all subsequences for all cascades. Such a temporal representation $\mathcal{N}$ helps us in avoiding disjoint subsequences for network analysis and replicates a sliding window approach. We denote the first network in $\mathcal{N}$ containing $\tau'_{inhib}$ as $N_{inhib}$. We note that $\tau'_{inhib}$ is not necessarily the last subsequence in the cascade, as there may be few more reshares before the cascade finally dies down but since we are interested in the subsequences before $t_{inhib}$, we discard the rest of the subsequences after $\tau'_{inhib}$ from our analysis. Along similar lines, we denote the first network containing $\tau'_{steep}$ as $N_{steep}$. Figure~\ref{fig:net_evolve} gives a visual depiction of the method of analysis performed on the cascades using the subsequences. We perform our network analysis on each $N$ in sequence of formation, under the representation described above until we reach the $\tau'_{inhib}$ as shown in the figure. We fix $|V^{\tau'}|$ to 40 for all the cascades. Following this, $|V^{N}|$ $\leq$ 80 although $|E^{N}|$ would vary for each $N$. Table\ref{tab:table_inhib} shows the statistics of the temporal network $N_{inhib}$ for the cascades in our study.

\section{Network Motifs} \label{sec:motifs}
Network motifs are recurring patterns of interconnections occurring in complex networks that appear significantly
higher in count than those in random networks (significance here being determined by $z$-scores with the null model being motifs extracted from a random network with the same degree sequence as the original network, as described in \cite{Milo824}).

\noindent (\textbf{Network Motif}): A $k$-size network motif is defined by a set of nodes $V_M$ = $\{v_1, v_2, \ldots v_k\}$ and a set of edges(node tuples) $E_M$ = $\{(v_1, v_2), \ldots (v_i, v_j)\}$, $\forall v_i \in V_M$ where the ordering of  nodes in tuples is ignored by the set operation, that is to say $(v_1, v_2)$ = $(v_2, v_1)$. An induced network motif $M$ is a subgraph of an undirected network $G = (V, E)$ such that $v \in V$, $\forall v \in V_M$ and $e = (v_i, v_j)$ $\in E$, $v_i, v_j$ $\in V_M$. 

Given a network motif of size $k$, the number of possible patterns with all possible combinations of undirected edges is given by $(2^{d+1}-1)^{\frac{n}{d+1}} + n$ where $d$ is the maximum vertex degree which is equal to $n-1$. \cite{Bjorklund2012}. However the number of induced network motif patterns would depend on the edges of the original graph $G$, therefore as $k$ increases, so does the number of motif patterns. In this paper, for each temporal network $N$, we extract undirected graph motifs of node size $k$ equal to 5 and we perform feature analysis using 5-sized motifs although we later check the transition pattern from 4-sized motifs to 5-sized motifs. We choose 5 nodes over smaller size motifs as the number of possible patterns occurring from the permutations on edges for smaller node sized motifs as described above, are too low to conclude anything significant. One of the motivations behind using undirected network motifs in this research is as follows: while direction of information flow is definitely one of ways to understand the cascade progress, we investigate how users form these micro-scale patterns that constitute the bigger network in these stages. So in a sense understanding these motifs occurrences helps us understand the overall structural organization of the cascades instead of just looking at the flow of information. Additionally, due to the canonical labeling method used for finding isomorphic patterns, we would have too many patterns for directed 5-sized motifs which would grow out of proportion to conclude anything concrete. \\

\noindent \textbf{Motif Detection:} We use the FANMOD algorithm in \cite{Wer2006}  and it can detect network motifs
up to a size of eight vertices using a novel algorithm called
RAND-ESU. The 3 main steps for motif extraction proceed as following: (1)  finding what patterns are possible in that network (2) finding which of these patterns are topologically equivalent(isomorphic classes) and grouping them together and calculating the count of instances in each class (3) determining which of these subgraph classes occur at much higher frequencies or have a high $z$-score than random graphs under a specified random graph model. Since the second step of computing subgraph isomorphism is NP-Hard \cite{Ullmann1976}, the algorithm is computationally expensive for large graphs. However, since we limit the size of $V^N$ in our paper, therefore we avoid such overheads and it takes relatively less time for motif computation which can be distributed for the cascades.Following this strategy, we implemented all the experiments using the multi-threaded environment in Python on 256 GB RAM Intel machine with 24 cores.  The 5-node patterns that occurred in some stage of the cascade lifecycle in our dataset is shown in Table~\ref{tab:motif_patterns}.

\begin{table}[!t]
	\centering
	\caption{Motif patterns of size 4} \label{tab:motif_patterns_4}
	\begin{tabular}{|p{1cm}|p{1.5cm}|p{0.8cm}|p{1.2cm}|p{1cm}|p{1.5cm}|p{0.8cm}|p{1.2cm}|}
		\hline \textbf{Motif ID} & \textbf{Pattern} & \textbf{Edges} & \textbf{Density} & \textbf{Motif ID} & \textbf{Pattern} & \textbf{Edges} & \textbf{Density}\\
		\hline $M1$ & \parbox[c]{1em}{
			\includegraphics[width=0.3in]{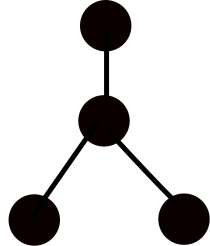}} & 3 & 0.5
		& $M4$ & \parbox[c]{1em}{
			\includegraphics[width=0.3in]{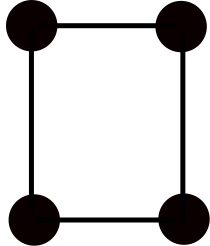}} & 4 & 0.67
		\\
		\hline $M2$ & \parbox[c]{1em}{
			\includegraphics[width=0.3in]{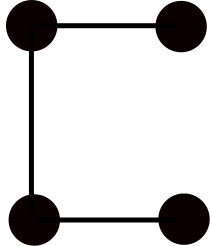}} & 3 & 0.5 
		& $M5$ & \parbox[c]{1em}{
			\includegraphics[width=0.3in]{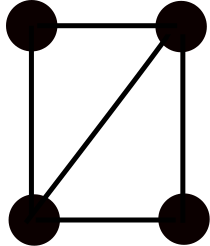}} & 5 & 0.83
		\\
		\hline $M3$ & \parbox[c]{1em}{
			\includegraphics[width=0.3in]{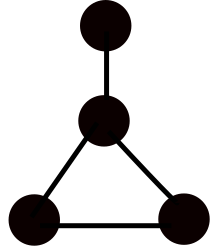}} & 4 & 0.67 
		& $M6$ & \parbox[c]{1em}{
			\includegraphics[width=0.3in]{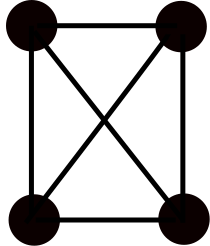}} & 6 & 1
		\\
		\hline
	\end{tabular}
	\renewcommand{\arraystretch}{4}
\end{table}

\begin{table}[]
	\centering
	\caption{Motif patterns of size 5} \label{tab:motif_patterns}
	\begin{tabular}{|p{1cm}|p{1.5cm}|p{0.8cm}|p{1cm}|p{1.2cm}|p{1.5cm}|p{0.8cm}|p{1.2cm}|}
		\hline \textbf{Motif ID} & \textbf{Pattern} & \textbf{Edges} & \textbf{Density} & \textbf{Motif ID} & \textbf{Pattern} & \textbf{Edges} & \textbf{Density}\\
		\hline $M1$ & \parbox[c]{1em}{
			\includegraphics[width=0.3in]{M2.png}} & 4 & 0.4 
		& $M12$ & \parbox[c]{1em}{
			\includegraphics[width=0.3in]{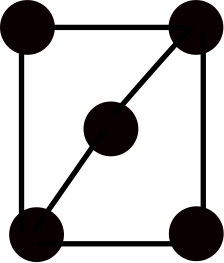}} & 6 & 0.6
		\\
		\hline $M2$ & \parbox[c]{1em}{
			\includegraphics[width=0.3in]{M1.png}} & 4 & 0.4 
		& $M13$ & \parbox[c]{1em}{
			\includegraphics[width=0.3in]{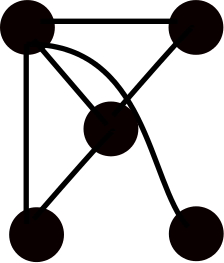}} & 6 & 0.6
		\\
		
		\hline $M3$ & \parbox[c]{1em}{
			\includegraphics[width=0.3in]{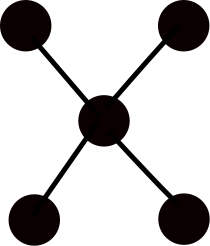}} & 4 & 0.4 
		& $M14$ & \parbox[c]{1em}{
			\includegraphics[width=0.3in]{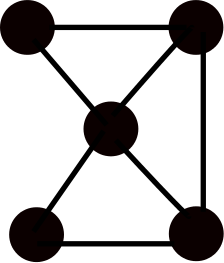}} & 7 & 0.7
		\\
		
		\hline $M4$ & \parbox[c]{1em}{
			\includegraphics[width=0.3in]{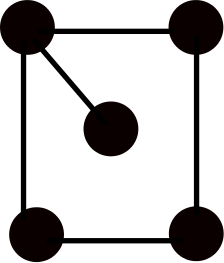}} & 5 & 0.5 
		& $M15$ & \parbox[c]{1em}{
			\includegraphics[width=0.3in]{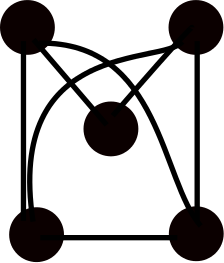}} & 7 & 0.7
		\\
		
		\hline $M5$ & \parbox[c]{1em}{
			\includegraphics[width=0.3in]{M3.png}} & 5 & 0.5 
		& $M16$ & \parbox[c]{1em}{
			\includegraphics[width=0.3in]{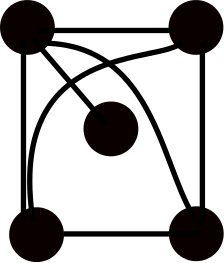}} & 7 & 0.7
		\\
		
		\hline $M6$ & \parbox[c]{1em}{
			\includegraphics[width=0.3in]{M13.png}} & 5 & 0.5 
		& $M17$ & \parbox[c]{1em}{
			\includegraphics[width=0.3in]{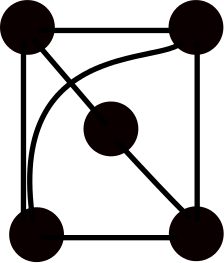}} & 7 & 0.7
		\\
		
		\hline $M7$ & \parbox[c]{1em}{
			\includegraphics[width=0.3in]{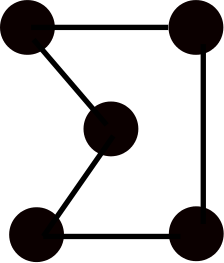}} & 5 & 0.5 
		& $M18$ & \parbox[c]{1em}{
			\includegraphics[width=0.3in]{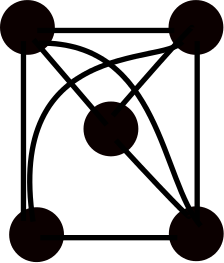}} & 8 & 0.8
		\\
		
		\hline $M8$ & \parbox[c]{1em}{
			\includegraphics[width=0.3in]{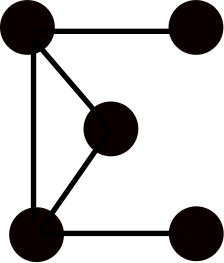}} & 5 & 0.6 
		& $M19$ & \parbox[c]{1em}{
			\includegraphics[width=0.3in]{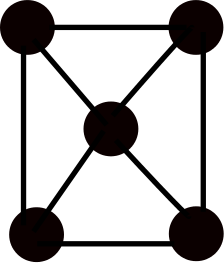}} & 8 & 0.8
		\\
		
		\hline $M9$ & \parbox[c]{1em}{
			\includegraphics[width=0.3in]{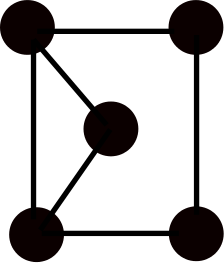}} & 6 & 0.6
		& $M20$ & \parbox[c]{1em}{
			\includegraphics[width=0.3in]{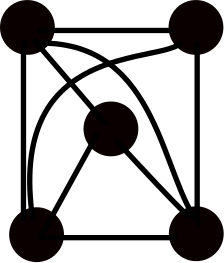}} & 9 & 0.9
		\\
		
		\hline $M10$ & \parbox[c]{1em}{
			\includegraphics[width=0.3in]{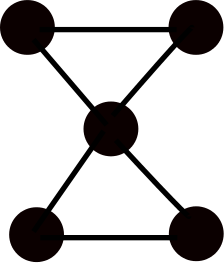}} & 6 & 0.6 
		& $M21$ & \parbox[c]{1em}{
			\includegraphics[width=0.3in]{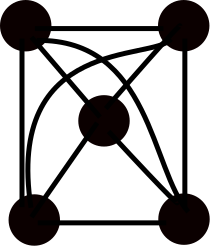}} & 10 & 1
		\\ 
		\hline $M11$  & \parbox[c]{1em}{
			\includegraphics[width=0.3in]{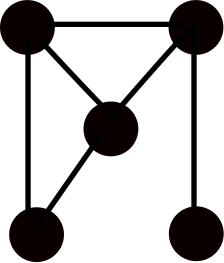}} &  6 & 0.6 
		& &  & & 
		\\
		\hline 
	\end{tabular}
	\renewcommand{\arraystretch}{4}
\end{table}

\section{Methods for Network Analysis}
\label{sec:motif_feat}
In this paper, we use network motifs to characterize the evolution of the cascade network structure in terms of interaction patterns and how repeatedly individuals group together in a similar pattern. Let $\mathcal{M}_{k}$ denote the global set of all motif patterns of motif size $k$. $\mathcal{M}_{k, C}^N $ represents the list of motifs of size $k$ found in the network $N $ of cascade $C$. Similar to our previous approach, we will drop $C$ and $N$ from all these features when we generalize them over all the networks in all the cascades. We will use the notation $M$, where $M \in \mathcal{M}_{k} $ to denote a particular pattern of size $k$ within the list of motif patterns. We use the shorthand notation $m$ to denote a particular instance of a pattern $M$ that appears in a network $N$. We use two elements of motifs: the frequency of their appearances and the way they transition into new patterns to understand how cascades progress. These are described in details in this section.

\subsection{Motifs Counts}
To examine the occurrence of particular motifs in specific time phases preceding $\tau'_{inhib}$ in the cascade, we observe the count of motifs of size $k$ of each pattern $M$ $\in$ $\mathcal{M}$, that occurs in network $N_j$, $\forall $ $j \in [steep, inhib]$ , which we denote by $MC_{k, M}^{N_j}$. We attempt to understand the following questions with respect to motif appearances: \\

\textit{RQ 1: Does the appearance of certain motif patterns uniquely characterize the approaching inhibition stage and do they uniquely characterize the approaching inhibition stage?} \\

For evaluating the motif appearance frequencies in the intervals leading to $\tau'_{inhib}$, we consider the last 20 networks in $\mathcal{N}_C$ preceding $N_{inhib}$ since as mentioned before, the index $inhib$ corresponding to the subsequence containing $t_{inhib}$ would vary for each cascade $C$ and therefore a particular subsequence $N_i$, $i \in [1, \mathcal{Q}]$ cannot be selected as $N_{inhib}$ for all cascades. Following this, it is not possible to select one particular network $N$ for analysis. So we resort to this analysis of the preceding intervals leading to $N_{inhib}$ for each cascade separately. This is also to ensure that we do not miss out on any time subsequences that may be early signs of an approaching $inhibition$ interval. We come across
\begin{figure}[t]
	\centering
	\begin{minipage}{0.3\textwidth}%
		\includegraphics[width=5cm, height=3.7cm]{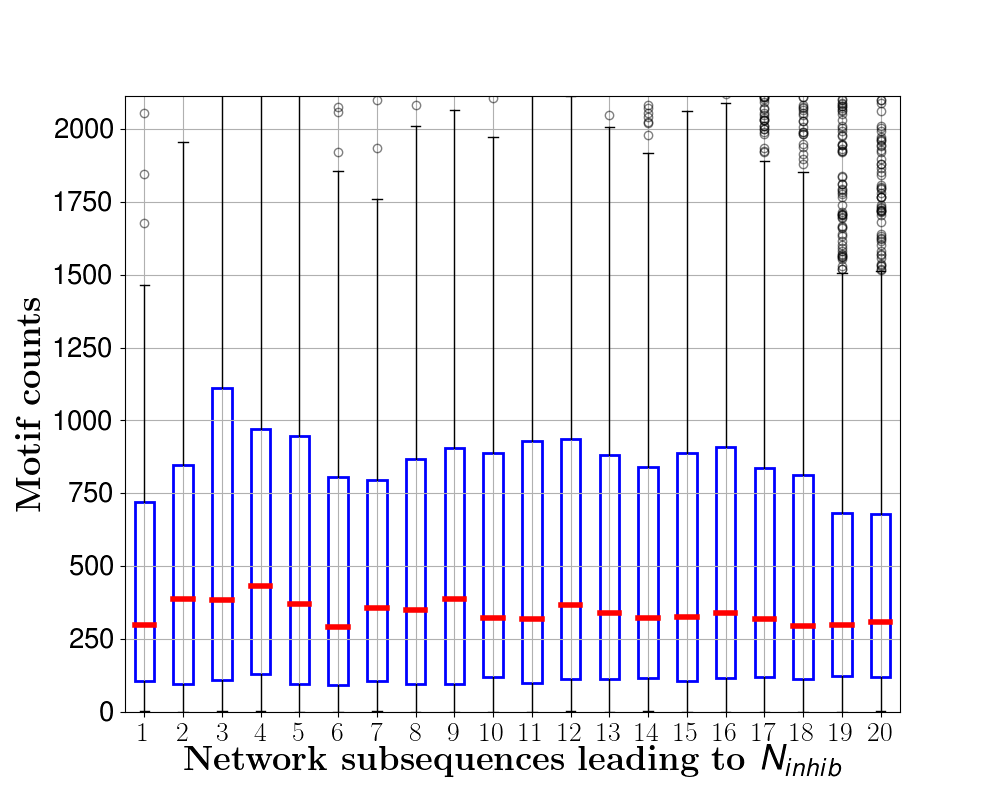}
		\subcaption{M1: \ \includegraphics[width=0.4cm]{M2.png}}
		\label{(a)}
	\end{minipage}
	\hfill
	\begin{minipage}{0.305\textwidth}
		\includegraphics[width=5cm, height=3.5cm]{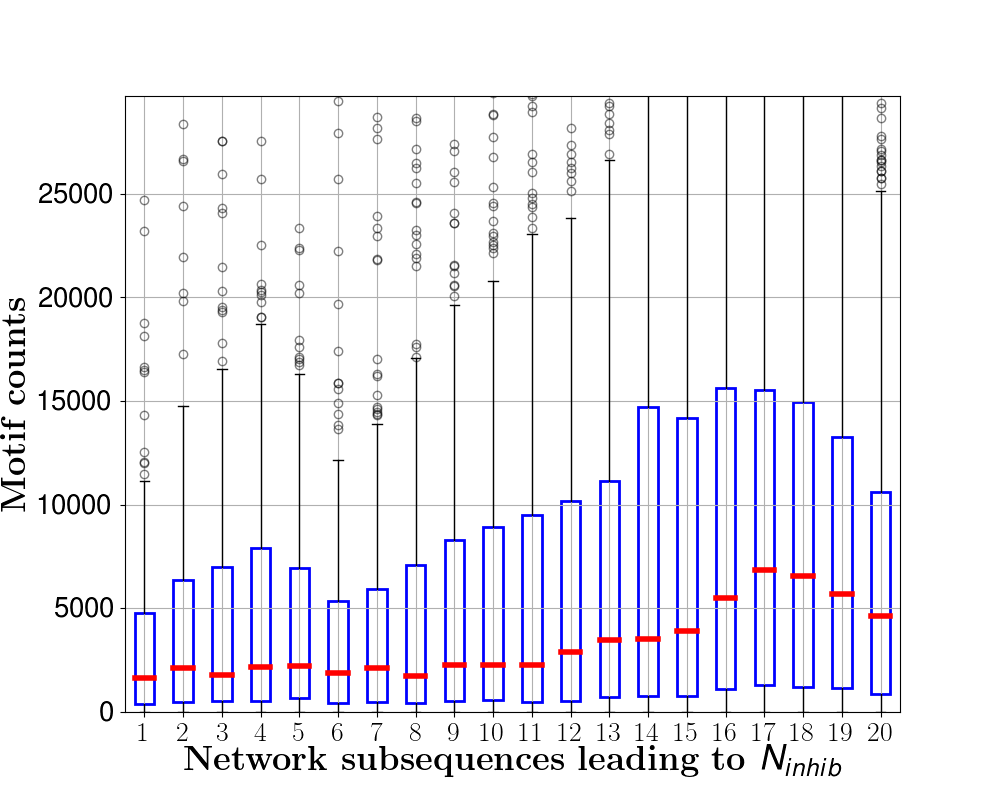}
		\subcaption{M2: \ \includegraphics[width=0.4cm]{M1.png}}
		\label{(b)}
	\end{minipage}
	\hfill
	\begin{minipage}{0.3\textwidth}
		\includegraphics[width=5cm, height=3.7cm]{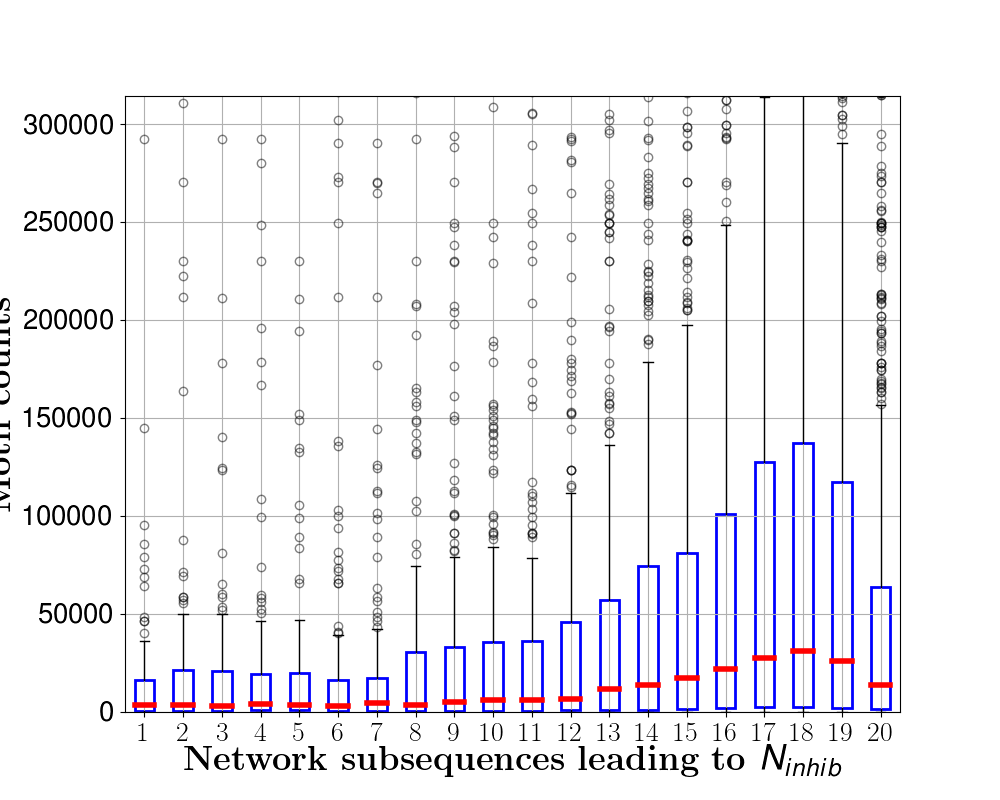}
		\subcaption{M3: \ \includegraphics[width=0.4cm]{M0.png}}
		\label{(c)}
	\end{minipage}
	\hfill
	\\
	\begin{minipage}{0.3\textwidth}%
		\includegraphics[width=5cm, height=3.7cm]{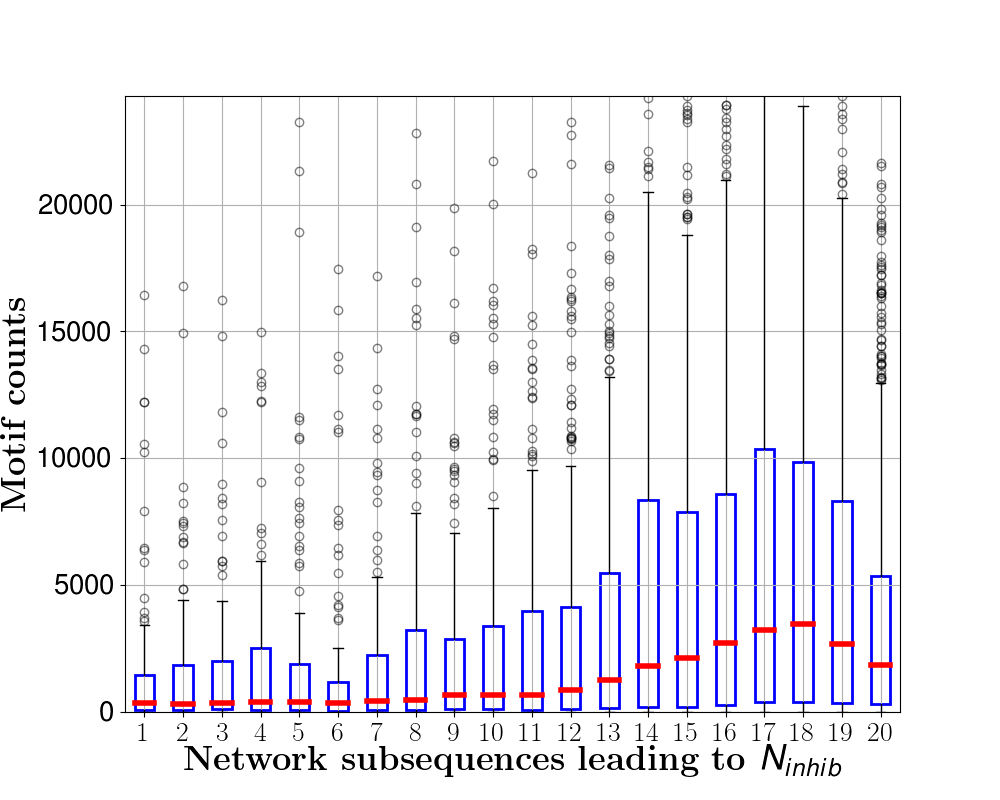}
		\subcaption{M5: \ \includegraphics[width=0.4cm]{M3.png}}
		\label{(d)}
	\end{minipage}
	\hfill
	\begin{minipage}{0.305\textwidth}
		\includegraphics[width=5cm, height=3.7cm]{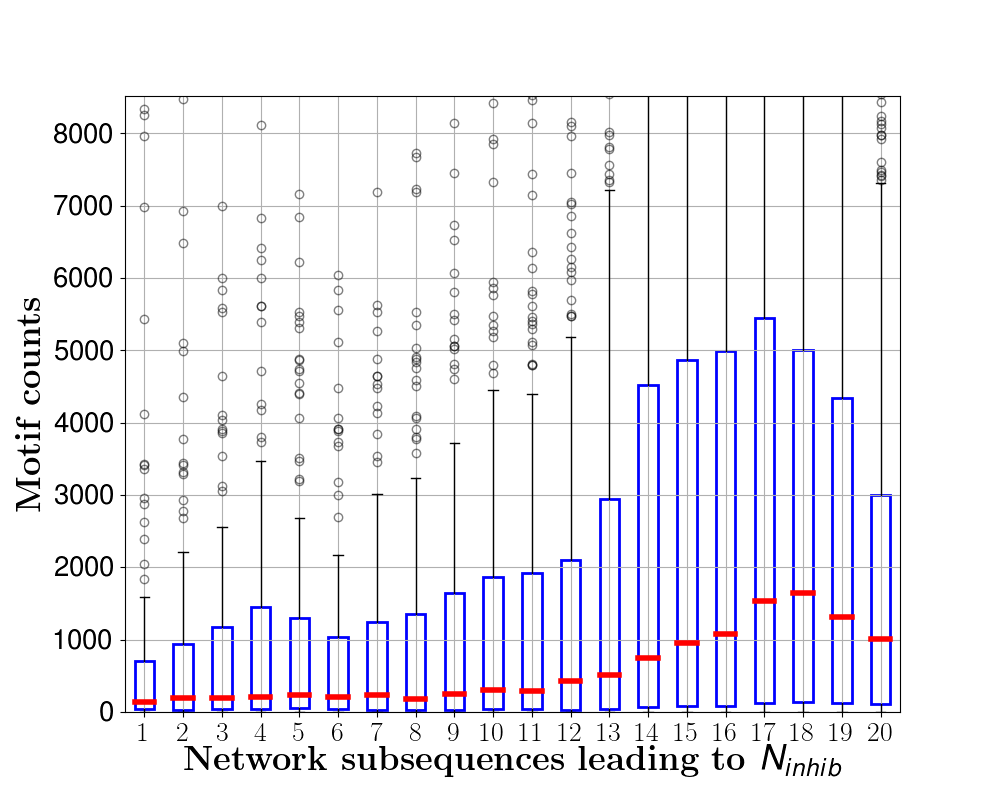}
		\subcaption{M6: \ \includegraphics[width=0.4cm]{M13.png}}
		\label{(e)}
	\end{minipage}
	\hfill
	\begin{minipage}{0.3\textwidth}
		\includegraphics[width=5cm, height=3.7cm]{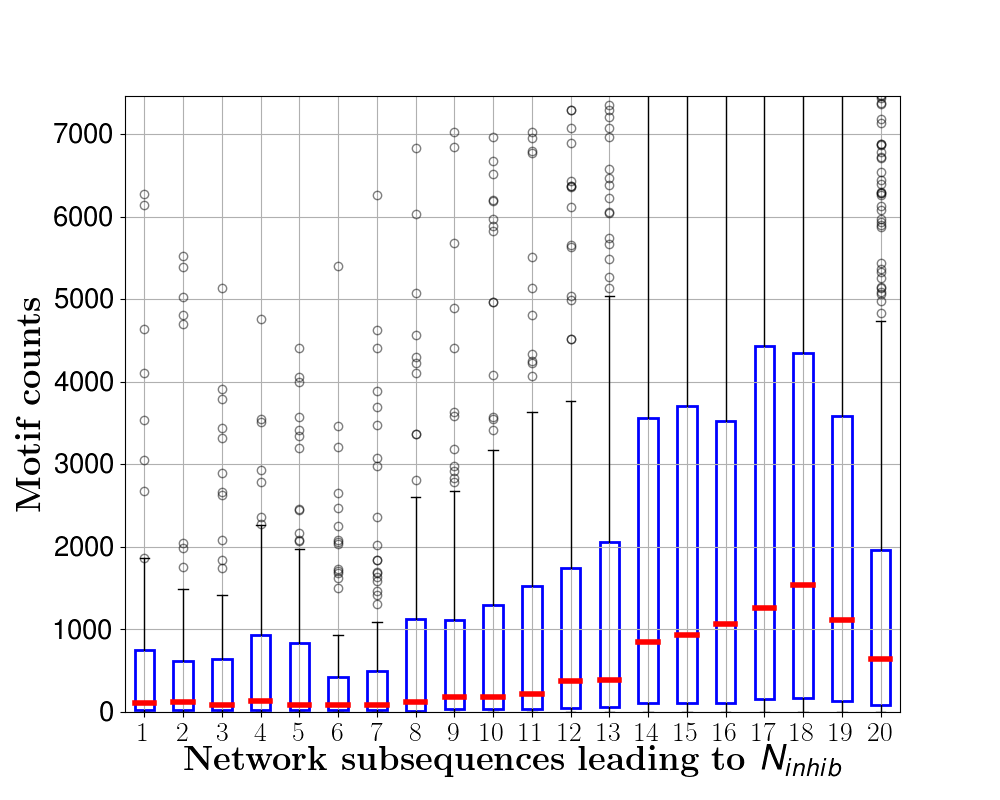}
		\subcaption{M11: \ \includegraphics[width=0.4cm]{M12.png}}
		\label{(f)}
	\end{minipage}
	\hfill
	\caption{Plots of motif counts. The intervals are increasing from left to right leading to $N_{inhib}$. }
	\label{fig:mcounts_1}
\end{figure}
two important observations with respect to motif counts: first about the dynamics of the count in the networks  preceding $N_{inhib}$. While there is a steady increase followed by a decrease in counts towards the end of $N_{inhib}$ for the acyclic patterns \ \includegraphics[scale=0.04]{M1.png} \ and \  \includegraphics[scale=0.04]{M0.png} \ , we do not see such a trend for the linear chain pattern \ \includegraphics[scale=0.04]{M2.png}. In relation to this, the plot in Figure~\ref{fig:mcounts_1} \subref{a} reveals that the formation of such linear patterns of path length 4 is not favored denoted by lower median counts marked in red compared to higher values in Figures~\ref{fig:mcounts_1} \subref{b} and \subref{c}. We observe that \  \includegraphics[scale=0.04]{M0.png} \ exhibited the highest count among all the patterns denoted by the medians marked in red in Figure~\ref{fig:mcounts_1} \subref{(c)}.  Among the patterns with cycles, we find that the patterns \ \includegraphics[scale=0.04]{M3.png} \, 
\includegraphics[scale=0.04]{M13.png} \ and \ \includegraphics[scale=0.04]{M12.png} 
exhibit the highest counts which are significantly higher than patterns with squares and higher edge densities where significance is measured by the difference between the medians.
Such trends in motif counts suggest the tendency of individuals to interact in such a way that form groups of 3 nodes or resharing from central nodes in \includegraphics[scale=0.04]{M0.png} \ but as such there are no clear signs of nodes forming chains of long paths since find that patterns such as \includegraphics[scale=0.04]{M15.png} which also contains a linear chain of size 5 is not very common.

\begin{figure}[!t]
	\centering
	\begin{minipage}{0.3\textwidth}%
		\includegraphics[width=5cm, height=3.7cm]{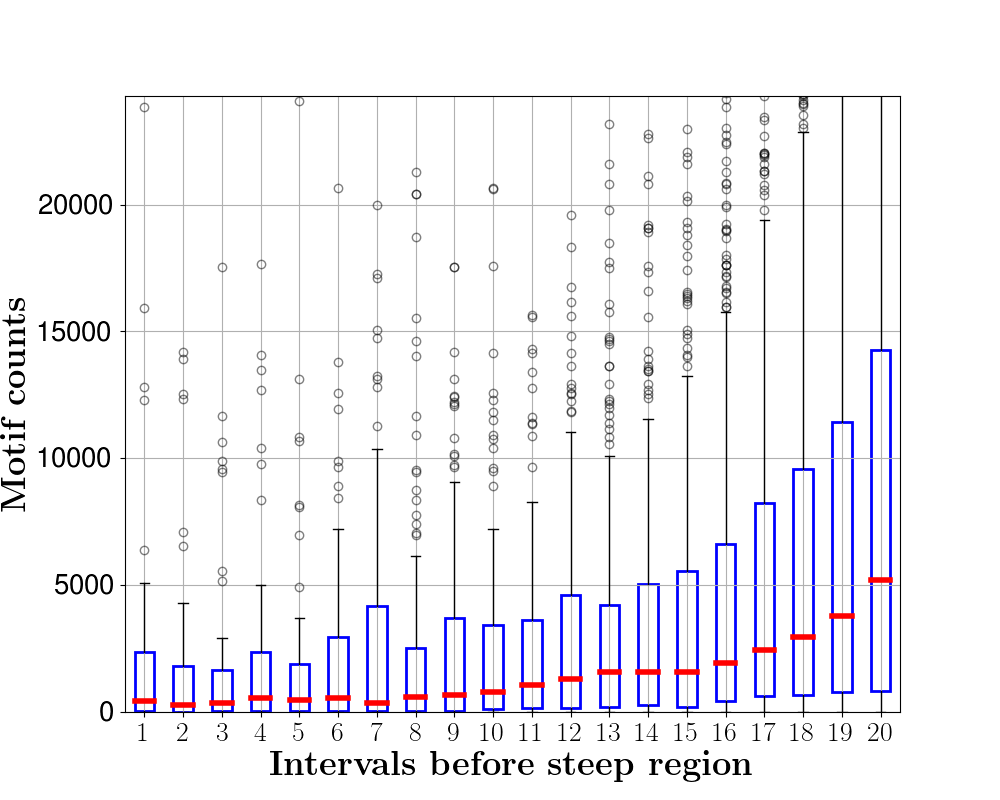}
		\subcaption{M1: \ \includegraphics[width=0.4cm]{M2.png}}
		\label{(a)}
	\end{minipage}
	\hfill
	\begin{minipage}{0.305\textwidth}
		\includegraphics[width=5cm, height=3.7cm]{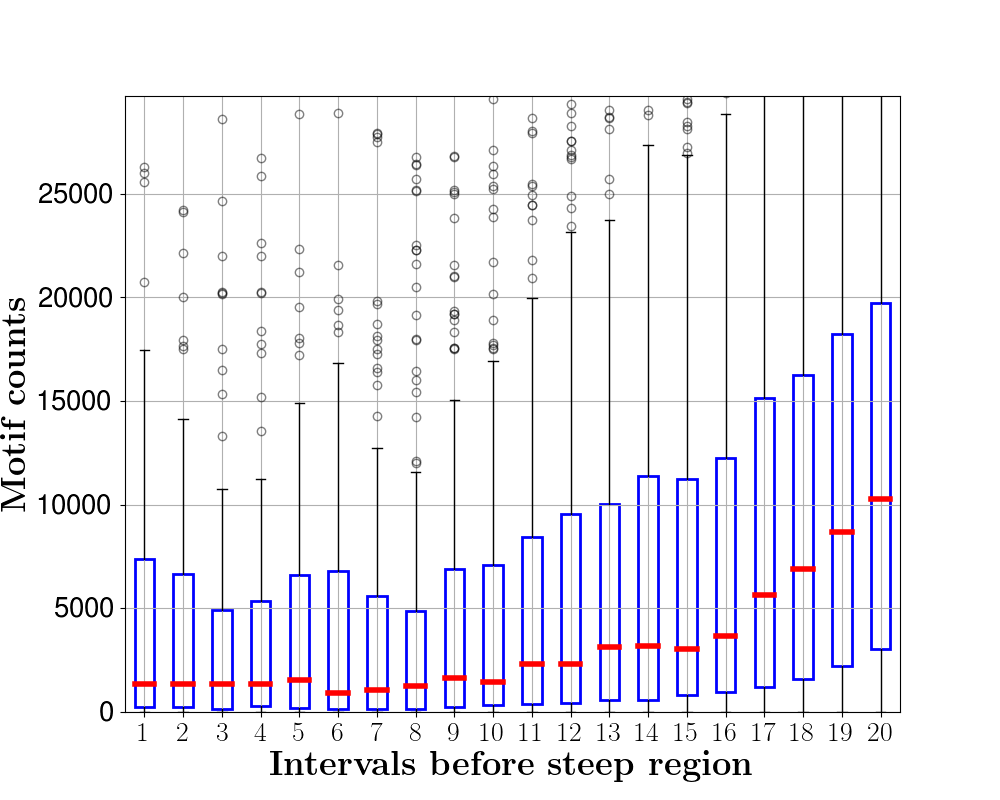}
		\subcaption{M2: \ \includegraphics[width=0.4cm]{M1.png}}
		\label{(b)}
	\end{minipage}
	\hfill
	\begin{minipage}{0.3\textwidth}
		\includegraphics[width=5cm, height=3.7cm]{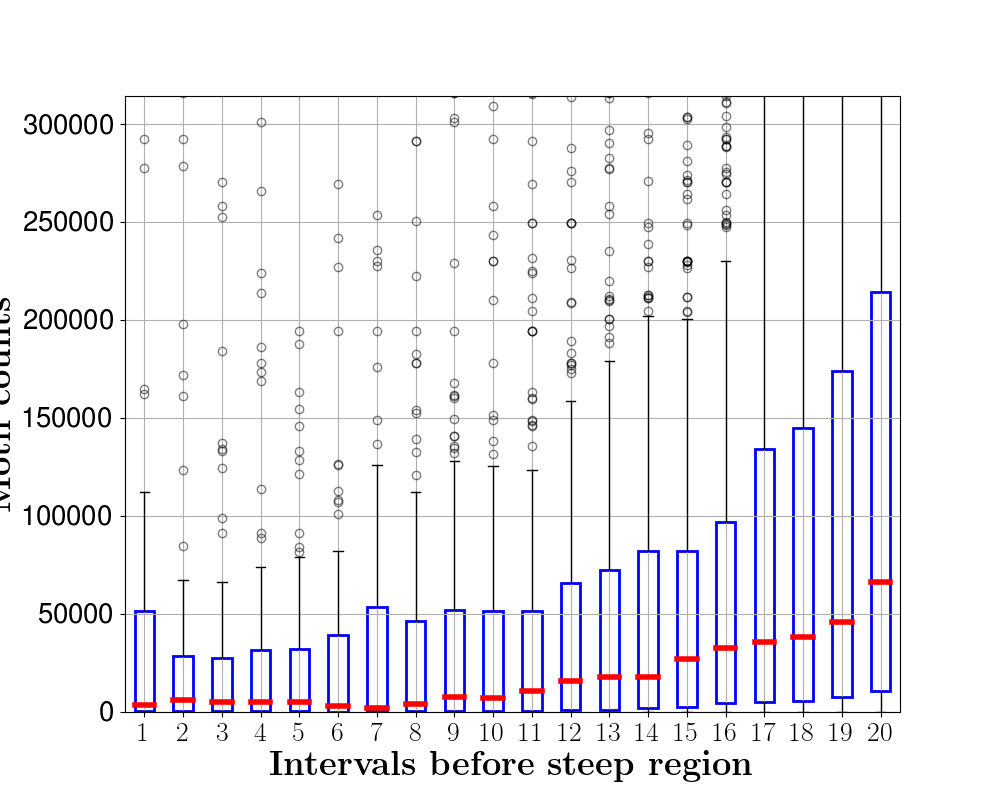}
		\subcaption{M3: \ \includegraphics[width=0.4cm]{M0.png}}
		\label{(c)}
	\end{minipage}
	\hfill
	\\
	\begin{minipage}{0.3\textwidth}%
		\includegraphics[width=5cm, height=3.7cm]{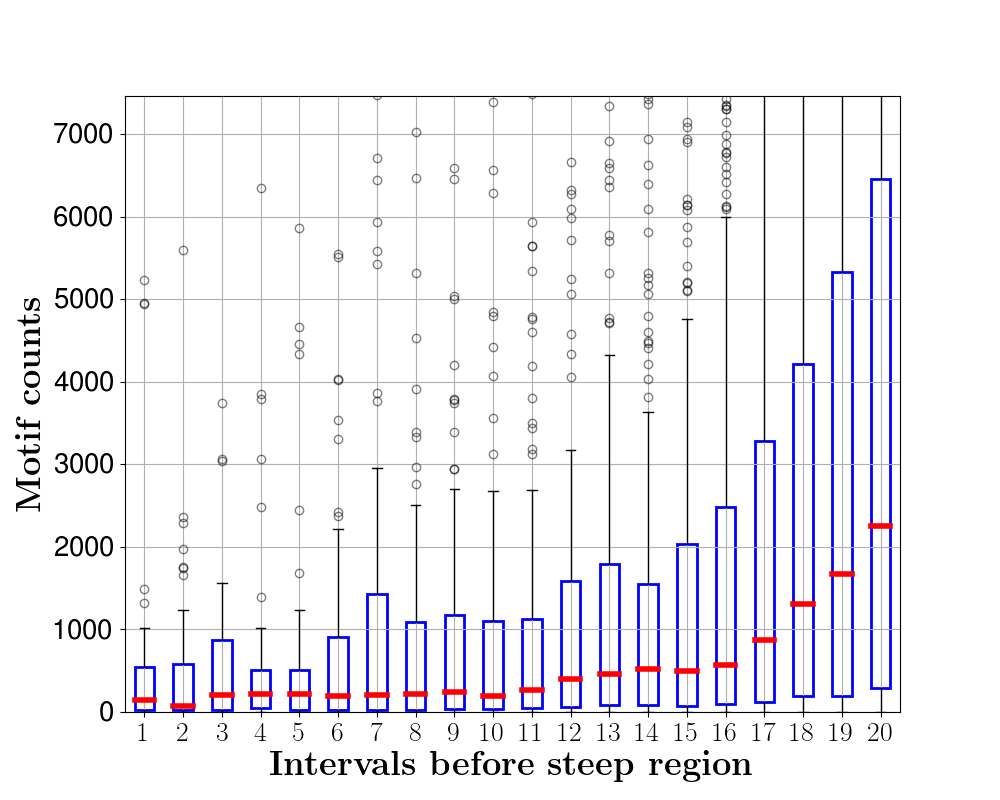}
		\subcaption{M5: \ \includegraphics[width=0.4cm]{M3.png}}
		\label{(d)}
	\end{minipage}
	\hfill
	\begin{minipage}{0.305\textwidth}
		\includegraphics[width=5cm, height=3.7cm]{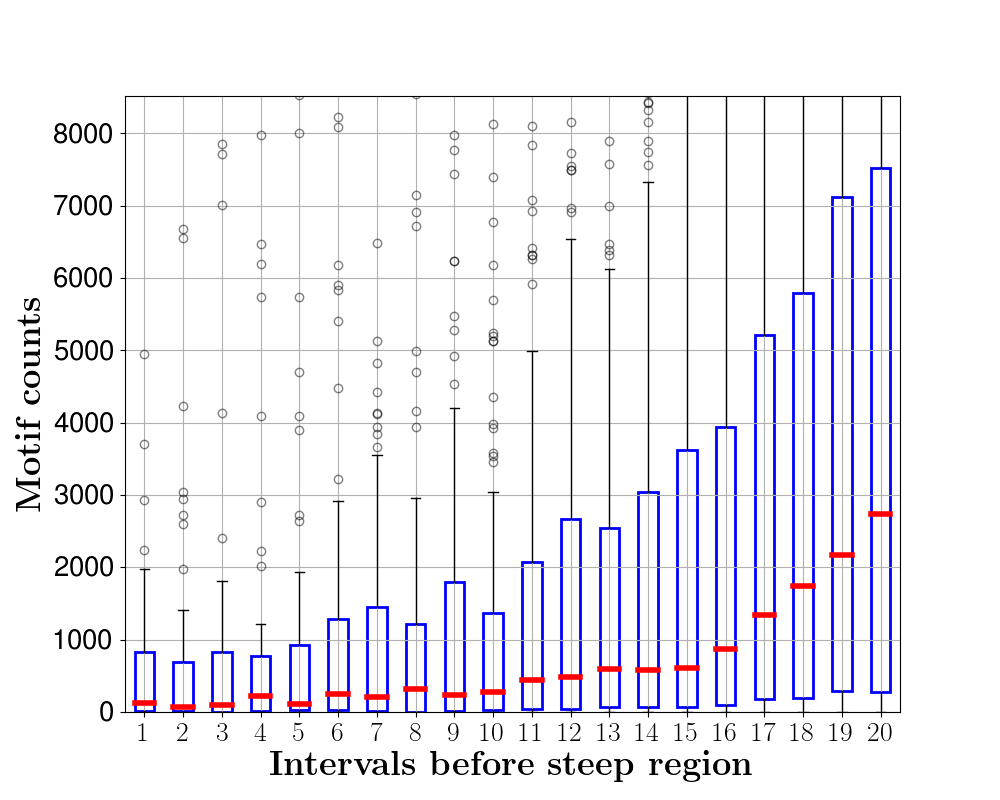}
		\subcaption{M6: \ \includegraphics[width=0.4cm]{M13.png}}
		\label{(e)}
	\end{minipage}
	\hfill
	\begin{minipage}{0.3\textwidth}
		\includegraphics[width=5cm, height=3.7cm]{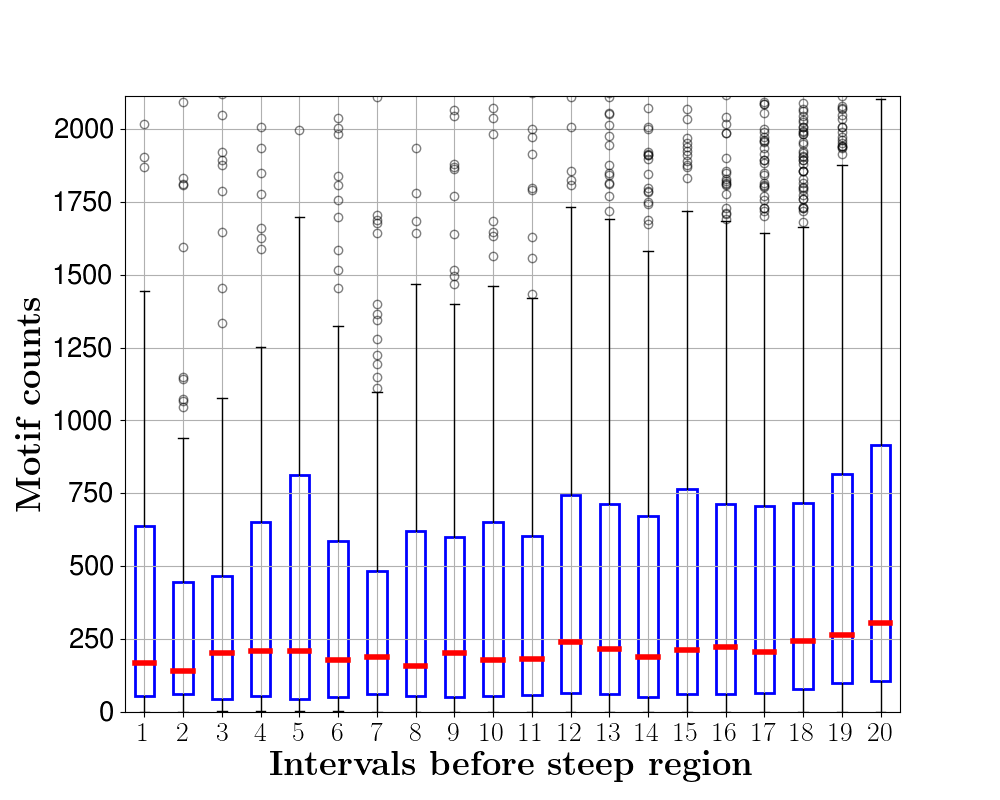}
		\subcaption{M11: \ \includegraphics[width=0.4cm]{M12.png}}
		\label{(f)}
	\end{minipage}
	\hfill
	\caption{Plots of motif counts. The intervals are increasing from left to right leading to $N_{steep}$. }
	\label{fig:mcounts_2}
\end{figure}
In order to understand whether the occurrence of these patterns are unique to the inhibition stage, we plot the motif counts in the stages preceding $\tau'_{steep}$ in Figure~\ref{fig:mcounts_2}. We observe two things here: (1) for the patterns \  \includegraphics[scale=0.04]{M3.png} \ and \includegraphics[scale=0.04]{M12.png} \ the counts for the stages preceding inhibition is an order-of-magnitude (almost 3 times for the median statistics) higher than the corresponding stages preceding the steep phase. It suggests that individuals tend to interact in a way that completes the patterns to form triads near the inhibition stage, (2) for the pattern \  \includegraphics[scale=0.04]{M2.png}, we find that counts are higher for the stages near steep phase. This suggests that there are more linear chains that drive the cascade progress faster during the steep phase, whereas the peer influence adding to the loops near the end of the cascade causes its final inhibition.

\begin{figure}[!t]
	\centering
	\hfill
	\begin{minipage}{0.4\textwidth}%
		\includegraphics[width=6cm, height=4cm]{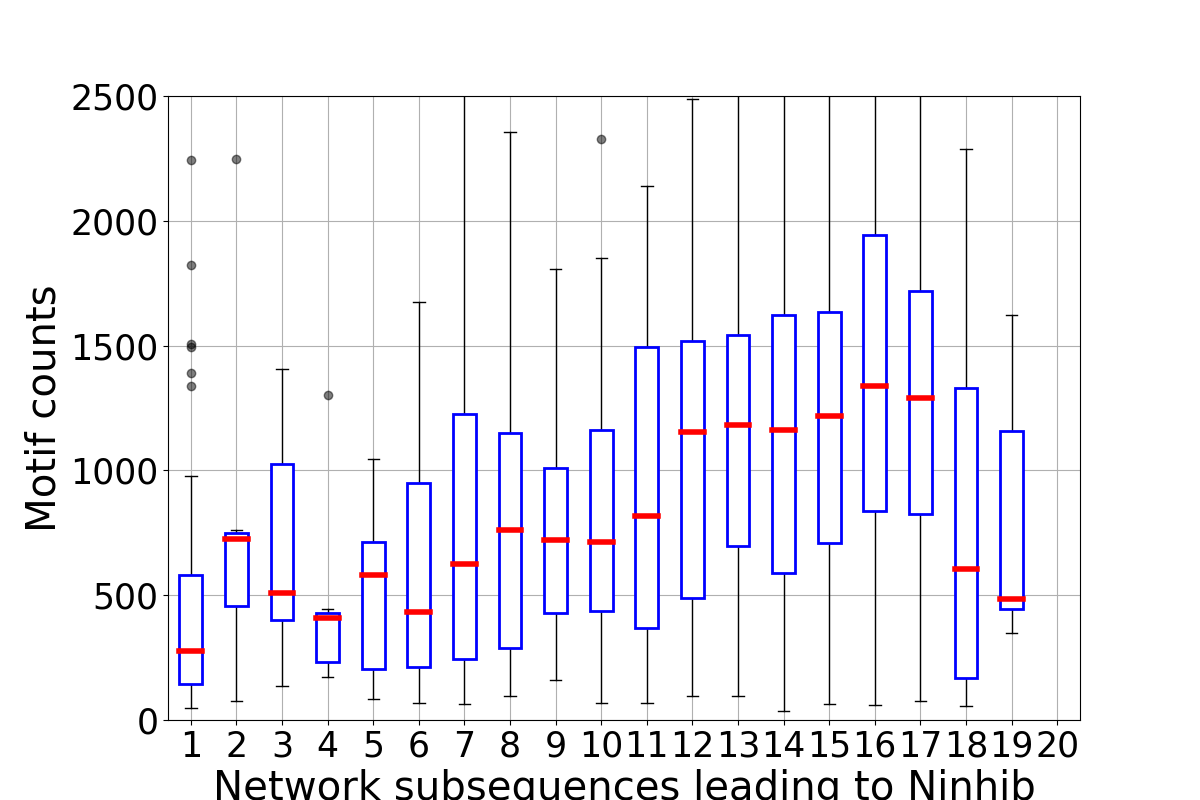}
		\subcaption{M0: \ \includegraphics[width=0.4cm]{M0.png}}	\label{a}
	\end{minipage}
	\hfill
	\begin{minipage}{0.4\textwidth}
		\includegraphics[width=6cm, height=4cm]{Motif_0.png}
		\subcaption{M0: \ \includegraphics[width=0.4cm]{M0.png}}
		\label{b}
	\end{minipage}
	\hfill
	\\
	\hfill
	\begin{minipage}{0.4\textwidth}%
		\includegraphics[width=6cm, height=4cm]{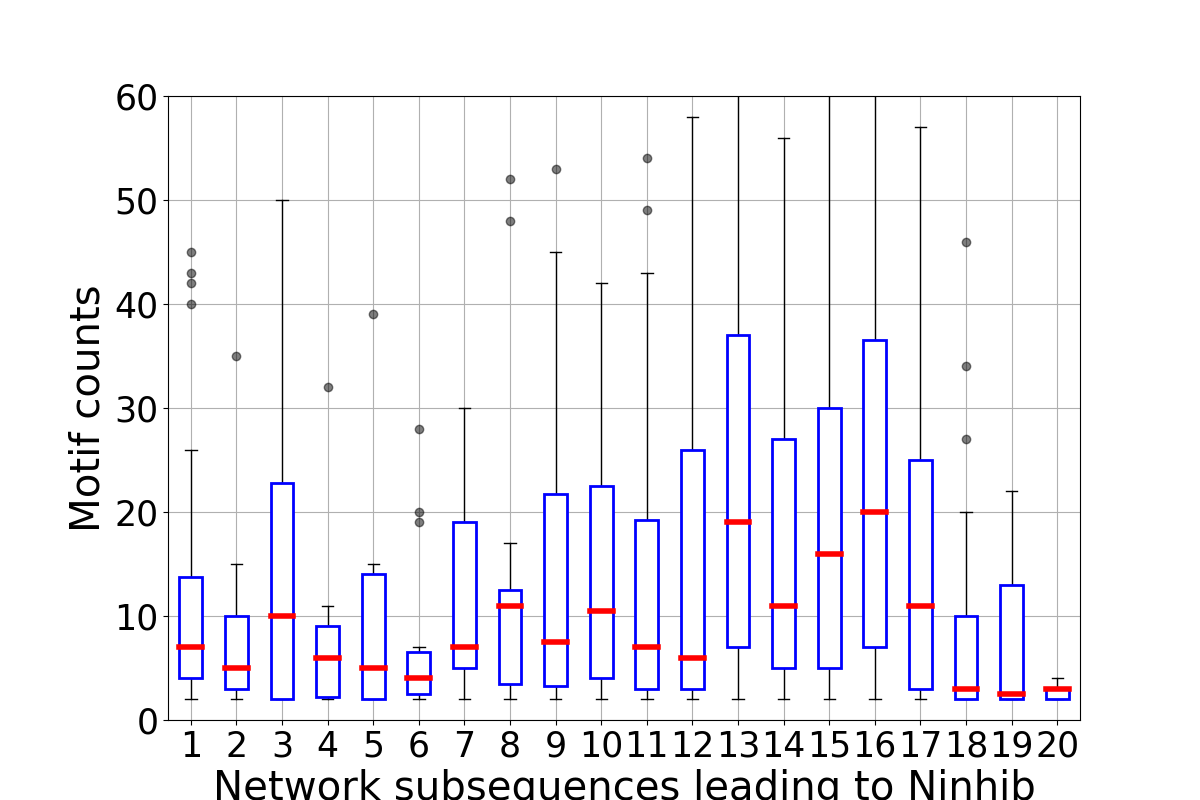}
		\subcaption{M1: \ \includegraphics[width=0.4cm]{M1.png}}
		\label{c}
	\end{minipage}
	\hfill
	\begin{minipage}{0.4\textwidth}
		\includegraphics[width=6cm, height=4cm]{Motif_1.png}
		\subcaption{M1: \ \includegraphics[width=0.4cm]{M1.png}}
		\label{d}
	\end{minipage}
	\caption{Plots of motif counts of size 3 over time. The intervals are increasing from left to right leading to $N_{steep}$ and $N_{inhib}$. }
	\label{fig:mcounts_3}
\end{figure}

In order to understand whether the dynamics of the larger size motifs, specifically the ones with triangles, are rooted in the formation of the 3-node motif, we empirically observe the counts of the 3-node motifs shown inf Figure~\ref{fig:mcounts_3}. With only two possible 3-node patterns: \  \includegraphics[scale=0.04]{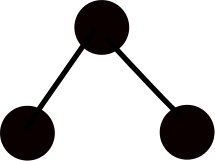} \  and \  \includegraphics[scale=0.04]{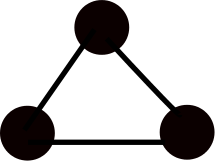} \ we observe the following trend: for both the patterns, the intervals preceding the inhibition stage exhibit higher motif counts. This can be attributed to the way in which motifs are counted. Although the canonical labeling procedure for graph isomorphism \cite{canonical} ensures that every set of nodes is counted once for a pattern, we note that for every additional triangle that is added, there are more linear chains that are counted. For example, considering the 5-node loop A-B-C-D-E-A, when we attach a chord A-C to form a single triangle A-B-C. However, we get 3 new chains of size 3: A-B-C, B-C-A and C-A-B. Therefore one of the possible reasons for higher counts of motifs in general is that there are more loops towards the inhibition period.  The plots in Figures~\ref{fig:mcounts_3} \subref{b} and \subref{d} demonstrate this.\\

\textit{RQ 2: Does the higher appearance of patterns with triads explain the slowing down of the cascade growth?} \\

\begin{figure}[!t]
	\centering
	\includegraphics[width=8cm, height=4cm]{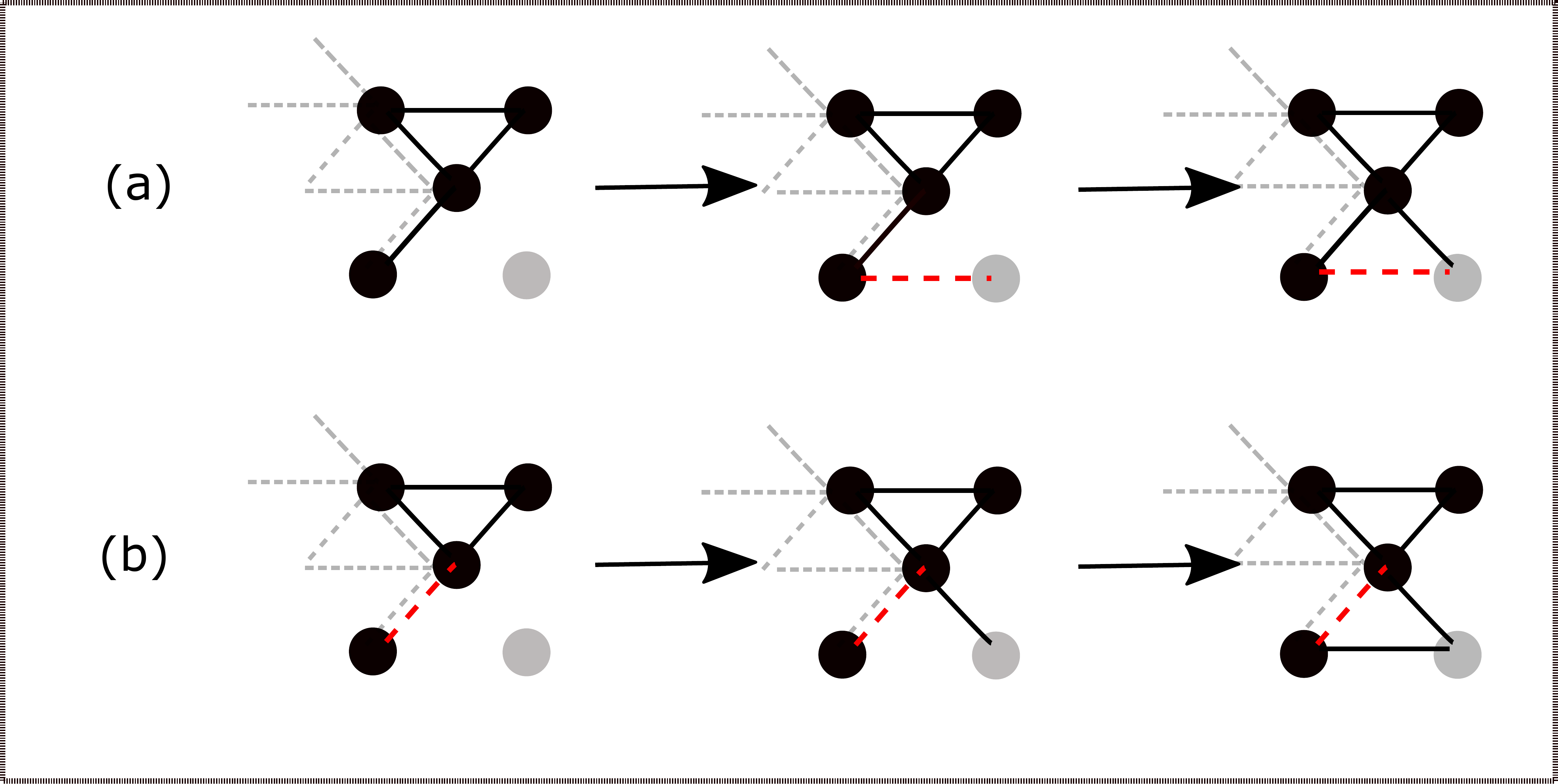}
	\caption{Visual illustration to show how past influences and preferences from social network or historical traces leads a motif to form loops. }
	\label{fig:cause_example}
\end{figure}

To understand how the impact of past influence creates more loops and therefore higher motif count in general, we use the following example illustrated in Figure~\ref{fig:cause_example} where in  subfigures (a) and (b), we consider a 4-node motif at some stage in the cascade with the node colored in gray being the  next potential node to adopt the cascade. Observe that to form a loop in a cascade, one of the edges has to belong  to a social network (who-follows-whom / mutual friend) or a historical diffusion network.  This can be observed from the assumption of diffusion in cascades where there is only one parent from which a node can reshare information leading to the implicit tree structure of cascades. But when a historical link exists in the cascade network, over which there is no information propagation for the current cascade (denoted by the red edges in Figure~\ref{fig:cause_example}, a loop is created (otherwise if information had been propagated over that link, then the social network link would be removed and only the diffusion link would be present, since we do not consider parallel edges in the networks). This gives rise to two very important observations that explains the slowing down of the cascade growth: (1) presence of loop indicates that a node has received multiple exposures (or multiple rounds of the same information) from its friends before it finally adopts. It explains that when users need multiple exposures or rounds of social reinforcement for them to adopt an information, information starts spreading slowly thereby indicating the death of the cascade (2) secondly, when loops are created due to existence of users with past close connections who have engaged in spreading information among themselves, there is a strong chance of diffusion inhibition, since now the information has lesser chances to move out of the loop and the information stays within, a phenomena which has been studied before \cite{David:2010:NCM:1805895}. We also note that these observations about the comaparison between the steep phase and the inhibition stage are independent on the network size, since each temporal network size $|V^N|$ is same throughout. \\

\textit{RQ3: Does the significance of the motif patterns explain the structural organization of the network leading to the inhibition stage?} \\

To quantify the extent to which these patterns are indigenous to the temporal networks preceding inhibition, the occurrence of motif patterns that constitute the motif profile must be followed by comparison with the null model and subsequent significance test metrics. We follow the methods described in the work done in \cite{motif_princ,Milo1538} and as has been the standard method to evaluate the significance of motifs profile in these studies. In this paper, we adopt the $z$-score metric for hypothesis testing where the alternate hypothesis is the frequency of motif patterns obtained from the original network and the null model is the frequency obtained from ensemble of random networks with specific attributes.  
Let a size-$k$ motif pattern $M_k \in \mathcal{M}_k$ occur $MC_{k, M}^{input}$ times in the actual or $input$ network. We consider a set of random networks under the null model for evaluating the significance of the obtained motif profile in the original network. The random network has the same degree sequence as the original network and the randomization procedure involves switching a pair of edges $a$ $\rightarrow$ $b$ and $c$ $\rightarrow$ $d$ to form new edges $a$ $\rightarrow$ $d$ and $b$ $\rightarrow$ $c$.  Using this choice of the null model, we computed the set of motifs from these random networks and use the $z$-scores as a measure of significance. Let a size-$k$ motif pattern $M_k$ occur $MC_{k, M}^{random}$ times in the $random$ network. Then the $z$-score for $M_k$ is defined as follows:

\begin{equation}
z(M_k) = \frac{MC_{k, M}^{input} - MC_{k, M}^{random}}{\sqrt{\sigma_{random}^2}}
\end{equation}

\begin{figure}[t!]
	\centering
	\begin{minipage}{0.3\textwidth}%
		\includegraphics[width=5cm, height=3.7cm]{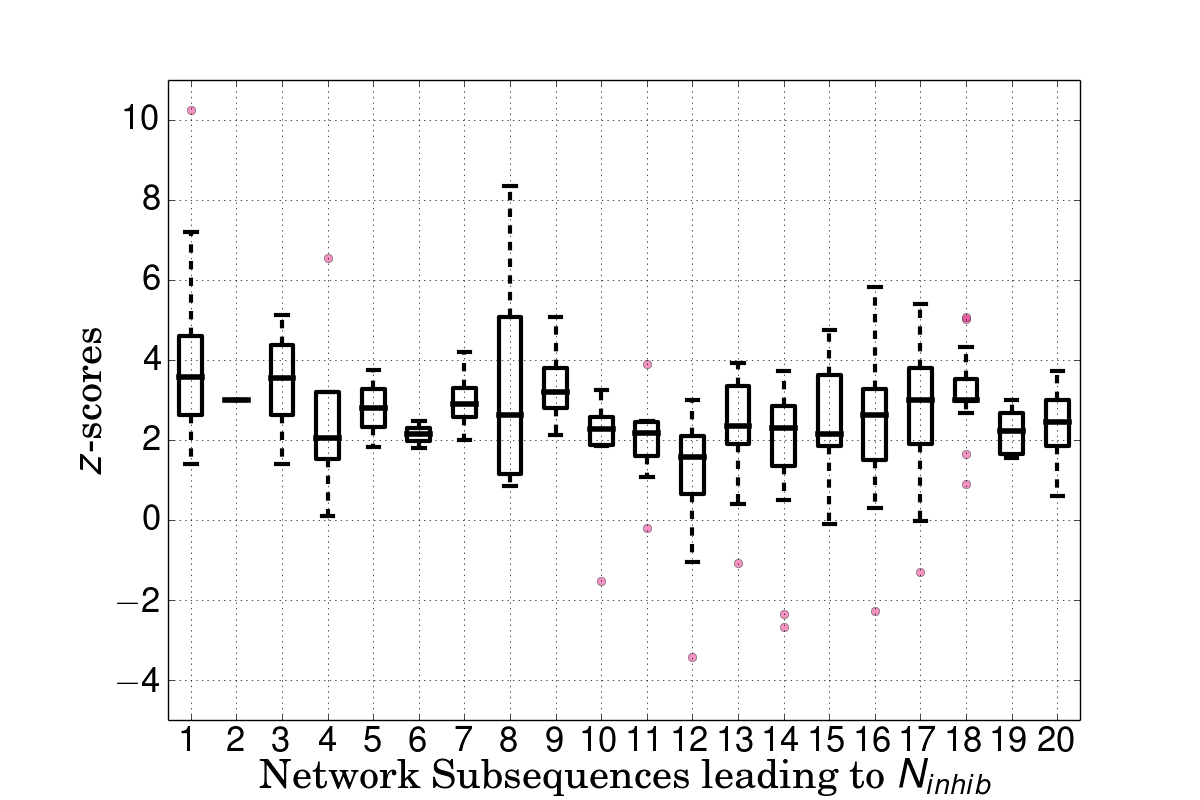}
		\subcaption{M1: \ \includegraphics[width=0.4cm]{M2.png}}
		\label{(a)}
	\end{minipage}
	\hfill
	\begin{minipage}{0.305\textwidth}
		\includegraphics[width=5cm, height=3.5cm]{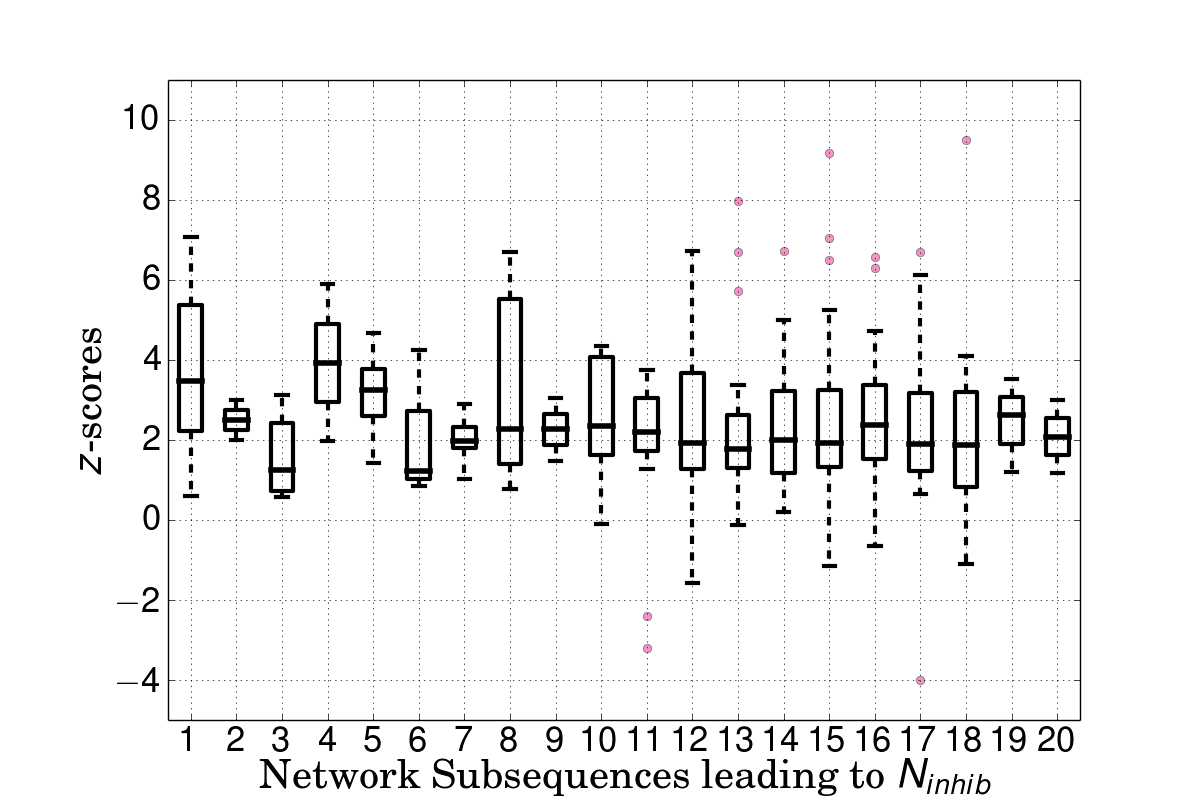}
		\subcaption{M2: \ \includegraphics[width=0.4cm]{M1.png}}
		\label{(b)}
	\end{minipage}
	\hfill
	\begin{minipage}{0.3\textwidth}
		\includegraphics[width=5cm, height=3.7cm]{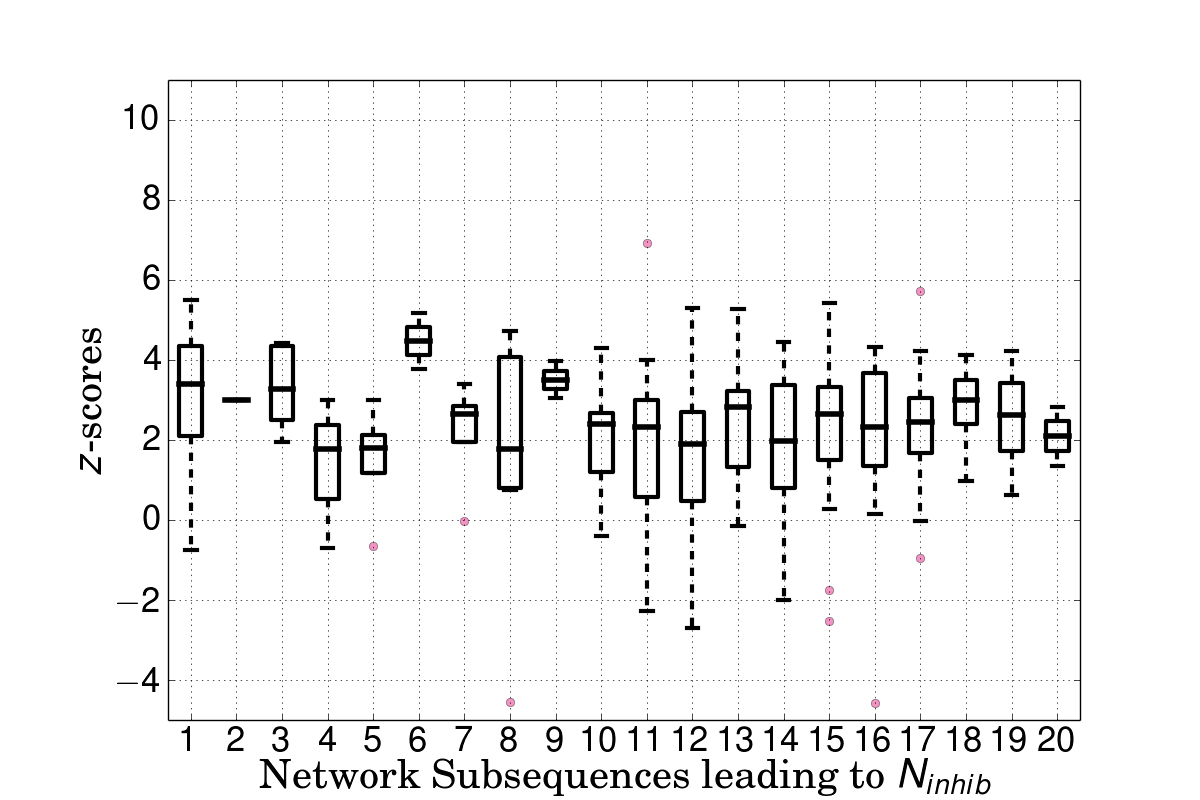}
		\subcaption{M3: \ \includegraphics[width=0.4cm]{M0.png}}
		\label{(c)}
	\end{minipage}
	\hfill
	\\
	\begin{minipage}{0.3\textwidth}%
		\includegraphics[width=5cm, height=3.7cm]{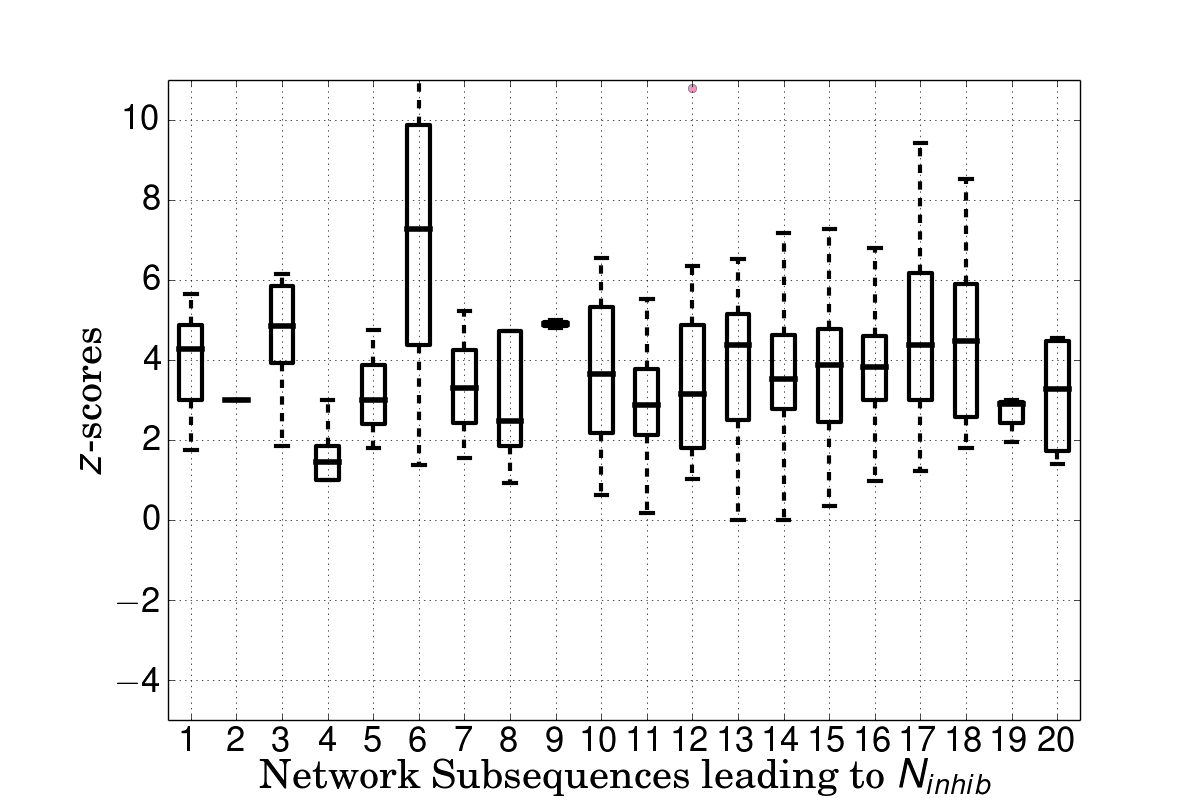}
		\subcaption{M5: \ \includegraphics[width=0.4cm]{M3.png}}
		\label{(d)}
	\end{minipage}
	\hfill
	\begin{minipage}{0.305\textwidth}
		\includegraphics[width=5cm, height=3.7cm]{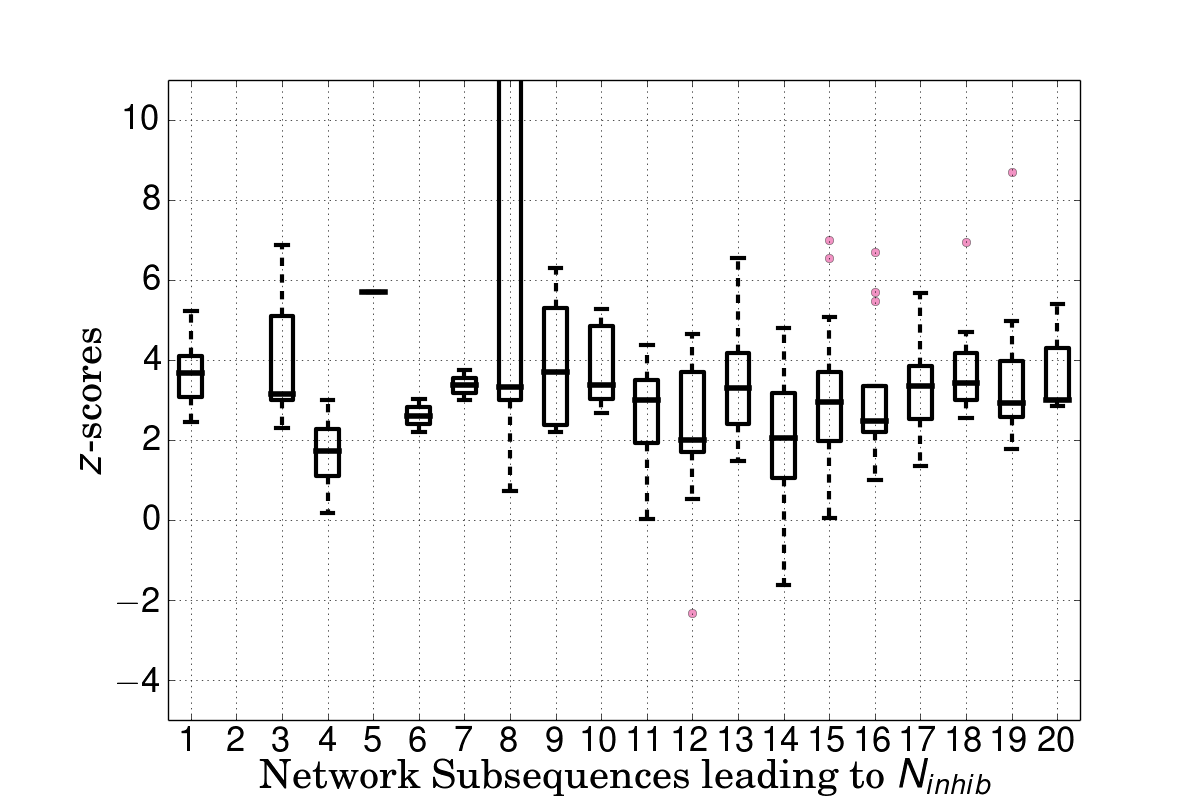}
		\subcaption{M6: \ \includegraphics[width=0.4cm]{M13.png}}
		\label{(e)}
	\end{minipage}
	\hfill
	\begin{minipage}{0.3\textwidth}
		\includegraphics[width=5cm, height=3.7cm]{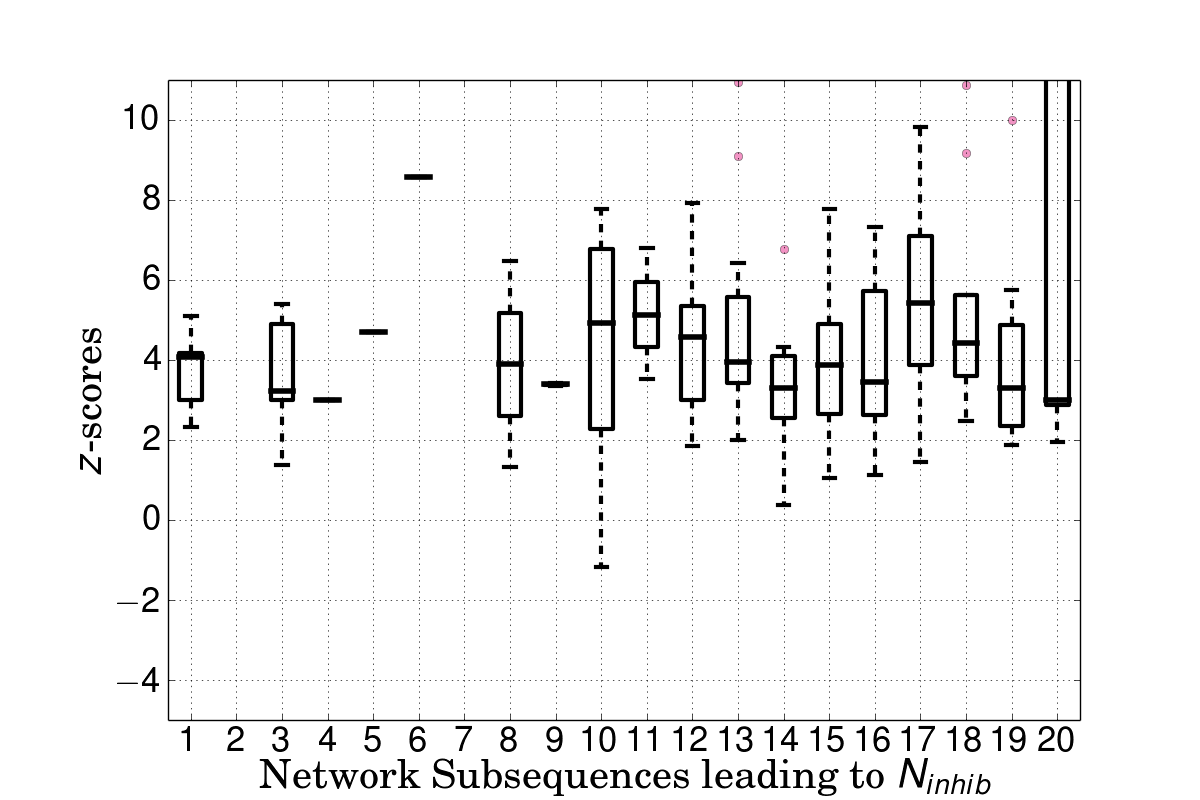}
		\subcaption{M11: \ \includegraphics[width=0.4cm]{M12.png}}
		\label{(f)}
	\end{minipage}
	\hfill
	\caption{Plots of $z$-scores for the respective motif patterns. The intervals are increasing from left to right leading to $N_{inhib}$. }
	\label{fig:zscores_5}
\end{figure}

where $\sigma_{random}$ denotes the standard deviation of the motif counts obtained from the ensemble of random networks chosen for the null model. The p-value in this case represents the probability  of a motif to appear an equal or greater number of times in a random network than in the actual network. In general, a motif is usually considered significant if the associated p-value is less than 0.01 or $z$-score is greater than 2.0. Although the network size of each $N$ in our case is fixed, that is $V^N$ is chosen manually and kept constant throughout for all cascades, the $z$-score largely depends on the organization of the network structure in the temporal networks which we discuss in detail in the experimental sections. In this work, we consider an ensemble of 100 networks and apply the switching strategy.
\begin{figure}[!t]
	\centering
	\includegraphics[width=6.5cm, height=6cm]{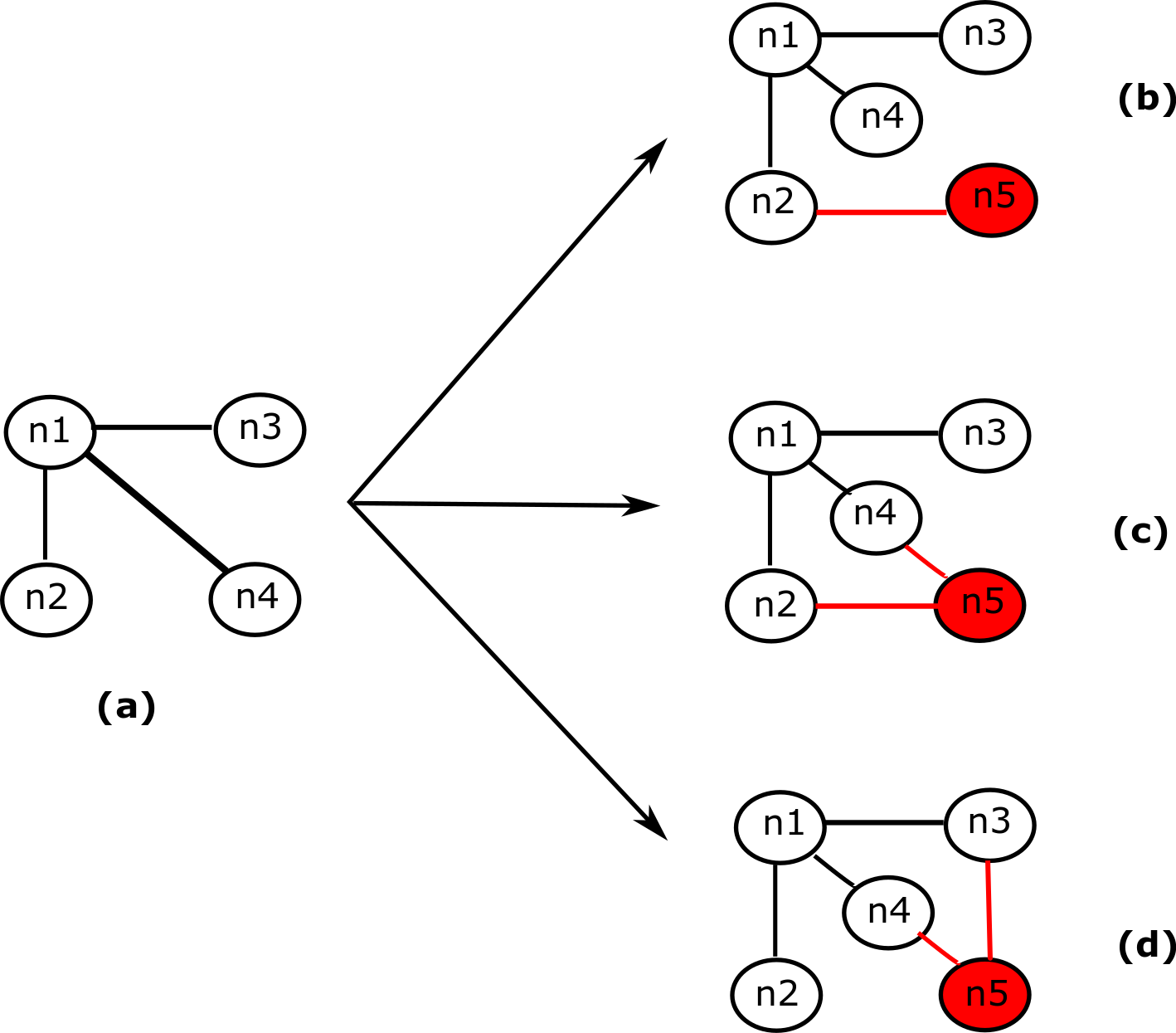}
	\caption{An example of motif transition. Figure (a) denotes a 4-size motif in some network $N$ for a cascade. Figures (b), (c) and (d) show 3 possible transition patterns into 5-sized motifs. Observe that in this ideal transition, all the nodes in motif in (a) are present in (b), (c) and (d) which may not always be the case. The red node and edges denote new additions. Note, here we are not just focusing on the information flow direction (since we disregard directed edges), but how in the ecosystem of emerging motifs, the new node fits in. That is, whether a 4-node motif triggers new activity within susceptible nodes (red edges can be in any direction). }
	\label{fig:motifs_transition}
\end{figure}

We plot the distribution of these $z$-scores over the intervals shown in Figure~\ref{fig:zscores_5} where we plot the scores considering the  null model with the \textit{degree-correlated} random networks.  We observe two important aspects of the motif abundance of acyclic patterns and those with loops: (i) first, the appearance of patterns with loops like \  \includegraphics[scale=0.04]{M12.png} \, \  \includegraphics[scale=0.04]{M13.png} \ and \  \includegraphics[scale=0.04]{M3.png} \ shown in Figures~\ref{fig:zscores_5} \subref{d} to \subref{f} are more significant than the acyclic patterns like \includegraphics[scale=0.04]{M2.png} \, \  \includegraphics[scale=0.04]{M1.png} \ and \  \includegraphics[scale=0.04]{M0.png} \ shown in Figures~\ref{fig:zscores_5} \subref{a} to \subref{c} respectively, observed from the median $z$-scores over all the cascades. For the interval $N_{inhib}$, we obtain the median $z$-score values of 3.2, 3.1 and 3.0 for the patterns  \  \includegraphics[scale=0.04]{M3.png} \, \  \includegraphics[scale=0.04]{M13.png} \ and \  \includegraphics[scale=0.04]{M12.png} \ respectively denoting these patterns are statistically significant. For the pattern \  \includegraphics[scale=0.04]{M0.png} \, we observe  a lower median $z$-score of  1.99 for $N_{inhib}$ showing that the abundance of these motifs might not always suggest that these patterns are statistically significant with respect to the network structure but might just be over-represented due to the high degrees of a few nodes. However, we observe that the $z$-scores do not follow the same dynamics as the motif counts in Figure~\ref{fig:mcounts_1}, and this can be attributed to the fact that the $z$-scores represent the unnormalized difference in motif counts in the two network models. Since for the null model we consider degree correlated networks as the original networks (which implies a stronger testing setup), the absence of similar dynamics reflects the structural organization of the networks. In networks which exhibit patterns with triads, the chances of having triads would also be higher for the null model due to degree correlation, but the extent to which the differences are significant is what $z$-scores denote. Thus the appearance of triads is due to change in edge density of the networks which are an effect of adding historical communication links to the networks. Note that we do not control how many edges are added from the historial social network to the current cascade topology - they are dictated by evidences from data and in that sense we do not control the network densities either.

\begin{algorithm}[H]
	\KwIn{Sequence of temporal networks $G^N_C$, $ N \in\mathcal{N}$ for cascade $C$, $steep$ and $inhib$ subsequence indices for $C$}
	\KwOut{Transition array $\mathbf{T}$.}
	
	$i$ = $inhib$ \tcp*{start subsequence for analysis is $\tau'_{inhib}$}
	
	\While{$i$ is not $steep$}{
		\tcc{Define 2 storage dictionary variables holding the [motif pattern<->(count, instances)] pairs}
		$\mathcal{M}_4$ $=$ dictionary mapping 4-node patterns to their frequencies and vertex maps \;
		$\mathcal{M}_5$ $=$ dictionary mapping 5-node patterns to their frequencies and vertex maps \; 
		
		$\mathcal{M}_4$ $\leftarrow$ $Compute\_motif\_counts(G^{N_{i-1}}, 4$) \tcp*{Extract  and store patterns of 4-sized motifs and their count along with vertex maps in the network   $N_{i-1}$}
		$\mathcal{M}_5$ $\leftarrow$ $Compute\_motif\_counts(G^{N_{i}}, 5$) \tcp*{Extract and store patterns of 5-sized motifs and their count along with vertex maps in the network   $N_{i}$}
		\tcc{The steps in the following iterations count the number of instances of size 4 of a prticular pattern that transition into instances of size 5 having common nodes and belonging to the same or other patterns}
		\tcc{Iterate over the 4-node patterns}
		\For{each $M_4$ $\in \mathcal{M}_4.keys()$}{ 
			\tcc{The motif counts for the current 4-node pattern should cross a threshold frequency}
			\If{$\mathcal{M}_4[M_4].values().count$ $<$ threshold\_count\_1}{
				Skip this motif and continue \tcp*{Filtering by motif count}			
			}
			\tcc{Iterate over the 5-node patterns}
			\For{each $M_5$ $\in \mathcal{M}_5.keys()$}{ 
				
				\tcc{The 4-node pattern must be a subgraph of the 5-node pattern and the motif counts for the current 5-node pattern should cross a threshold frequency}
				\If{$M_4$ is not a subgraph of $M_5$ \textbf{and} 
					$\mathcal{M}_5[M_5].values().count$ $<$ threshold\_count\_2 }{
					Skip $M_5$ and continue with other motifs \tcp*{Filtering by subgraph patterns and counts}
				}
				$\mathbf{T}$[$M_4$][$M_5$] = 0 \;
				\tcc{Count the number of common nodes between a motif instance of size 4 of pattern $M_4$ and an instance of size 5 of pattern $M_5$}
				\For{each $m_4$ $\in$ $\mathcal{M}_4[M_4].values().instances$}{
					\For{each $m_5$ $\in$ $\mathcal{M}_5[M_5].values().instances$}{
						\tcc{Condition to check: whether all nodes of $m_4$ are in $m_5$}
						\If{\textit{length}($V^{m_4}$ $\cap$ $V^{m_5}$) $==$ 4}{
							$\mathbf{T}$[$M_4$][$M_5$] = $\mathbf{T}$[$M_4$][$M_5$] + 1 					\tcp{Increment and store the counts in this transition matrix}
						}
					}
				}
				
			}
		}
		$i$ = $i$ - 1 \tcp{We iterate backwards from $inhib$ to $steep$}
	}
	return $\mathbf{T}$
	
	\caption{Motif Transition Algorithm}
	\label{Algo:MTransAlgo}
\end{algorithm}

\subsection{Motifs Transitions}
One of the ways to understand how information spreads through the stages of the cascade lifecycle and what leads to inhibition is to understand how these motifs transition from one pattern to another pattern as the cascade progresses. Since motifs characterize the network structure,  in a way understanding this transition would also uncover the process in which the network grows and how individuals share information. To this end, we consider the transition of $k-1$-node motifs in a temporal network $N_{i-1}$ into $k$-node motifs in the succeeding temporal network $N_i$, $\forall i \in [steep, \mathcal{Q}]$. This would enable us to understand whether particular $k-1$ motif patterns \textit{trigger} the $k^{th}$ new node to adopt the cascade thus forming the $k$ node motif. We consider all transition permutations between motif patterns of size $k-1$ and of size $k$ in each pair of consecutive networks and compute the number of $k-1$-node motif instances of some pattern $M_{k-1}$ that changed into $k$-node motif instances. An example of motif transition is shown in Figure~\ref{fig:motifs_transition}.  For any two patterns $M_{k-1} \in \mathcal{M}_{k-1}$ and $M_k \in \mathcal{M}_k$, we extract $k$-node motif instances having pattern $M_k$ in the current temporal network and $k-1$ node motif instances having pattern $M_{k-1}$ in the preceding network. A successful transition of a $k-1$-node instance into a $k$-node instance follows these two conditions: (1) the $k-1$-node instance should be a subgraph of the $k$-node instance (2) all the $k-1$ nodes from the $k-1$ pattern in$N_{i-1}$  should be present in the $k$  node-motif in $N_{i}$ (only one new node that has adopted the cascade in the the $i^{th}$ subsequence should be present).
In our case, $k$ = 5, that is we consider transitions from 4-node to 5-node motifs. Figure~\ref{fig:motifs_transition} shows an example of transition from 4-node to 5-node motifs. Notice that while there is always one new node added in the 5-node motif, the number of triggering edges might be greater than 1, depending on the influence from past history i.e. whether the new node had interactions with the other activated nodes in the past. This is demonstrated in transitions from (a) to (c) and (a) to (d).

A detailed algorithm on computing the pattern transition count is given in Algorithm 1. Briefly, for a given cascade, for any two consecutive temporal networks $N$, we compute and store the mapping of 4-node and 5-node patterns along with the count of the motif instances belong to the patterns and the vertex maps for those instances found in the networks. The \\ $Compute\_motif\_counts(G^N, \ k)$ which performs this operation takes as input the size of the motifs $k$ to be extracted and the temporal network $G^N$. It then uses the FANMOD algorithm to return a dictionary whose keys are the motif patterns of size $k$ (structures) and for each pattern the dictionary maps the count and the vertex maps in $N$ to that pattern. The algorithm then goes on to find the number of transitions following the constraints mentioned and returns a transition array $T$ that stores the count of motif transitions for 2 given 4-node and 5-node patterns for each $N$. However in order to increase efficiency and speed up computation, we ignore all transitions from 4-node to 5-node patterns where the motif counts for either pattern in a pair does not cross a threshold count for the temporal networks in consideration. We select these threshold counts keeping a different threshold for each pattern among the 21 5-node patterns and 5 4-node patterns by observing the counts of a carefully selected set of cascades.  We do not consider $k$ node to $k$ node transitions in successive networks since with these transitions, there would be no addition of any node and hence such transition would not capture the incremental diffusion process.

\begin{figure}[!t]
	\centering
	\begin{minipage}{0.305\textwidth}
		\includegraphics[width=5cm, height=3.7cm]{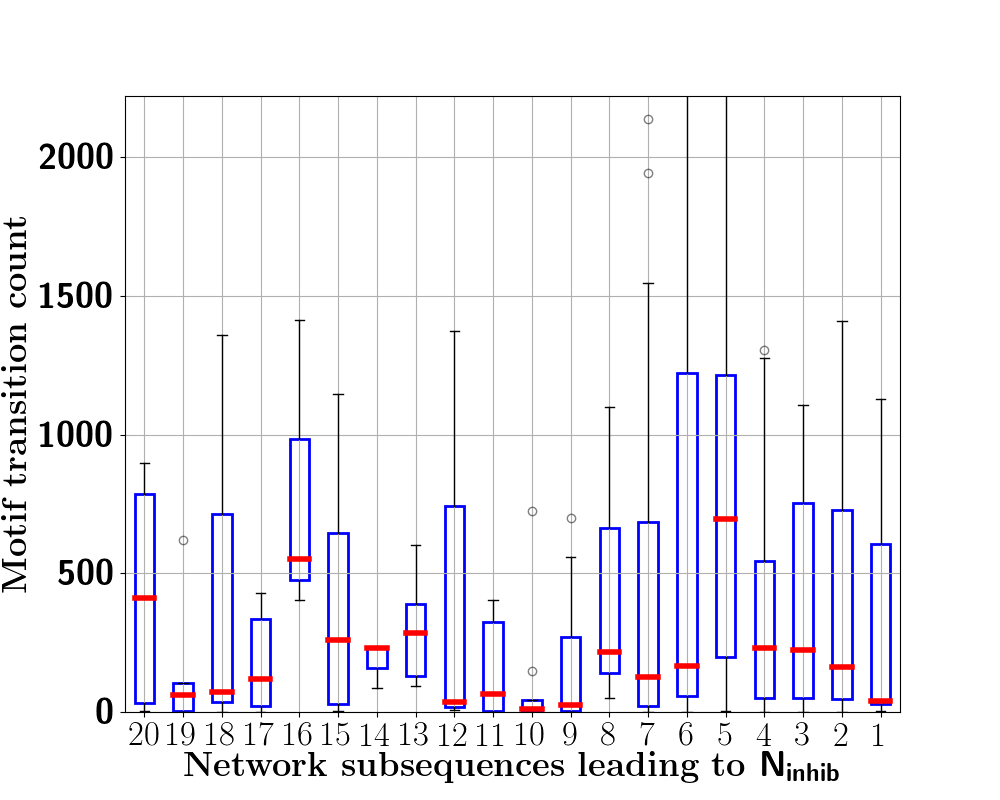}
		\subcaption{\includegraphics[width=0.4cm]{M1_4.png} \ \ \includegraphics[width=0.6cm, height=0.4cm]{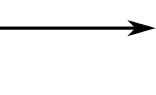} \  \  \includegraphics[width=0.4cm]{M2.png}}
		\label{mt:b}
	\end{minipage}
	\hfill
	\begin{minipage}{0.3\textwidth}%
		\includegraphics[width=5cm, height=3.7cm]{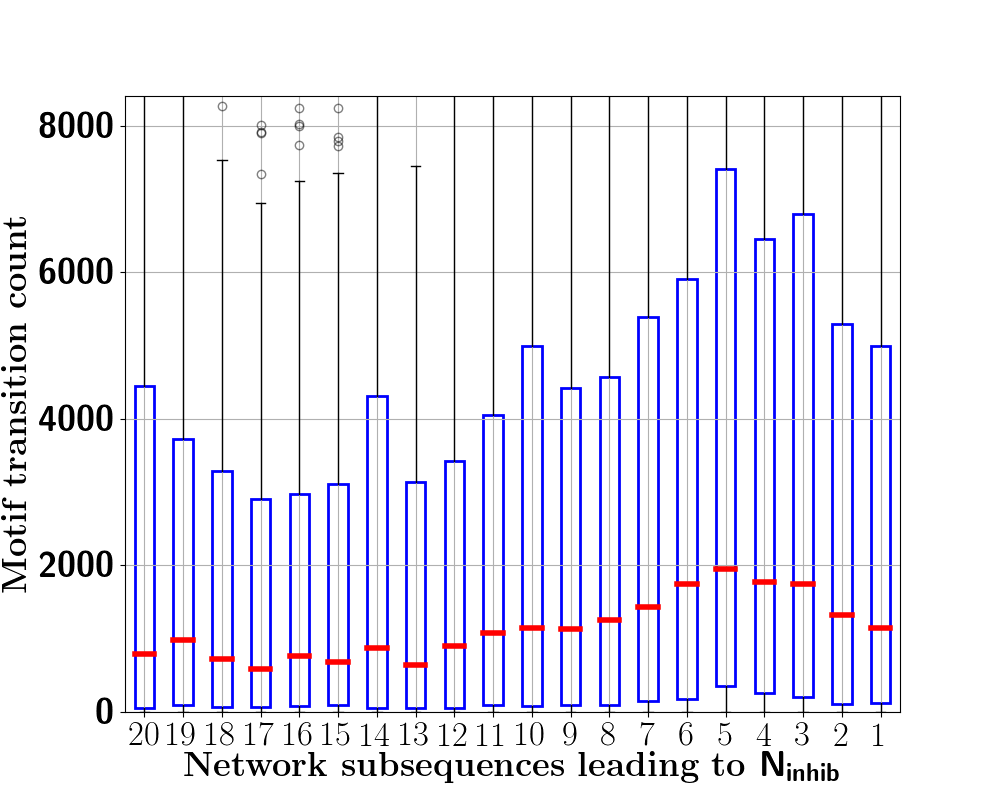}
		\subcaption{\includegraphics[width=0.4cm]{M1_4.png} \ \ \includegraphics[width=0.6cm, height=0.4cm]{arrow.png} \   \  \includegraphics[width=0.4cm]{M1.png}}
		\label{mt:a}
	\end{minipage}
	\hfill
	\begin{minipage}{0.3\textwidth}
		\includegraphics[width=5cm, height=3.7cm]{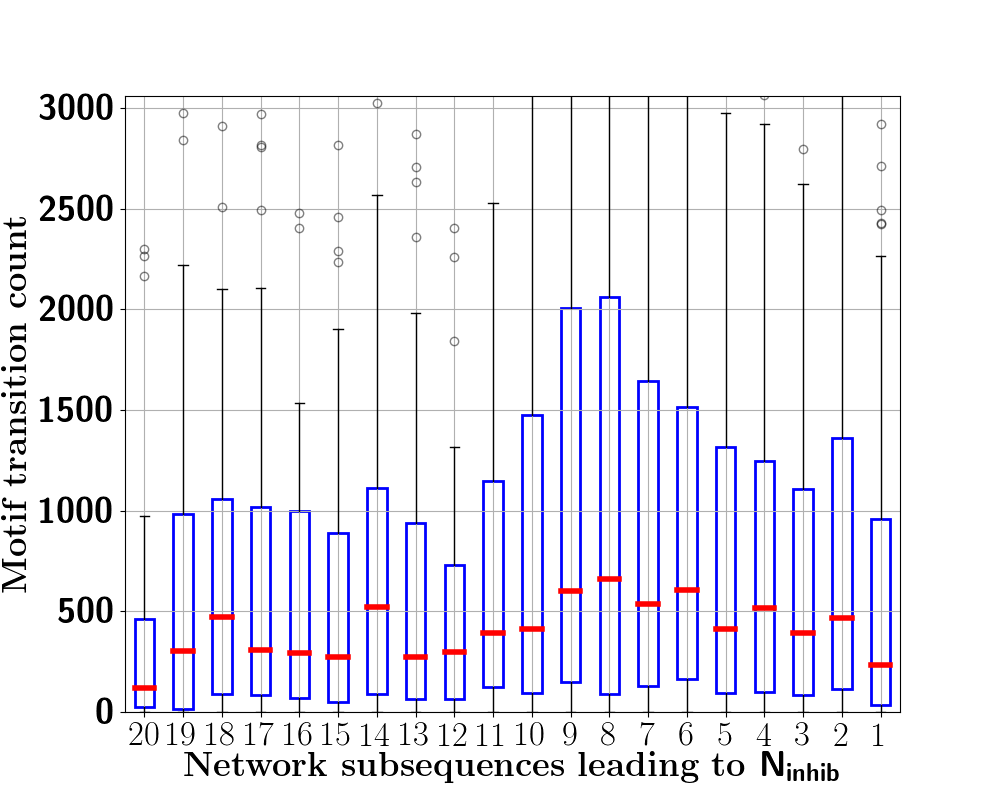}
		\subcaption{ \includegraphics[width=0.4cm]{M1_4.png} \ \ \includegraphics[width=0.6cm, height=0.4cm]{arrow.png} \   \  \includegraphics[width=0.4cm]{M13.png}}
		\label{mt:c}
	\end{minipage}
	\\
	\hfill
	\begin{minipage}{0.3\textwidth}%
		\includegraphics[width=5cm, height=3.7cm]{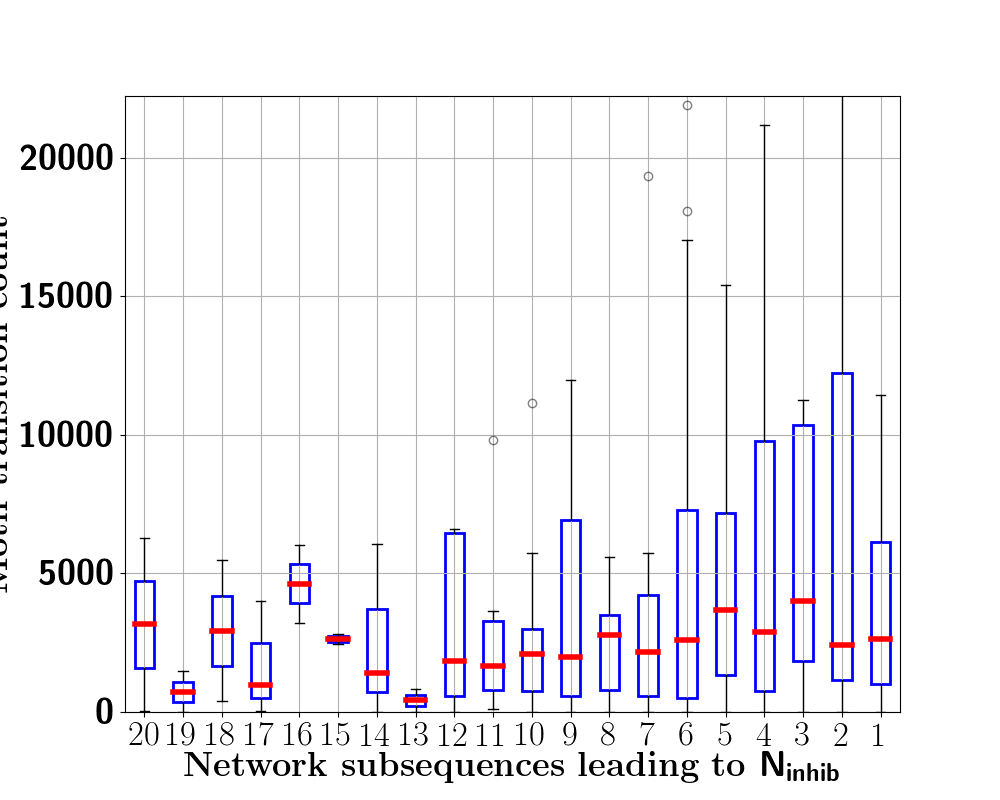}
		\subcaption{\includegraphics[width=0.4cm]{M0_4.png} \ \ \includegraphics[width=0.6cm, height=0.4cm]{arrow.png} \   \  \includegraphics[width=0.4cm]{M0.png}}
		\label{mt:d}
	\end{minipage}
	\hfill
	\begin{minipage}{0.305\textwidth}
		\includegraphics[width=5cm, height=3.7cm]{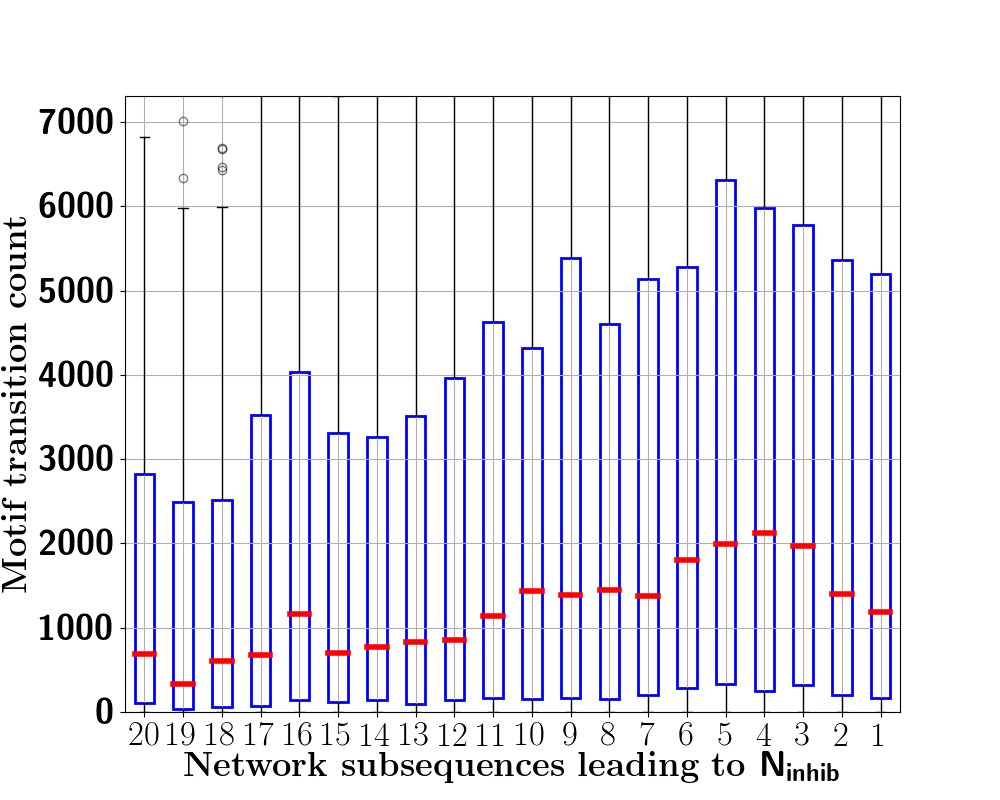}
		\subcaption{ \includegraphics[width=0.4cm]{M0_4.png} \ \ \includegraphics[width=0.6cm, height=0.4cm]{arrow.png} \   \  \includegraphics[width=0.4cm]{M1.png}}
		\label{mt:e}
	\end{minipage}
	\hfill
	\begin{minipage}{0.3\textwidth}
		\includegraphics[width=5cm, height=3.7cm]{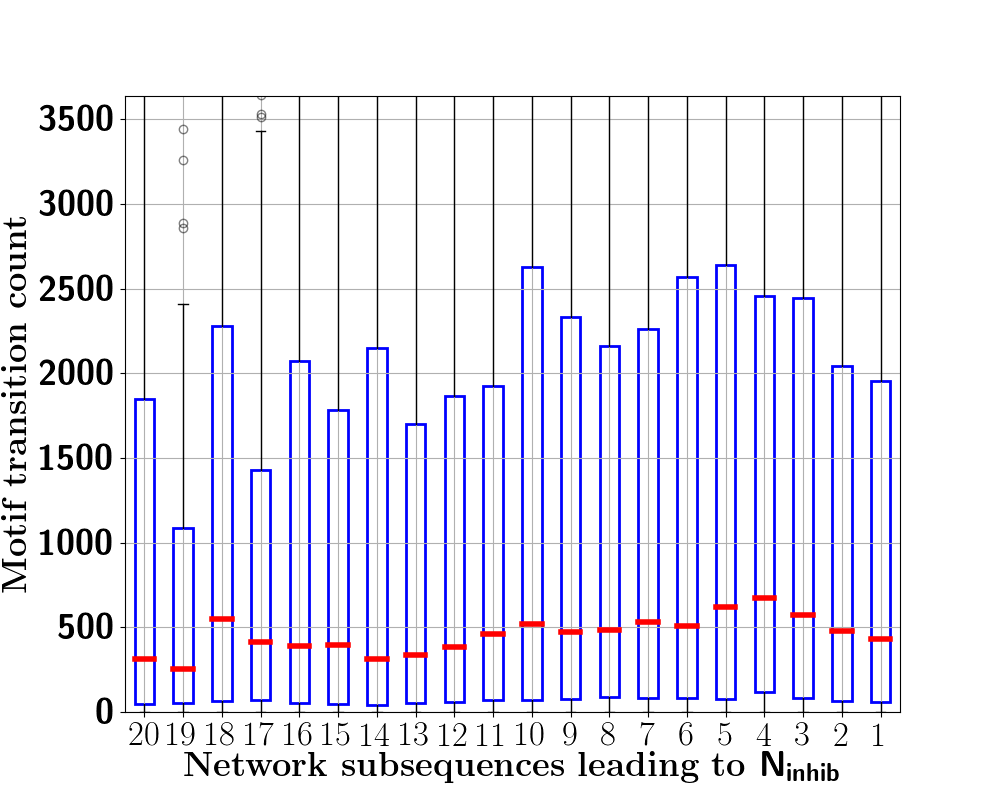}
		\subcaption{ \includegraphics[width=0.4cm]{M0_4.png} \ \ \includegraphics[width=0.6cm, height=0.4cm]{arrow.png} \  \  \includegraphics[width=0.4cm]{M3.png}}
		\label{mt:f}
	\end{minipage}
	\hfill
	\caption{Plots of motif transitions. The intervals are increasing from left to right leading to $N_{inhib}$. }
	\label{fig:motif_trans}
\end{figure}

We show the transitions from the patterns \  \includegraphics[scale=0.04]{M1_4.png} \ and \includegraphics[scale=0.04]{M0_4.png} \ in Figure~\ref{fig:motif_trans}. As shown in Figure~\ref{fig:motif_trans} \subref{mt:d}, this transition displays the highest median counts over the intervals although there is no clear trend of increasing or decreasing dynamics shown by the medians marked in red. When we compare the transitions \includegraphics[scale=0.04]{M0_4.png} $\rightarrow$ \includegraphics[scale=0.04]{M0_4.png} \ and \ \includegraphics[scale=0.04]{M0.png} $\rightarrow$ \includegraphics[scale=0.04]{M0_4.png} \ , we observe that the primary reason for higher counts for \includegraphics[scale=0.04]{M0.png} \ is that people tend to reshare from the central node more often in \includegraphics[scale=0.04]{M0_4.png}  \ than the edge node towards the inhibition region.  The transitions shown in Figure~\ref{fig:motif_trans} \subref{mt:e} and \subref{mt:f} exhibit lower counts and from the results we observed that these three transitions from \  \includegraphics[scale=0.04]{M0_4.png} \ had the highest counts. For the transition from \  \includegraphics[scale=0.04]{M1_4.png} , we find that the transitions shown in Figure~\ref{fig:motif_trans} \subref{mt:a} to \subref{mt:c} exhibit the highest count among all transitions restating our previous observations that linear chains are not favored towards the approaching inhibition stage and the transition \includegraphics[scale=0.04]{M1_4.png} $\rightarrow$ \includegraphics[scale=0.04]{M3.png} suggests the multiple exposure effect and the emergence of small circle diffusion loops.

\section{Prediction Study} \label{sec:prediction}
We end our analysis with a simple prediction study that demonstrates the effectiveness of motifs as informative features of networks. Predicting future links in temporal networks is a widely studied problem and it has some interesting applications in the real world such as predicting academic collaborations \cite{Liben2007}, terrorist links \cite{Ber2016}, recommendation systems \cite{Huang2005}. Informally, the problem of link prediction infers new pairs of interactions that are possible in future given the snapshot of the network structure at present. From the perspective of anticipating ``when'' the network would be experiencing a decayed growth in future, taking a step back into the problem of trying to predict the network structure during the terminal phase of the cascade can help us understand ahead of time

However, the main challenges of link prediction in temporal networks are:

\begin{enumerate}
	\item Unlike the traditional link prediction problem \cite{Liben2007} where a static snapshot is provided and the goal is to infer the future links in this static snapshot, in a temporal network, there are new nodes that adopt the cascade over time and for inferring the diffusion edges, one has to first infer the number of new nodes that would be added and their related node attributes which is itself is a challenging problem \cite{KimL11}. 
	
	\item Since the time required for formation of edges in a temporally evolving cascade increases with time, it becomes difficult to model the edge appearance probability given that most of these inference algorithms consider a fixed time interval for edge inference. 
	
\end{enumerate} 

Keeping these challenges in mind, since in a cascade setting, there is a constant appearance of new nodes, we instead try to predict the attributes of the network in the inhibition interval $N_{inhib}$. That is to say, using the motif features of pattern $M$ in the intervals preceding $N_{inhib}$, we target the following prediction problem : we predict $|E^{N_{inhib}}|$, that is the number of edges that would be formed in the final inhibition phase of the cascade. \\

\noindent \textbf{Why prediction:} We try to see whether we can predict what kind of network structure would be formed in the final stages ahead of time. This is a relaxation to predicting the exact edges. To understand the network structure, one could try to predict the network attributes that define the network: number of edges in the network (cascade propagation paired with social network/historical diffusion links), average degree of nodes, number of triangles formed and so on. The most intuitive and simplest among them is to see how many edges would be formed in that network. That would at least give us some information about the edge density during inhibition and in situations that require manipulating the network for viral marketing, appropriate measures can be taken. To attempt this prediction problem, we use the motif features as subgraphs. We use the motif features studied in Section~\ref{sec:motif_feat}, and consider the temporal networks preceding $N_{inhib}$  in intervals of two. That is, we consider separate learning models for features using the networks of $[N_{inhib-1}, N_{inhib-2}]$, $[N_{inhib-3}, N_{inhib-4}]$ and so on. The results would give us an idea about how far before the inhibition region can we actually start predicting the attributes of $N_{inhib}$. We use linear regression with LASSO regularization \cite{Tibshirani94}, where the cost function to minimize for the problems for motif pattern $M$  is:

\begin{equation}
Cost_M = \sum_{c=1}^{|C|} \Big\{ |E_c^{N_{inhib}}| - \sum_{i=s t}^{st+1} \sum_{f \in {MC, MT}} X^{N_i}_{c, f} * w_{f}^{N_i}  \Big\}^2 + \eta \sum_{i=st}^{st+1} \sum_{f \in {MC, MT}} {\{w_{f}^{N_i}\}}^2
\end{equation}

where the symbols $|E_c^{N_{inhib}}|$ denote the number of edges in $G^{N_{inhib}}$ for cascade instance $c$, $X$ denotes the feature value of $f$ in the set of features studied in Section~\ref{sec:motif_feat}, $st$ denotes the start of interval before the inhibition region belonging to the set $\mathcal{I}$ = $\{1, 3, 5, 7, 9\}$ for separate regression models.  We use LASSO regularization as a means of variable selection to tackle the issues that arise when the features are correlated or are sparse. 

\begin{figure}[t!]
	\centering
	\begin{minipage}{0.3\textwidth}%
		\includegraphics[width=5cm, height=3cm]{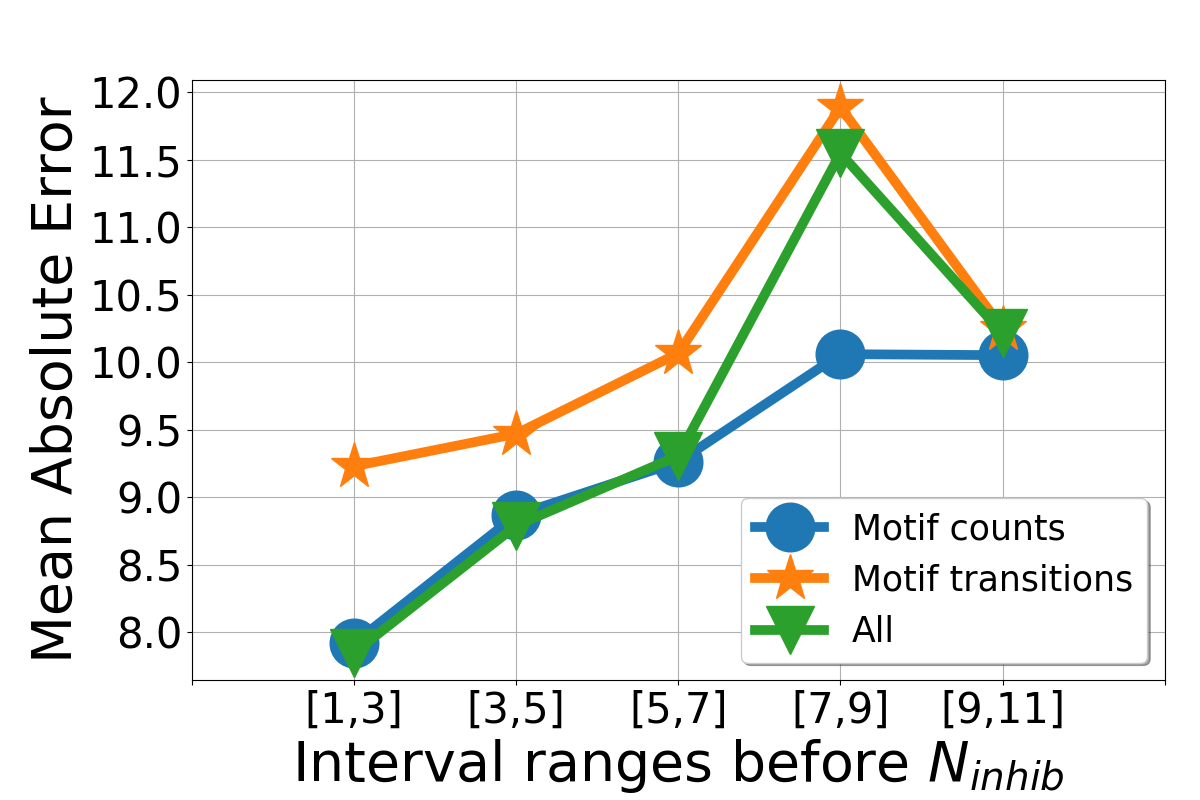}	
		\subcaption{M1: \ \includegraphics[width=0.4cm]{M2.png}}
		\label{a}
	\end{minipage}
	\hfill
	\begin{minipage}{0.32\textwidth}
		\includegraphics[width=5cm, height=3.5cm]{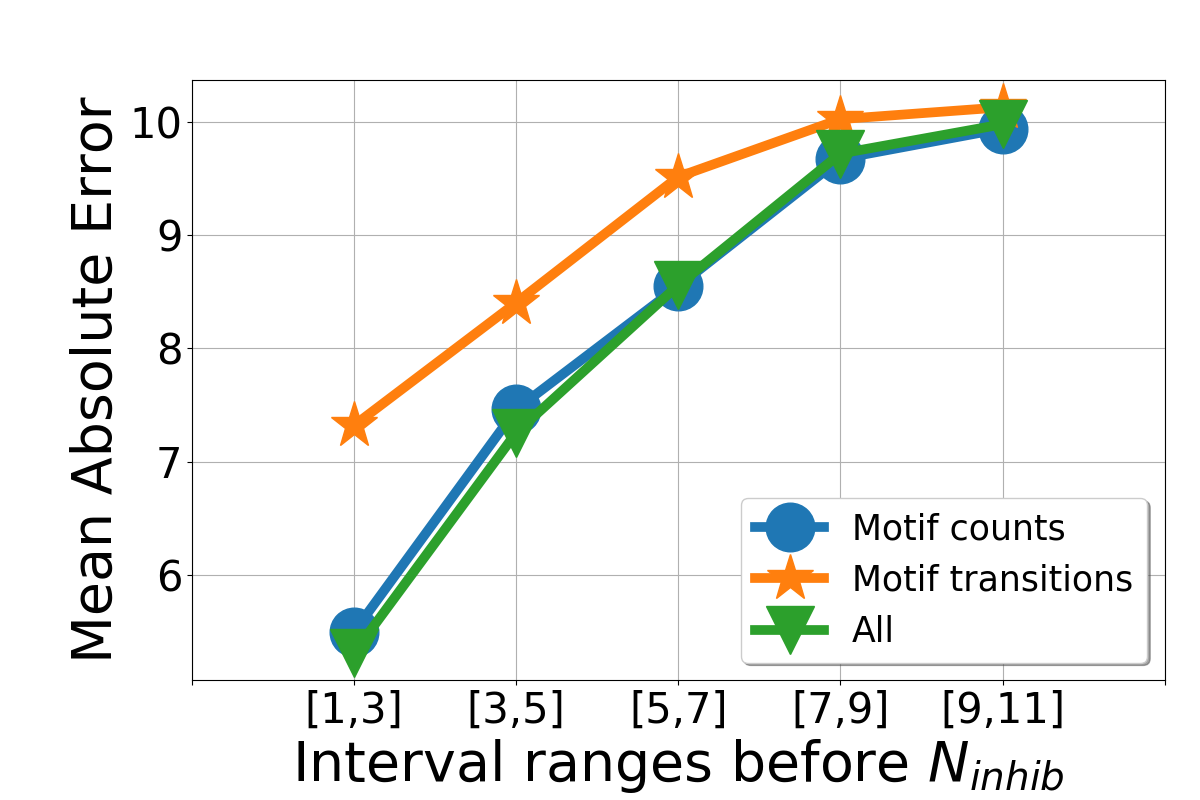}
		\subcaption{M2: \ \includegraphics[width=0.4cm]{M1.png}}
		\label{b}
	\end{minipage}
	\hfill
	\begin{minipage}{0.24\textwidth}
		\includegraphics[width=5cm, height=3.5cm]{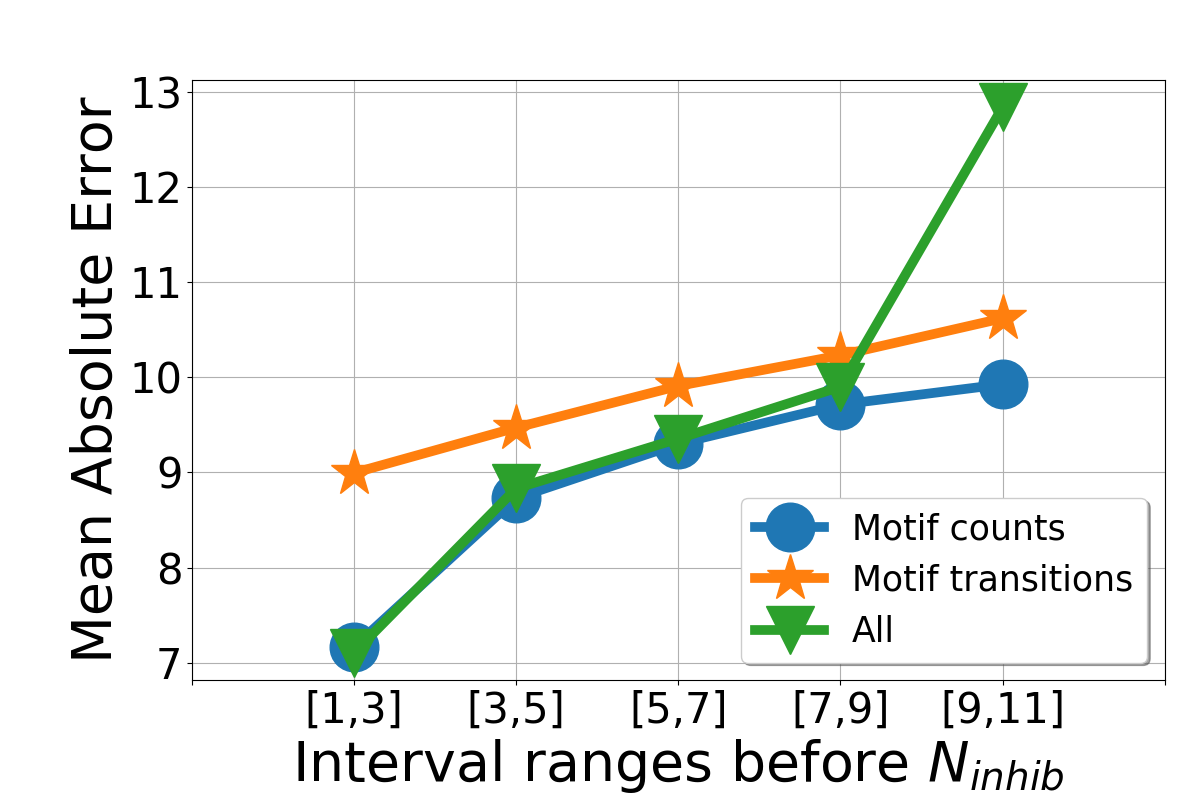}
		\subcaption{M3: \ \includegraphics[width=0.4cm]{M0.png}}
		\label{c}
	\end{minipage}
	\hfill
	\\
	\begin{minipage}{0.3\textwidth}%
		\includegraphics[width=5cm, height=3.5cm]{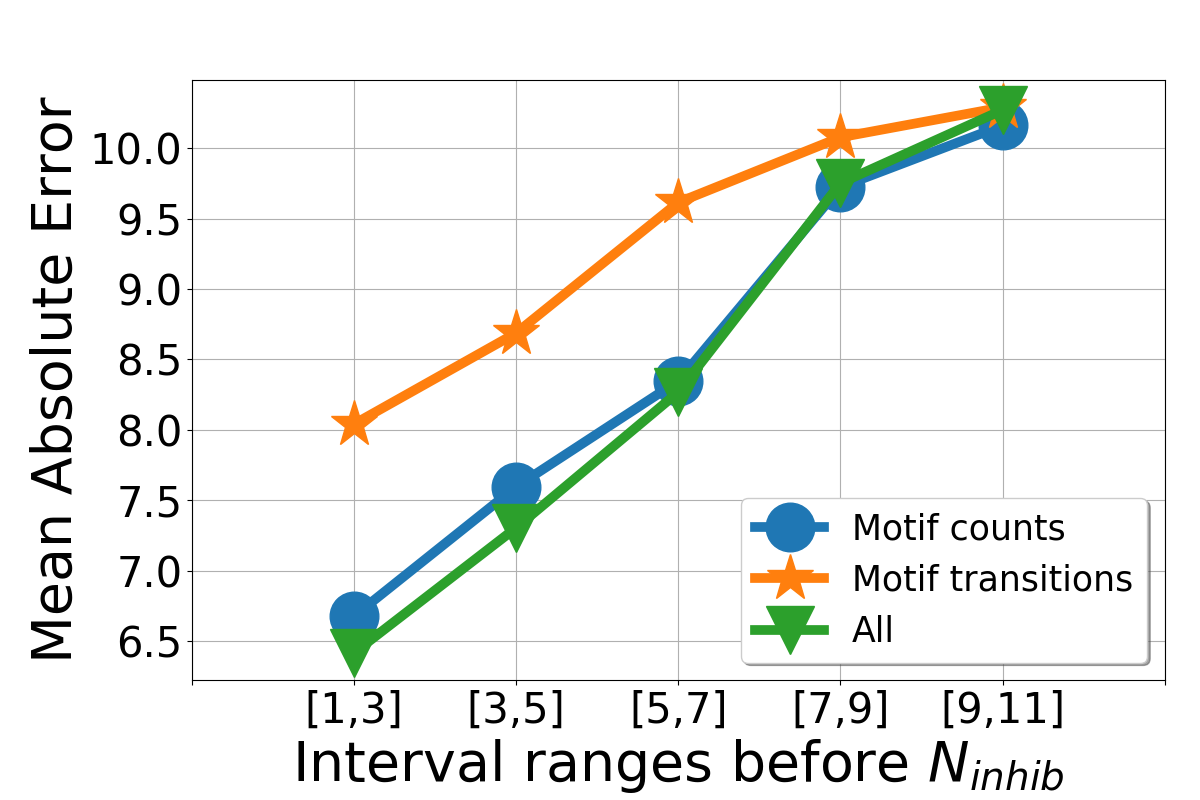}
		\subcaption{M5: \ \includegraphics[width=0.4cm]{M3.png}}
		\label{d}
	\end{minipage}
	\hfill
	\begin{minipage}{0.32\textwidth}
		\includegraphics[width=5cm, height=3.5cm]{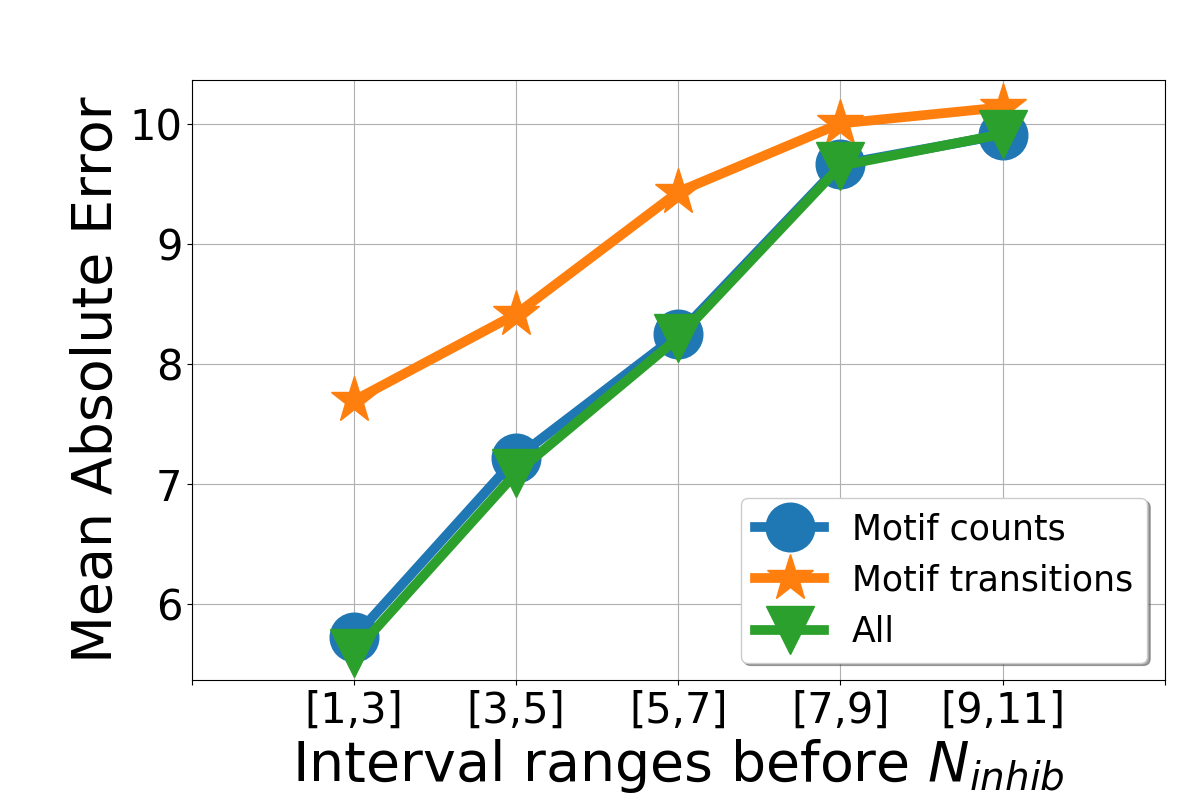}
		\subcaption{M6: \ \includegraphics[width=0.4cm]{M13.png}}
		\label{e}
	\end{minipage}
	\hfill
	\begin{minipage}{0.24\textwidth}
		\includegraphics[width=5cm, height=3.5cm]{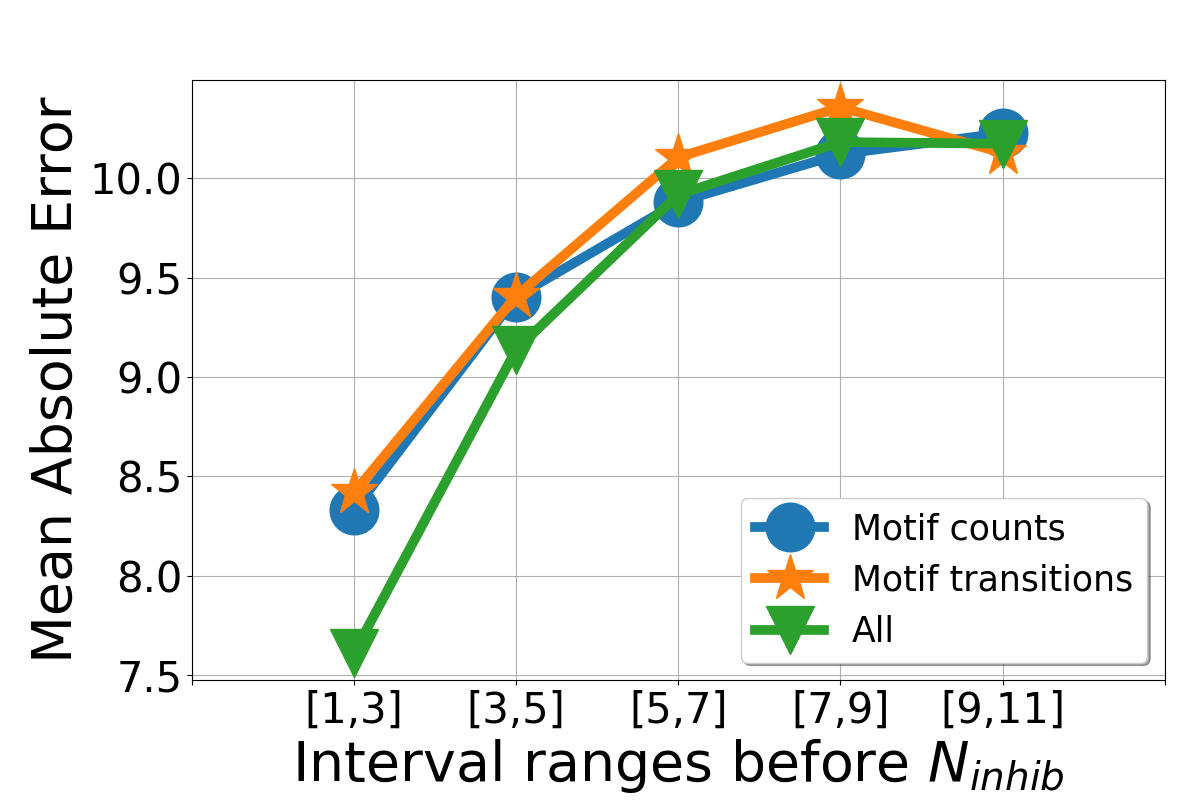}
		\subcaption{M11: \ \includegraphics[width=0.4cm]{M12.png}}
		\label{f}
	\end{minipage}
	\hfill
	\caption{Regression results for prediction problem $P1$ where we predict $|E^{N_{inhib}}|$, the number of edges in $G^{N_{inhib}}$.}
	\label{fig:edges_reg}
\end{figure}
\subsection{Baselines  for comparison}
To assess how useful are these motifs features in predicting $|E^{N_{inhib}}|$, we look at several network features that also study the network topology over time. Network centralities have been widely used to study the graph topology of social networks that ultimately govern the diffusion process in  social networks \cite{Kitsak10}.  We compute several centralities in-situ on these temporal networks instead of evaluating the historical networks or the social networks with the final goal of understanding how the node placements in the user networks affect the growth dynamics. We use the following centrality measures:

\begin{enumerate}
	\item Degree analysis: Nodal degree and Degree entropy
	\item Connectivity analysis: Clustering and Pagerank 
	\item Path analysis: Betweenness
\end{enumerate}

To use these centralities as features, we use the following procedure: we compute the centralities in each temporal network $N$ and consider the top 10 nodes ranked by the value of these measures. Thereon, for any given $N_i$, $i \in [1, \mathcal{Q}]$ formed by the networks in subsequences $\tau_i$ and $\tau_{i-1}$, we compute the mean of the measures with these 10 node values and use the mean as the feature for that interval $i$ in the cascade lifecycle. We use the same regression model as was used for the motifs, and evaluate the results based on  the MAE metric in Equation~\ref{eq:mae}. 

\begin{table}[!t]
	\centering
	\caption{Error values (MAE) for prediction of $|E^{N_{inhib}}|$. The lower values mean better performance. The values underlined indicate the top 15\% among all the models. $NA$ indicates that there were not significant number of cascades having that pattern for that feature.}
	\label{tab:edges_inhib}
	\begin{tabular}{|l|l|l|l|l|}
		\hline
		\textbf{Patterns} & \textbf{Edge density} & \textbf{Motif Counts} & \textbf{Motif Transitions}  & \textbf{All}   \\ \hline
		\includegraphics[scale=0.04]{M2.png}      & 0.4 & 8.01              & 9.2                              & \textbf{7.89}  \\ \hline
		\includegraphics[scale=0.04]{M1.png}      & 0.4 & 5.45                  & 7.26                         & \textbf{\underline{5.34}}  \\ \hline
		\includegraphics[scale=0.04]{M0.png}        & 0.4 & 7.22                & 8.98                            & \textbf{7.13}  \\ \hline
		\includegraphics[scale=0.04]{M8.png}       & 0.5 & 9.31                & 12.72                              & \textbf{10.17} \\ \hline
		\includegraphics[scale=0.04]{M3.png}        & 0.5 & 6.6        & 8.07                                        & \textbf{\underline{6.5}}          \\ \hline
		\includegraphics[scale=0.04]{M13.png}        & 0.5 & 5.82                 & 7.87                                         & \textbf{\underline{5.77}}  \\ \hline
		\includegraphics[scale=0.04]{M6.png}       & 0.5 & \textbf{11.88}        & 13.93                                       & 12.80          \\ \hline
		\includegraphics[scale=0.04]{M15.png}       & 0.5 & 7.52                  & 9.72                                        & \textbf{7.45}  \\ \hline
		\includegraphics[scale=0.04]{M7.png}       & 0.6 & 13.32                 & 16.73                                      & \textbf{13.19} \\ \hline
		\includegraphics[scale=0.04]{M18.png}       & 0.6 & 10.80                 & 12.62                                       & \textbf{10.43} \\ \hline
		\includegraphics[scale=0.04]{M12.png}       & 0.6 & 8.37                & 8.92                                       & \textbf{7.64} \\ \hline
		\includegraphics[scale=0.04]{M20.png}       & 0.6 & 16.38               & 18.25                      & \textbf{16.16}               \\ \hline
		\includegraphics[scale=0.04]{M4.png}        & 0.6 & 8.78                 & 9.77                        & \textbf{8.46} \\ \hline
		\includegraphics[scale=0.04]{M5.png}       & 0.7 & 12.78                & 13.68                                       & \textbf{12.30} \\ \hline
		\includegraphics[scale=0.04]{M17.png}        & 0.7 & 14.07                 & 14.16                               & \textbf{13.11}          \\ \hline
		\includegraphics[scale=0.04]{M9.png}      & 0.7 & 14.84                 & 16.69                         & \textbf{14.23} \\ \hline
		\includegraphics[scale=0.04]{M10.png}        & 0.7 & 15.02                 & 15.23                                  & \textbf{14.98} \\ \hline
		\includegraphics[scale=0.04]{M11.png}      & 0.8 & 11.47              & 13.67                                    & \textbf{11.18} \\ \hline
		\includegraphics[scale=0.04]{M19.png}       & 0.8 & \textbf{8.15}        & 9.36                                        & 8.16          \\ \hline
		\includegraphics[scale=0.04]{M14.png}       & 0.9 & \textbf{9.65}        & 12.21                                      & 10.03          \\ \hline
		\includegraphics[scale=0.04]{M16.png}       & 
		1.0 & 9.16         & NA                                         & \textbf{9.16}           \\ \hline
	\end{tabular}
\end{table}

\subsection{Evaluation Methods and Results}
We use 10-fold cross validation on the population of cascades for the prediction problems. For the LASSO term parameter in the regression cost function, we examine $\eta$ for all values in $\Lambda$ = $[0.01, 0.02, 0.03, 0.04]$ and for each model, we set $\eta $ to that value in $\Lambda$ which maximizes $R^2$, the coefficient of determination for the regression model \cite{Yuan06}. To evaluate the results of the prediction, we use \textit{mean absolute error(MAE)} as the performance metric defined as:

\begin{equation}
MAE = \frac{1}{|C|} \sum_{c=1}^{|C|} \left| |E_c| - \hat{|E_c|} \right|
\label{eq:mae}
\end{equation}

where the $|E_c|$ denotes the edges for $G^N$ and $|\hat{E_c}|$ denotes the estimated values from the regression models. Since all motif patterns do not occur in all intervals and in all cascades, especially the ones with higher densities and their appearance depends on the network structure, we replace the missing values of the features by the mean of the rest of the population as an imputation measure.\\

We perform 3 experimental evaluations on the prediction problem  and present the results from the intervals $[N_{inhib-1}, N_{inhib-2}]$ (and which gave the best results for all models) as follows: \\

\noindent \textbf{1. Individual Motifs:} We plot the results in Figure~\ref{fig:edges_reg} and we obtain the best results for the pattern \  \includegraphics[scale=0.04]{M1.png} \ with an MAE of 5.34, which is the best among all the models considered, reiterating the fact that the emergence of this pattern in the network process plays an important role in the evolution of the cascade network in terms of topology.  From Table~\ref{tab:edges_inhib}, we see that the best three results occur from the patterns \ \includegraphics[scale=0.04]{M1.png}, \ \includegraphics[scale=0.04]{M13.png} and  \ \includegraphics[scale=0.04]{M3.png} \ with MAE values of 5.34, 5.77 and 6.5 respectively. This also backs our previous observation of these patterns playing a prominent role in the way nodes reshare information. Furthermore, the performance drops for patterns with increasing edge densities as can be seen the increasing MAE values in Table~\ref{tab:edges_inhib}. Additionally, we also use a polynomial regression of order 2, to test the prediction results on individual motif patterns but we did not observe any significant improvement in terms of lower MAE. We present the results for the polynomial regression model in \textbf{Appendix A2}. \\

\begin{figure}[!t]
	\centering
	\hfill
	\begin{minipage}{0.4\textwidth}%
		\includegraphics[width=6cm, height=4cm]{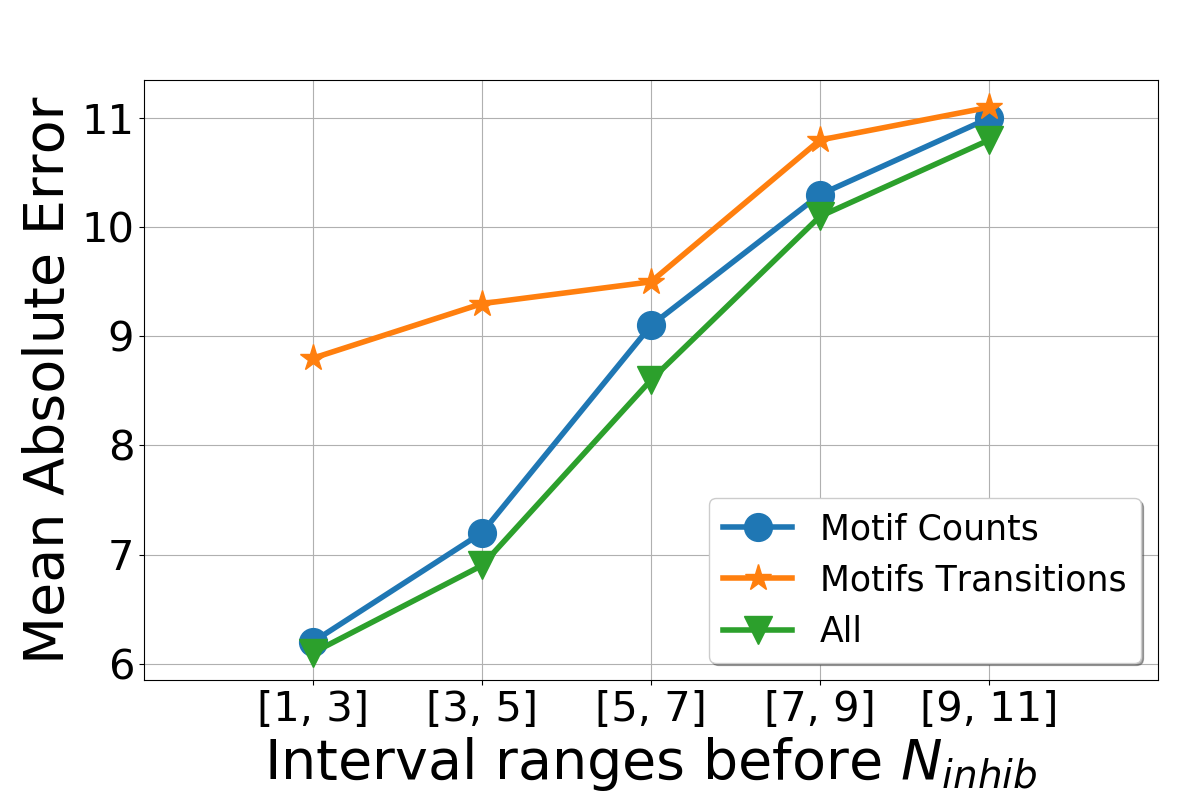}
		\subcaption{\ \includegraphics[width=0.4cm]{M1.png}, \ \includegraphics[width=0.4cm]{M2.png}, \ \includegraphics[width=0.4cm]{M0.png}}
		\label{a}
	\end{minipage}
	\hfill
	\begin{minipage}{0.4\textwidth}
		\includegraphics[width=6cm, height=4cm]{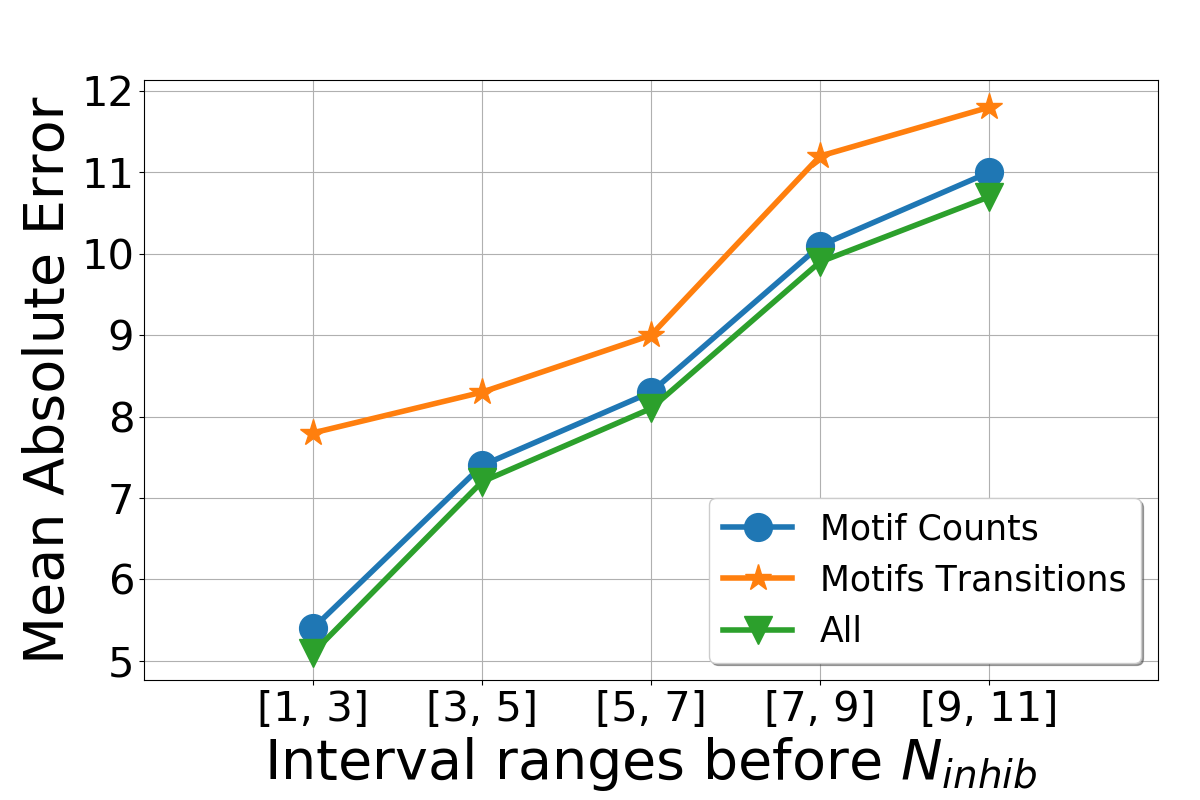}
		\subcaption{\ \includegraphics[width=0.4cm]{M13.png}, \ \includegraphics[width=0.4cm]{M15.png}, \ \includegraphics[width=0.4cm]{M16.png}}
		\label{b}
	\end{minipage}
	\hfill
	\caption{Regression results for prediction problem $P1$ where we predict $|E^{N_{inhib}}|$, the number of edges in $G^{N_{inhib}}$ using combinations of motifs.}
	\label{fig:motif_combine}
\end{figure}

\noindent \textbf{2. Motifs Combination:} Although it is not intuitively clear which motifs in combination would give the best prediction results, we perform an experiment in which we combine the 3 acyclic patterns here: \ \includegraphics[width=0.4cm]{M2.png}, \ \includegraphics[width=0.4cm]{M1.png} and \ \includegraphics[width=0.4cm]{M0.png}, and for the other patterns with loops, we consider the 3 best patterns which gave the best results: here we consider the patterns \ \includegraphics[width=0.4cm]{M13.png}, \ \includegraphics[width=0.4cm]{M15.png} and \ \includegraphics[width=0.4cm]{M16.png}. From the results shown in Figure~\ref{fig:motif_combine} we observe that for the set of acyclic patterns, we obtain the best results from the last couple of intervals with a MAE value of 6.1 considering both features together which is worse than  the best value of 5.34 obtained from the individual motifs. However, for the set of motifs with loops, we obtain the best MAE value of 5.15 considering all features and on the same last interval preceding $\tau_{inhib}$ which is better than the results from individual motifs. This suggests that the combination of motifs does help in improving the prediction results suggesting that these dense motifs can almost reconstruct the future temporal network edge cardinality ahead of time, although the best of combinations would require comparing all possible subsets which is not computationally feasible.  \\

\begin{figure}[!h]
	\centering
	\includegraphics[width=6cm, height=4cm]{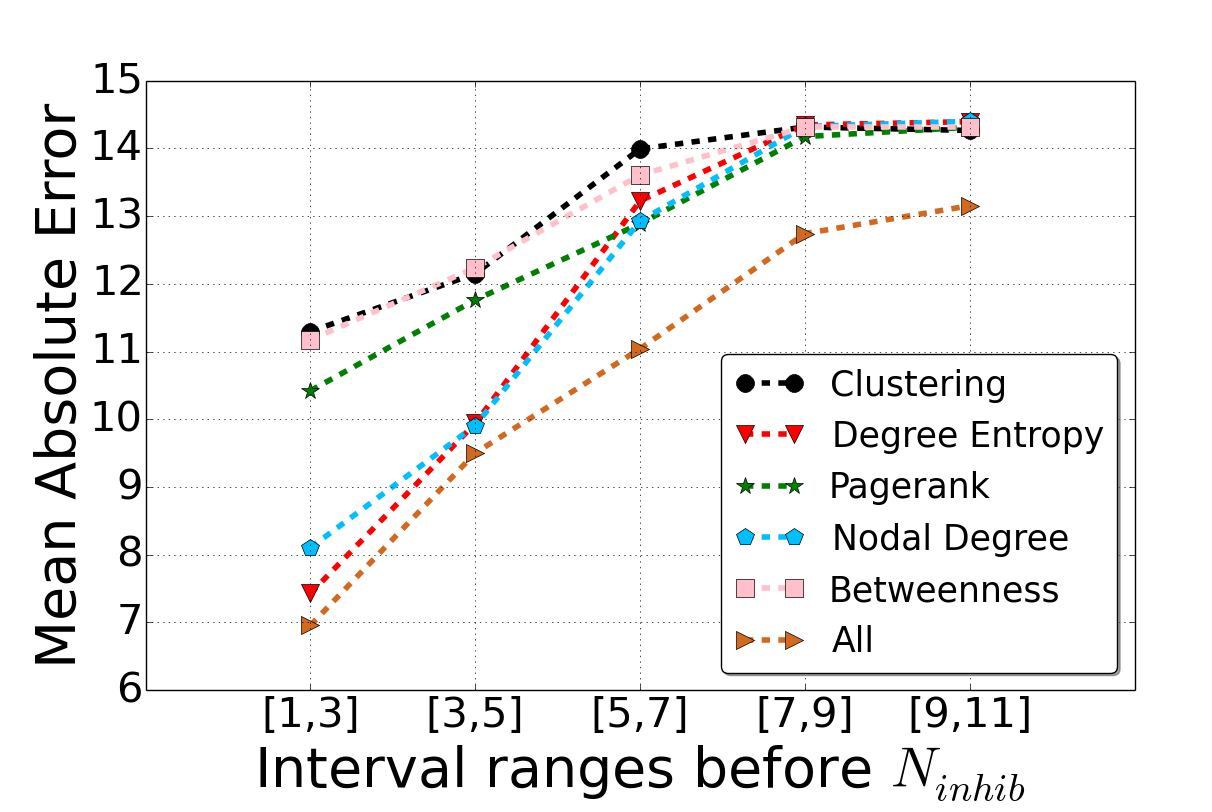}
	\caption{Regression results for prediction problem $P1$ where we predict $|E^{N_{inhib}}|$, the number of edges in $G^{N_{inhib}}$ using the centrality measures.}
	\label{fig:reg_cent}
\end{figure}

\noindent \textbf{3. Comparison with Centralities:} When we compare the results  obtained from the centrality features, we observe that the best results are obtained from the combination of all the features in the last interval preceding inhibition with an MAE value of 6.95. When we compare individual features, we find that the degree entropy measure and the nodal degree values have the best performance among the measures with MAE values of 7.42 and 8.10. Although the combination of motifs provide better results than the combination of these node features, the results from the motifs and the centralities hint at one direction: when evaluating the progress of the diffusion process through cascades and historical influence, it is better to look at the local network structure (the neighbors or 2 hop neighbors) surrounding a group of nodes instead of considering the entire network to assess the diffusive power, since we also observe that the betweenness measures and the clustering coefficient measures perform very poorly for this prediction problem.

\section{Experiments with Flixster movie ratings  data} \label{sec:flixster_exp}
To investigate whether the dynamics of motifs patterns hold in a different cascade setting, we use the Flixster dataset \cite{Rozenshtein:2016} that contains movie ratings by users over a period of time. The dataset comes along with the social network information of friends and a dataset containing which user rated a movie and at what time. Here we consider $G^D$, the historical network to be substituted by the social network which was available in Flixster. We consider cascades of movies where the information propagation process comprises how many users end up rating the movie. So the main idea is to see how users take up rating movies when their friends have already done so, and how does this reflect the popularity of the movie in terms of the number of ratings it received over time. We do not consider the ratings per se, but this is definitely a strong motivation to see what kind of motifs constitute positive ratings over time. However unlike the Wiebo resharing dataset, the ratings do not come with implicit information about who influenced whom to rate (thus the default cascade tree structure is not available). So we follow the lines of the work done in \cite{Rozenshtein:2016} to create a cascade in the following manner: for any user $u$ who has rated a movie at time $t_u$, we form the undirected cascade link $(u, v)$ if $t_u - t_v \leq 1$, that is if two users have rated the same movie within a period of 24 hours. In this case the cascade network does not exhibit the implicit tree structure.

We consider the period from January 2006 to December 2008 to gather all the movie ratings and we only retain those movies which have greater than 300 ratings over that time period. We obtained a total of 1799 cascades of movies and we show the statistics of the social network and the cascades in Table~\ref{tab:table_flixster}.

\begin{table}[!t]
	\centering
	\renewcommand{\arraystretch}{1}
	\caption{Properties of Social (Friends) Network and Cascades}
	\begin{tabular}{|p{5cm}|p{4cm}|}
		\hline 
		{\bf Properties} & {\bf Social Network}\\ 
		\hline\hline
		Vertices           & 7,86,936  \\
		\hline
		Edges & 58, 97, 324  \\
		\hline 
		Average Degree & 14.98  \\       
		\hline
	\end{tabular}
	\begin{tabular}{|p{5cm}|p{4cm}|}
		\hline 
		{\bf Properties} & {\bf Rating Network}\\ 
		\hline\hline
		Number of cascades (movies) & 48794 \\
		\hline
		Number of cascades (2006-2008) & 1799 \\
		\hline
	\end{tabular}
	\label{tab:table_flixster}
\end{table}
The number of cascades in total are significantly lesser than the resharing dataset due to the fact that they are limited by the number of movies in the Flixster database. The social network statistics in terms of number of edges and nodes are an order of magnitude lesser than the Wiebo dataset. The average degree of users in the Flixster friends network is 14.98 compared to 18.02 for users in the historical diffusion network of Wiebo suggesting a denser historical network for Wiebo. 
\begin{figure}[!t]
	\centering
	\hfill
	\begin{minipage}{0.48\textwidth}%
		\includegraphics[width=6cm, height=4cm]{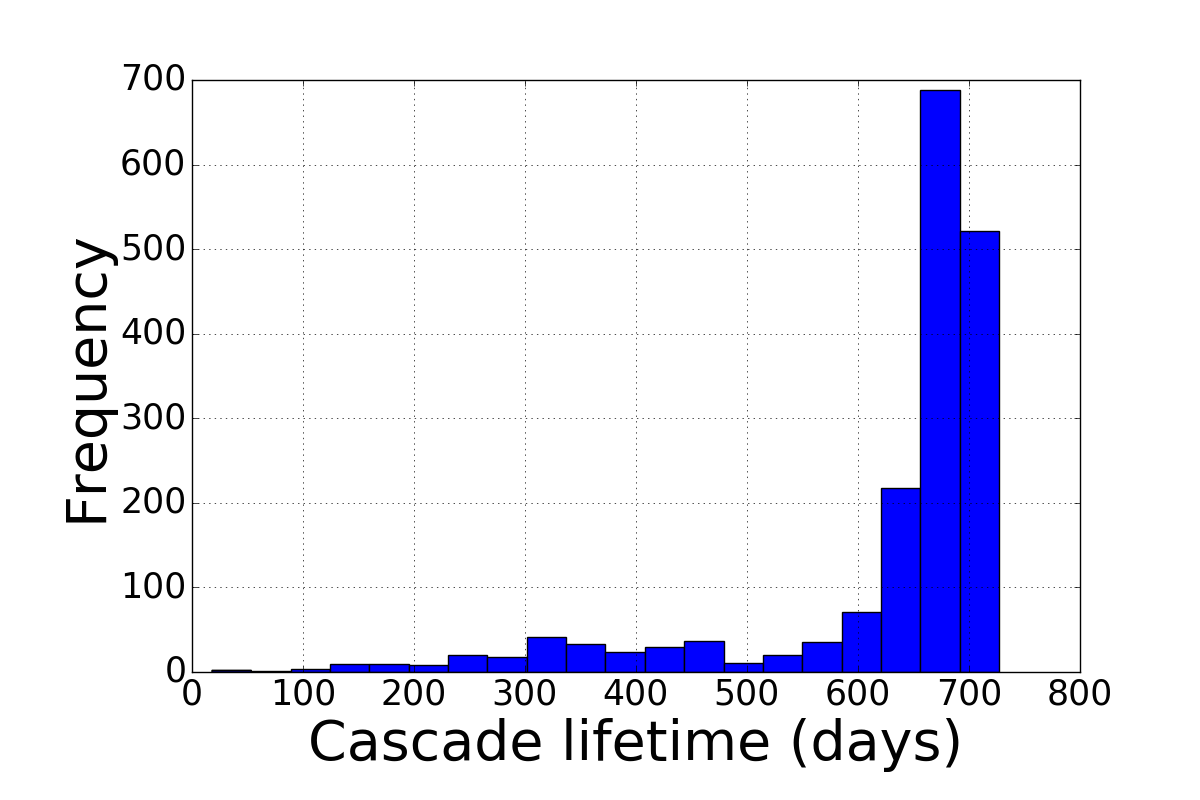}
		\subcaption{}
		\label{hist_net_flix:a}
	\end{minipage}
	\hfill
	\begin{minipage}{0.4\textwidth}
		\includegraphics[width=6cm, height=4cm]{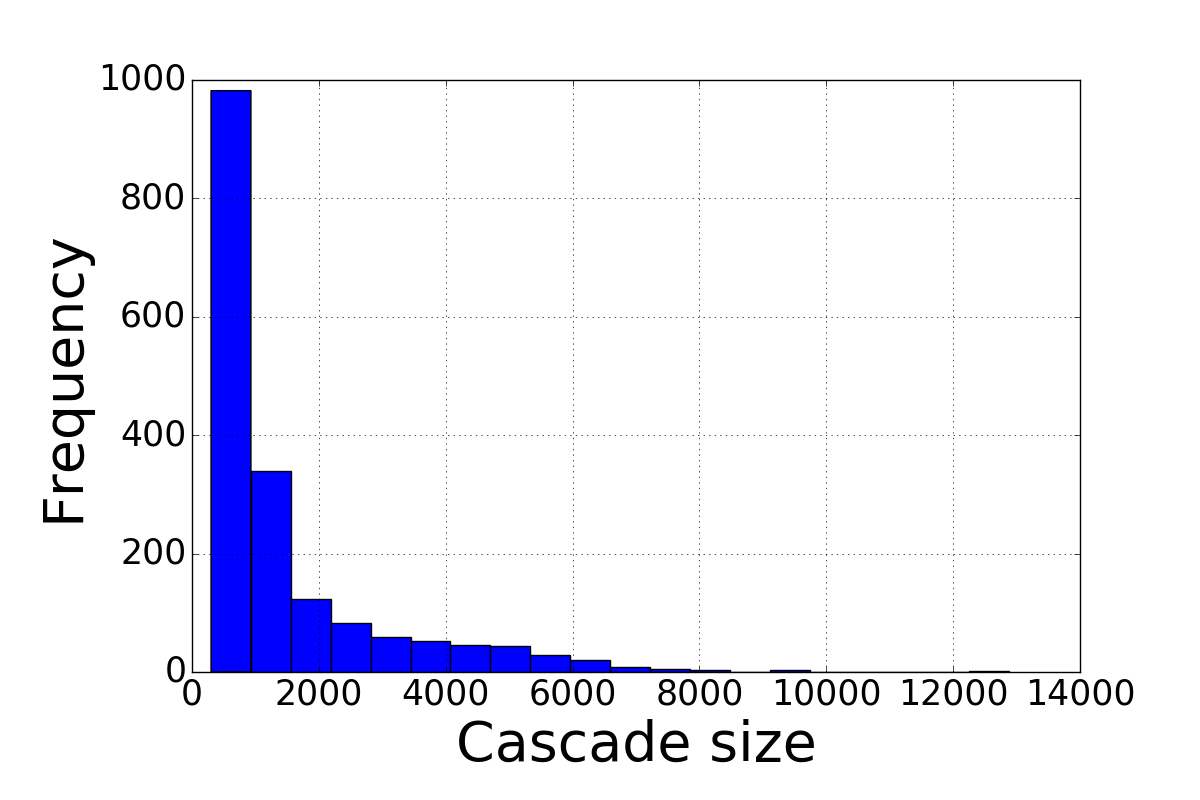}
		\subcaption{}
		\label{hist_net_flix:b}
	\end{minipage}
	\hfill
	\caption{Histogram of (a) Cascade lifetimes in minutes (b) Cascade size for Flixster cas}
	\label{fig:hist_net_flixster}
\end{figure}
We observe from Figure~\ref{fig:hist_net_flixster} \subref{hist_net_flix:a} that the distribution of cascade lifetime in this case is different than compared to the Wiebo dataset due to different dynamics between a resharing cascade and a rating cascade. On the other hand, we find that the cascade size distribution remains the same. This suggests that the parameters to split the cascade into windows would change due to much larger lifetimes of the rating cascades.

\begin{figure}[!t]
		\centering
		\begin{minipage}{0.3\textwidth}%
			\includegraphics[width=5cm, height=3.7cm]{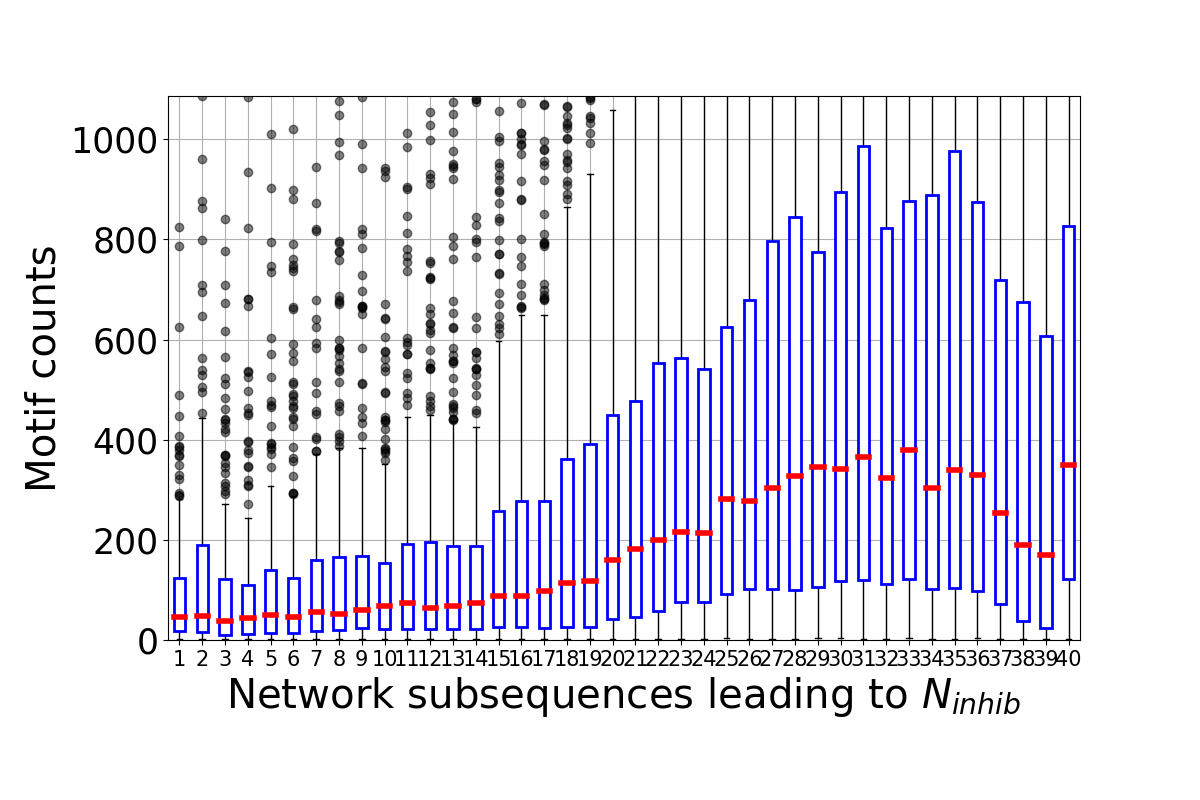}
			\subcaption{M1: \ \includegraphics[width=0.4cm]{M2.png}}
			\label{(a)}
		\end{minipage}
		\hfill
		\begin{minipage}{0.305\textwidth}
			\includegraphics[width=5cm, height=3.5cm]{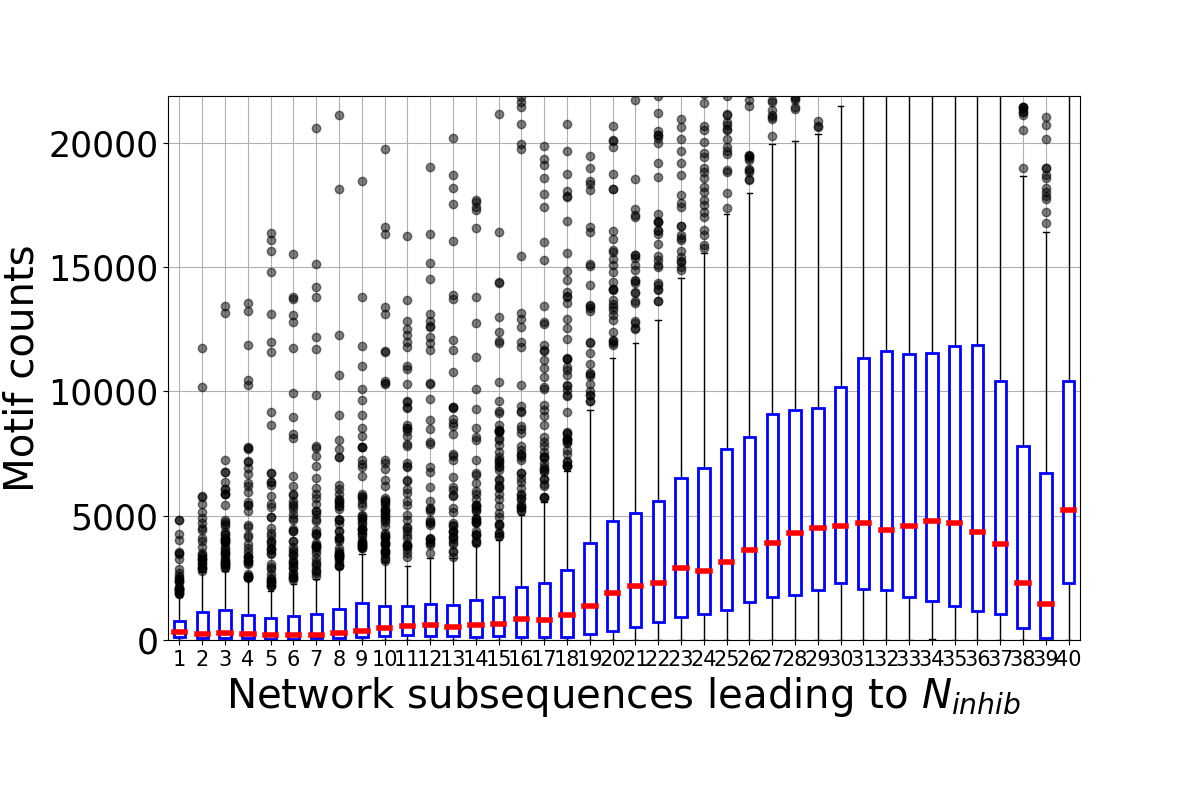}
			\subcaption{M2: \ \includegraphics[width=0.4cm]{M1.png}}
			\label{(b)}
		\end{minipage}
		\hfill
	\begin{minipage}{0.3\textwidth}
		\includegraphics[width=5cm, height=3.7cm]{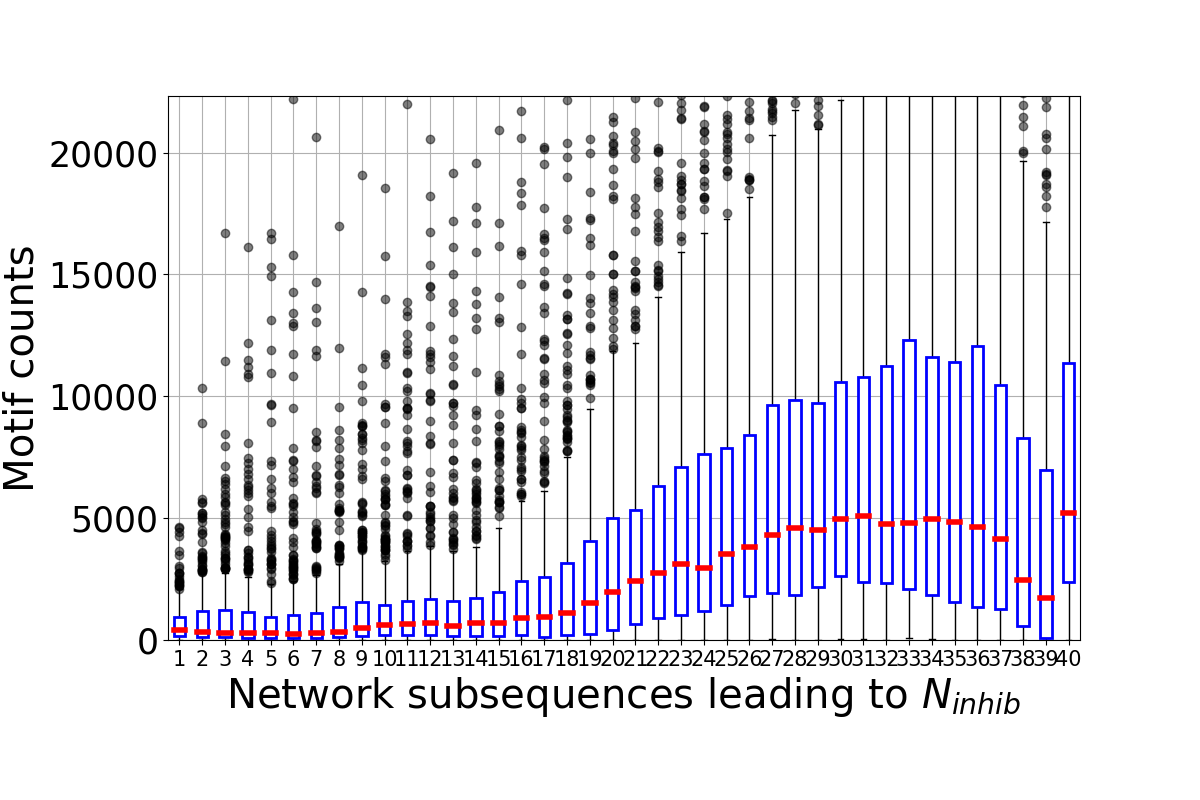}
		\subcaption{M3: \ \includegraphics[width=0.4cm]{M0.png}}
		\label{(c)}
	\end{minipage}
	\hfill
	\\
	\begin{minipage}{0.3\textwidth}%
		\includegraphics[width=5cm, height=3.7cm]{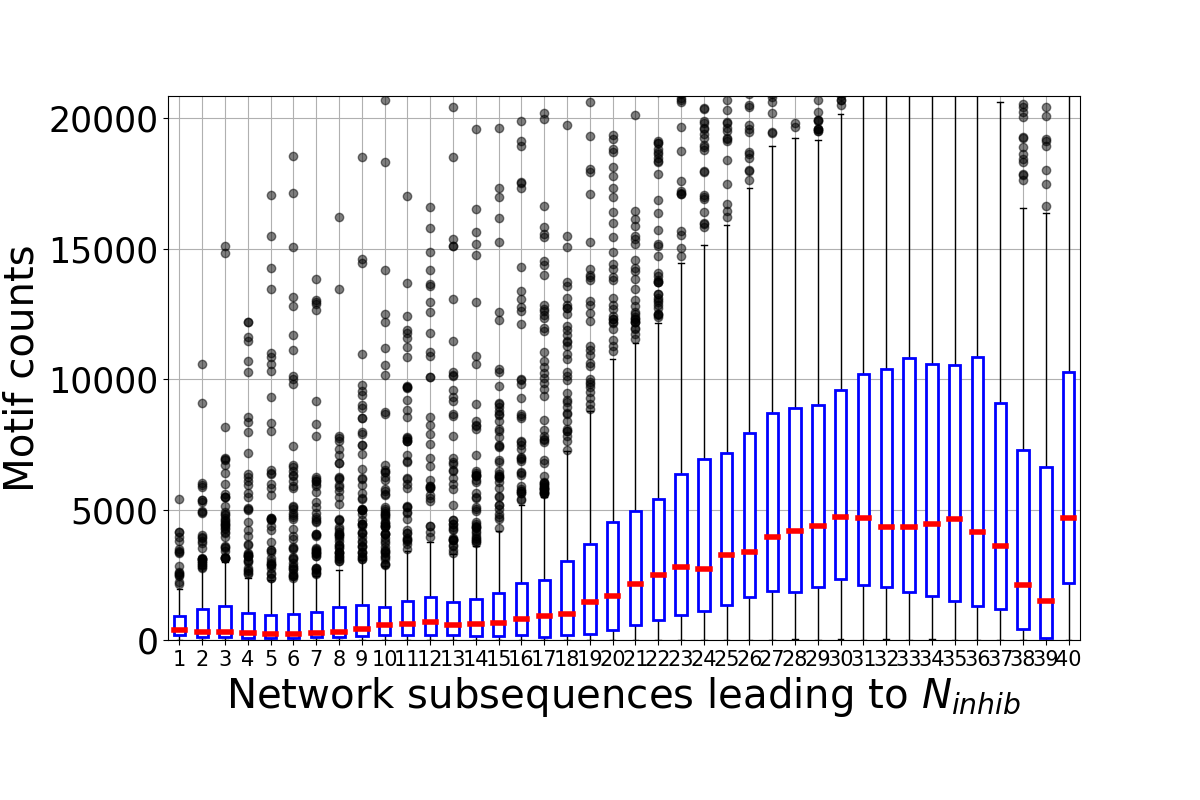}
		\subcaption{M5: \ \includegraphics[width=0.4cm]{M3.png}}
		\label{(d)}
	\end{minipage}
	\hfill
	\begin{minipage}{0.305\textwidth}
		\includegraphics[width=5cm, height=3.7cm]{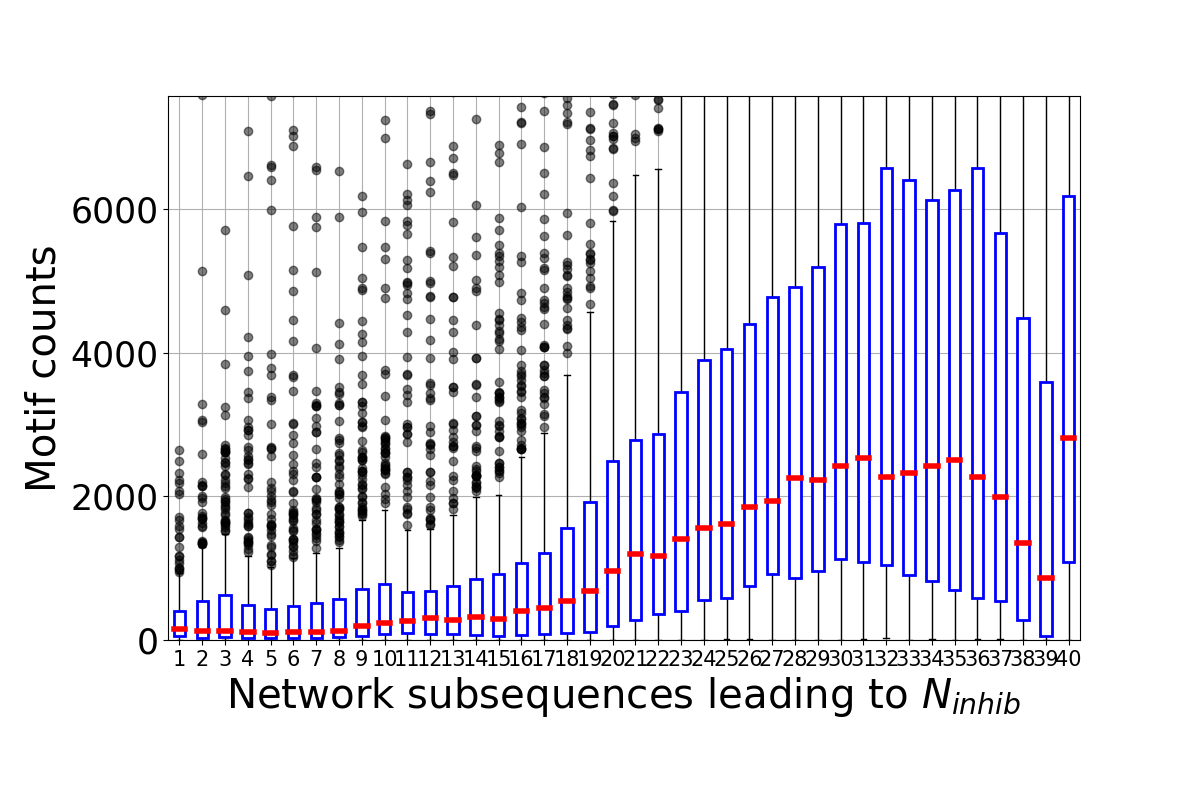}
		\subcaption{M6: \ \includegraphics[width=0.4cm]{M13.png}}
		\label{(e)}
	\end{minipage}
	\hfill
	\begin{minipage}{0.3\textwidth}
		\includegraphics[width=5cm, height=3.7cm]{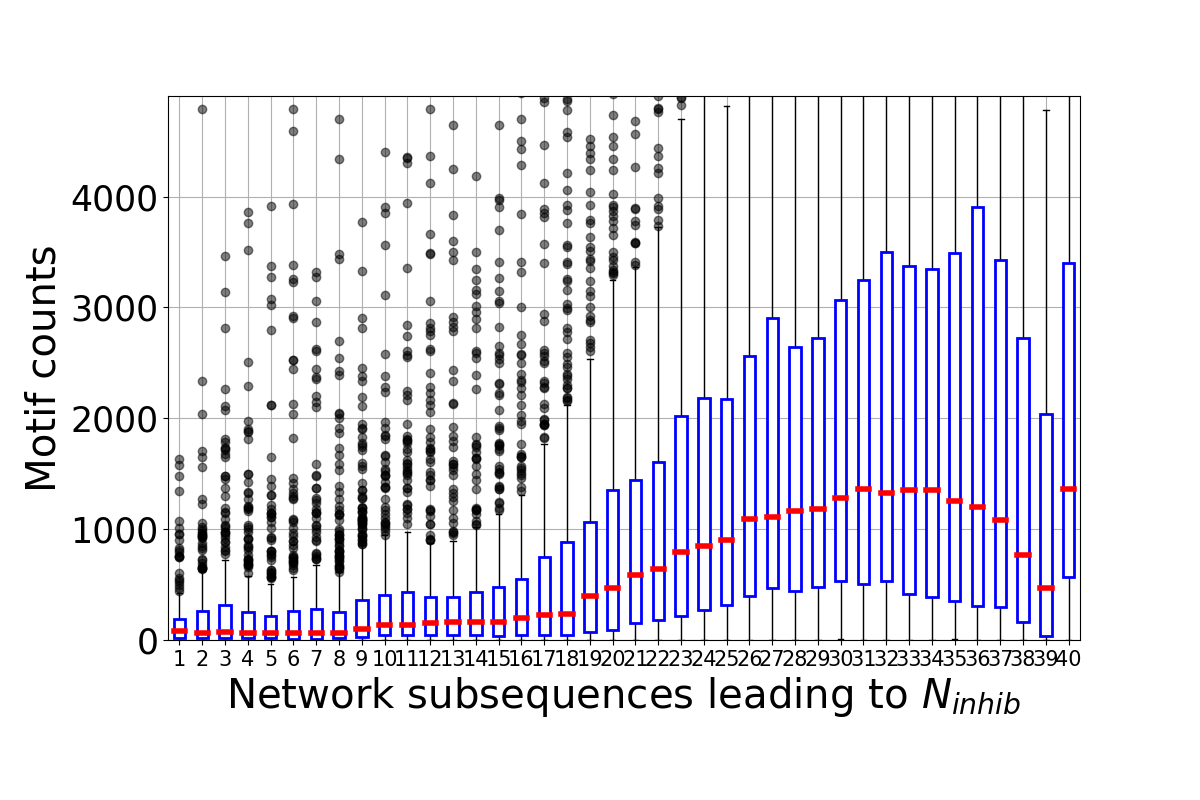}
		\subcaption{M11: \ \includegraphics[width=0.4cm]{M12.png}}
		\label{(f)}
	\end{minipage}
	\hfill
	\caption{Plots of motif counts for the Flixster dataset cascades. The intervals are increasing from left to right leading to $N_{inhib}$. }
	\label{fig:mcounts_flixster}
\end{figure}

We present the motif counts $MC$ for the patterns shown in Figure~\ref{fig:mcounts_flixster} and while we observe that the general dynamics of the counts for the patterns are similar to the patterns from the microblogging platform, we find the following notable differences: 
\begin{itemize}
	\item Unlike in the Weibo dataset where the pattern \ \includegraphics[scale=0.04]{M0.png} \ dominates the other patterns in terms of an order of magnitude higher proportion of motifs, we find that the pattern \includegraphics[scale=0.04]{M0.png}  \   shares similar dynamics with \includegraphics[scale=0.04]{M1.png}  \ and \includegraphics[scale=0.04]{M3.png}  \ in terms of the distribution of motif counts towards the inhibition period. The similar dynamics for  \includegraphics[scale=0.04]{M0.png}  \   and \includegraphics[scale=0.04]{M3.png}  \ suggests that the appearance of \includegraphics[scale=0.04]{M0.png} is heavily dependent on the presence of the pattern \includegraphics[scale=0.04]{M3.png} \ because of the way motifs are counted. This also suggests that the exposure effect is more prominent here and one of the reasons behind this is that here the default tree structure of the cascades no longer hold because of the way we construct cascades from the sequential ratings. Since one node can have multiple parents (within a days' rating period), the chances of being exposed to a rating is higher in such cascade scenarios.
	
	\item Secondly, we find that for the results for predicting $E^{N_{inhib}}$ shown in Table~\ref{tab:edges_inhib_flixster}, the MAE values are higher than the Weibo dataset, an this can be attributed to a denser network $G^{N_{inhib}}$ for FLixster data and higher variance in the edge cardinality since now each user has multiple parents. But as opposed to the \includegraphics[scale=0.04]{M1.png} \ shape which had the lowest MAE in Weibo dataset, we find that \includegraphics[scale=0.04]{M3.png} \  exhibits the lowest MAE (best performance) for the Flixster data. One of the reasons behind this observation is that the people tend to rate movies soon after their friends have watched and rated a movie - so in a sense the external stimuli for ratings is driven by what friends have seen and rated and even though personal opinions may vary, the will to rate is a factor of the social network.
\end{itemize}

\begin{table}[!h]
	\centering
	\caption{Error values (MAE) for prediction of $|E^{N_{inhib}}|$ on the Flixster cascades. The lower values mean better performance. }
	\label{tab:edges_inhib_flixster}
	\begin{tabular}{|l|l|l|l|l|}
		\hline
		\textbf{Patterns} & \textbf{Edge density} & \textbf{Motif Counts} & \textbf{Motif Transitions}  & \textbf{All}   \\ \hline
		\includegraphics[scale=0.04]{M2.png}      & 0.4 & 22.36                & 23.79                                 & 21.33  \\ \hline
		\includegraphics[scale=0.04]{M1.png}      & 0.4 & 16.29              & 18.26                            & 15.31  \\ \hline
		\includegraphics[scale=0.04]{M0.png}    & 0.4 & 16.93                 & 19.82                            & 16.28  \\ \hline
		\includegraphics[scale=0.04]{M3.png}      & 0.5 & 14.58                 & 16.92                        & \textbf{14.33} \\ \hline      
		\includegraphics[scale=0.04]{M13.png}        & 0.5 & 17.32                  & 17.89                                           & 17.77 \\ \hline
		\includegraphics[scale=0.04]{M12.png}       & 0.6 & 22.83                 & 23.98                                        & 21.36  \\ \hline
	\end{tabular}
\end{table}

\section{Conclusions and Future Work}
Using subgraphs to characterize static networks has been widely used before \cite{Sarajlic16,Katona11,Zhao2011,Leskovec2006} albeit not much for information diffusion in a temporal setting.  In this paper, we used network motifs in undirected graphs to assess the interaction patterns in cascades. Towards the end of the cascade lifecycle, we find that individuals have a tendency to reshare from high degree nodes instead of interacting in a way that would form large polygons. We find that the patterns with the highest count might not necessarily be the ones to have the best predictive power when forecasting network attributes. One of the areas where this work on motifs could be extended is by considering directed motifs and only considering specific patterns similar to the FFL and SIM motifs. For the problem of diffusion inhibition, one could see how these directed motifs influence the information diffusion process by considering latent variable models and measuring the underlying survival capability of these motifs. The one aspect of the network which has been ignored in this study has been the weight of the edges in the network, specifically to see how the effect of historical diffusion links in motifs vary from the cascade propagation links for its progress. That could also help us observe the extent to which the previous diffusion drives the current cascade progress. In the context of information diffusion which is the basis of this research, network motifs can help address the following:
\begin{enumerate}
	\item Seeking provenance of misinformation in social media can be a real challenge \cite{misinformation} since the misinformation can unintentionally diffuse or spread  like any other ordinary information involving well-meaning users until the inaccurate aspects catch the attention. Using network motifs, one can learn which kind of patterns are recurring significantly early on that are representative of such misinformation and that would help segment out the groups of users who are part of these motif profiles. These would help us to track the significant groups which cause these misinformation to spread. 
	
	\item Similarly, for commercial platforms like LinkedIn, where recommendation systems enable users to features like "People You May Know", significant subgraphs like motifs in content resharing diffusion can help understand the preferences of specific individuals towards other people to whom they are not connected. That way, they can recommend a specific set of users based on the motifs individuals engage in, instead of recommending all users with whom they share other forms of communication like emails exchanges or social media platforms.
	
	\item As mentioned before, one of the hypothesis we test in this paper revolves around the aspect of cascade adoption based on influence and adoption based on popularity or individual preferences. Specifically, we try to see whether motifs can explain the ways in which people reshare messages in the cascade and which ultimately paves the way towards inhibition.
\end{enumerate}

\noindent \textbf{Acknowledgements} Some of the authors are supported through the AFOSR Young Investigator Program (YIP) grant FA9550-15-1-0159, ARO grant W911NF-15-1-0282, and the DoD Minerva program grant N00014-16-1-2015.




\end{document}